\documentclass[useAMS,usenatbib,letterpaper]{mn2e} 

\usepackage{times}
\usepackage{graphicx}
\usepackage{amssymb}

\newcommand{\HI}{H\,{\sevensize I}}

\newcommand{\mNHI}{N\rm _{H\,{\sevensize I}}}
\newcommand{\NHI}{$N\rm _{H\,{\sevensize I}}$}
\newcommand{\NH}{$N\rm _H$}

\newcommand{\HeII}{He\,{\sevensize II}}
\newcommand{\SiII}{Si\,{\sevensize II}}
\newcommand{\SiIIl}{Si\,{\sevensize II}$\lambda$}

\newcommand{\lya}{Ly$\alpha$}
\newcommand{\fnx}{$f(N_{\rm \HI},X)$}

\newcommand{\lox}{$\ell(X)$}

\newcommand{\sfr}{M$_{\odot}$ yr$^{-1}$}
\newcommand{\cmm}{cm$^{-2}$}
\newcommand{\msun}{M$_\odot$}
\newcommand{\SNII}{SN\,{II}}
\newcommand{\SNIa}{SN\,{Ia}}
\newcommand{\rvir}{$R_{\rm vir}$}

\newcommand{\ngal}{7}

\title{Absorption line systems in simulated galaxies fed by cold streams}

\author[Fumagalli et al.]{Michele Fumagalli,$^{1}$\thanks{E-mail: mfumagalli@ucolick.org} 
J. Xavier Prochaska,$^{1,2}$ Daniel Kasen,$^{3,4}$
\newauthor 
Avishai Dekel,$^{5}$ Daniel Ceverino,$^{5}$ and Joel R. Primack$^{6}$\\
$^{1}$Department of Astronomy and Astrophysics, University of California, 
1156 High Street, Santa Cruz, CA 95064\\ 
$^{2}$UCO/Lick Observatory, University of California, 
1156 High Street, Santa Cruz, CA 95064\\
$^{3}$Department of Physics, University of California,
366 LeConte, Berkeley, CA 94720\\
$^{4}$Nuclear Science Division, Lawrence Berkeley National Laboratory,
Berkeley, CA 94720\\
$^{5}$Racah Institute of Physics, The Hebrew University,
Jerusalem 91904, Israel\\
$^{6}$Department of Physics, University of California, 
1156 High Street, Santa Cruz, CA 95064.}

\begin{document}

\date{Accepted 2011 August 08. Received 2011 June 29; in original form 2011 March 09}

\pagerange{\pageref{firstpage}--\pageref{lastpage}} \pubyear{xxxx}

\maketitle

\label{firstpage}

\begin{abstract}
Hydro cosmological simulations reveal that massive galaxies at high redshift 
are fed by long narrow streams of merging galaxies and a smoother component of 
cold gas.  We post-process seven high-resolution simulated galaxies with 
radiative transfer to study the absorption characteristics of the gas in 
galaxies and streams, in comparison with the statistics of observed 
absorption-line systems. We find that much of the stream gas is ionized by UV 
radiation from background and local stellar sources, but still optically thick 
($N_{\rm HI}> 10^{17}$ \cmm) so that the streams appear as Lyman-limit systems (LLSs). 
At $z>3$, the fraction of neutral gas in streams becomes non-negligible, 
giving rise to damped Lyman-$\alpha$
(DLA) absorbers as well. The gas in the central and incoming galaxies remains 
mostly neutral, responsible for DLAs.  Within one (two) virial radii, the 
covering factor of optically thick gas is $<25\%$ (10\%) for LLSs and $<5\%$ 
(1\%) for DLAs, slowly declining with time following the universal expansion. 
Nevertheless, galaxies and their cold streams in the studied mass range, 
$M_{\rm vir}=10^{10}-10^{12}$ \msun, account for $>30\%$ of the observed 
absorbers in the foreground of quasars, the rest possibly arising from smaller 
galaxies or the intergalactic medium. The mean metallicity in the streams is 
$\sim1\%$ solar, much lower than in the galaxies. The simulated galaxies 
reproduce the Ly$\alpha$-absorption equivalent widths observed around 
Lyman-break galaxies, but they severely underpredict the equivalent widths 
in metal lines, suggesting that the latter may arise from outflows.  
We conclude that the observed metal-poor LLSs are likely detections of 
the predicted cold streams. Revised analysis of the observed LLSs kinematics 
and simulations with more massive outflows in conjunction with the inflows 
may enable a clearer distinction between the signatures of the various gas 
modes.
\end{abstract}


\section{Introduction}

Cold neutral hydrogen that fuels star formation is an important driver 
of galaxy evolution at all times. Recent progress in observations and simulations
has remarkably improved our view of gas in the early Universe. 
Molecular gas in emission is detected in normal star forming galaxies to $z\sim2.3$ 
\citep{tac10,dad10}. Large imaging and spectroscopic surveys have increased by orders of 
magnitude the samples of quasars and galaxy pairs 
useful for probing in absorption the interstellar medium (ISM), the circumgalactic 
medium (CGM) and the intergalactic medium (IGM) \citep[e.g.][]{ade03,hen06,pro09,not09,pro10,ste10}.
Similarly, the increased resolution of numerical simulations has made
it possible to reproduce the morphology of galaxies especially at high redshift 
\citep[e.g.][]{age09,dek09b,cev10}.

Significant progress is being made on the basic issue of how gas is accreted 
from the IGM into galaxies. A consistent theoretical picture is emerging from 
both analytic theory \citep{bir03,dek06} and numerical simulations 
\citep{ker05,ocv08,ker09b,dek09}. High-redshift galaxies are predicted to acquire 
the bulk of their gas mass at temperatures of a few $10^4$ K via long narrow streams 
that consist of a smooth component and small merging galaxies. 
However, indisputable observational evidence for this predicted dominant mode 
of gas input is still lacking.
 
Ly$\alpha$ emission due to cooling radiation has been proposed 
as a signature of cold gas accretion that powers Ly$\alpha$ blobs 
\citep[e.g.][]{hai00,far01,fur05,dij09,goe10,mat10}, but the resonant nature of the Ly$\alpha$ line 
and the different physical mechanisms that influence the total Ly$\alpha$ 
luminosity \citep[][Kasen et al., in prep.]{fau10} prevent us from unambiguously connecting the 
observed fluxes with the presence of cold streams. Observations of gas 
in absorption against background sources are, in principle, ideal to probe the CGM 
to uncover infalling gas. However, it has been suggested that the low covering factor of 
the cold filaments reduces the probability of detecting them in absorption \citep{fau10b,stw10},
especially in the presence of large scale outflows \citep{ste10}.

The subject of the present paper is the problem of detecting streams of cold 
gas in absorption against background sources.
We formulate detailed theoretical predictions
of the covering factor, the cross section, the column density distribution, the kinematics, and 
the metallicity of the neutral gas in seven galaxies, simulated with 
a high-resolution AMR code in a cosmological context and analyzed in the redshift range
$z=1.4-4$. To facilitate the comparison between our theoretical predictions and 
observations, we adopt the same methodology used in observational
studies of absorption line systems (ALSs) in background quasars and galaxy pairs. 
This work focus on the visibility of cold stream and by 
comparing and contrasting these predictions with observations, we aim to 
find the observables that are best suited to probe this mode of accretion.
A companion paper (Kasen et al., in prep.) will focus on the Ly$\alpha$ emission 
properties using the same simulations.

The paper consists of three parts. In the first two sections, we briefly 
discuss the numerical simulations (Section \ref{sec:sim}) and the radiative transfer 
calculation used to estimate the hydrogen ionization state (Section \ref{sec:rt}). 
This part is complemented with Appendix \ref{rtcomp}, 
where we provide detailed comparisons of 
the neutral hydrogen masses and column density distributions from 
three different radiative transfer models.
In the second part, we quantify the neutral hydrogen cross section 
and covering factor as functions of the gas column density and redshift 
(Section \ref{sec:crsec}). In the third part, we compare more directly model 
predictions with observations, focusing in particular on the column density distribution and incidence of 
absorption line systems (Section \ref{alsingal})  as well as their metallicity and  kinematics
(Section \ref{alsingal2}). A summary and conclusions follow in Section \ref{sec:sumcon}. 

Throughout this work, we assume the WMAP5 values of the cosmological 
parameters \citep{kom09}, 
consistent with the cosmology adopted in the simulations 
($\Omega_{\rm m}=0.27$, $\Omega_{\Lambda}=0.73$ and $H_{\rm 0}=70$ km s$^{-1}$ Mpc$^{-1}$).


\begin{figure*}
\begin{tabular}{c c}
  \includegraphics[scale=0.3,angle=90]{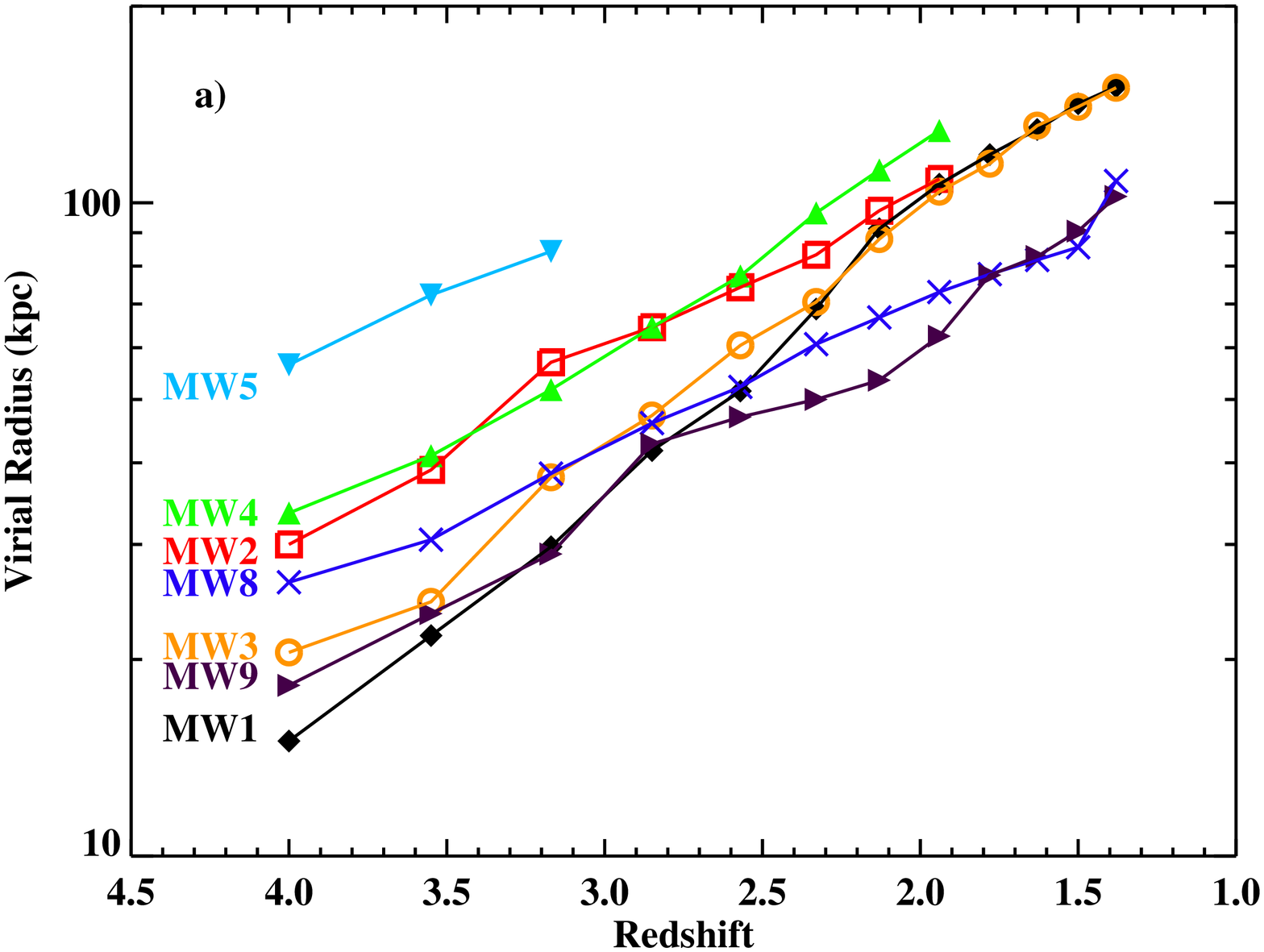}&
  \includegraphics[scale=0.3,angle=90]{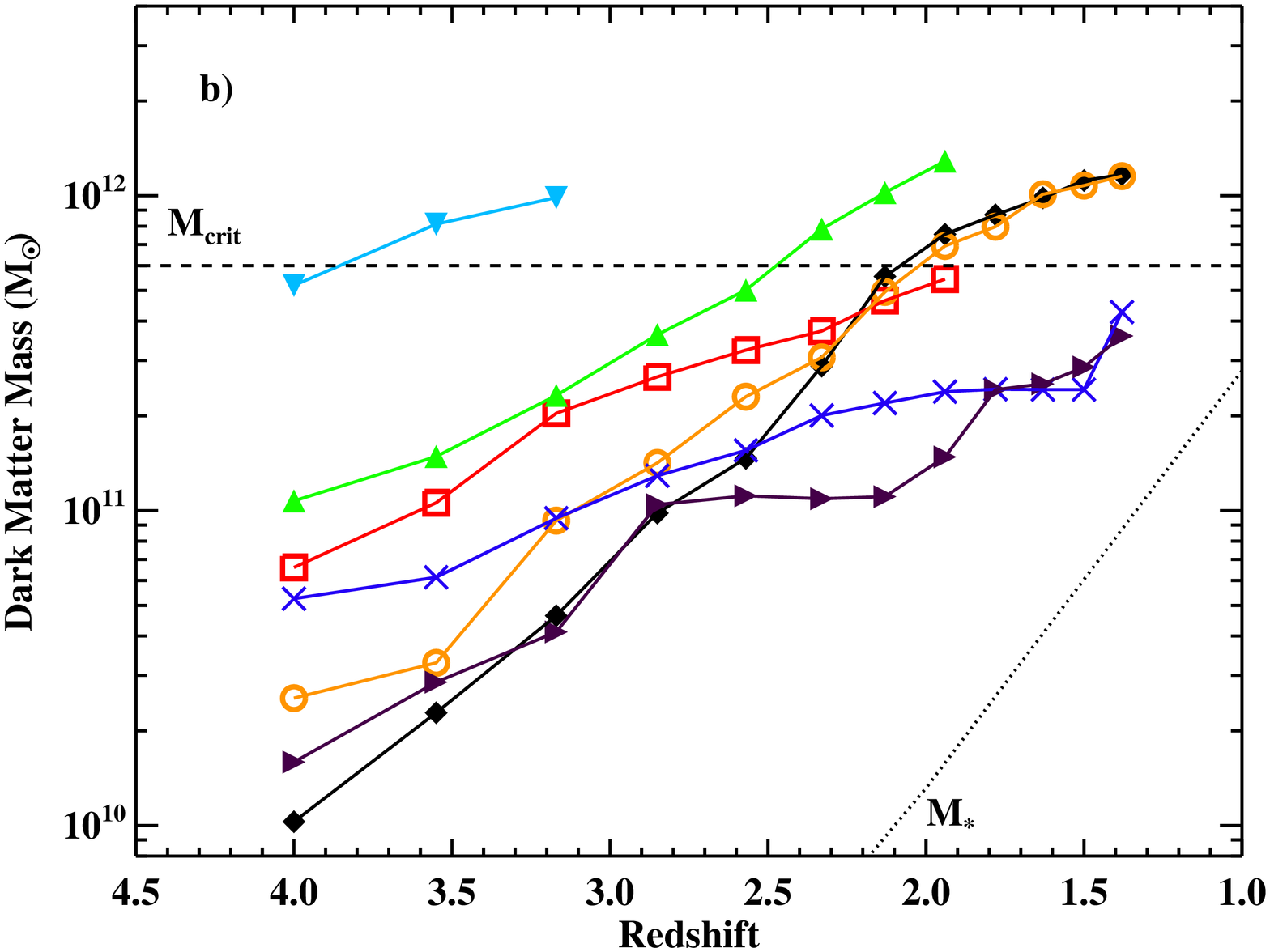}\\
  \includegraphics[scale=0.3,angle=90]{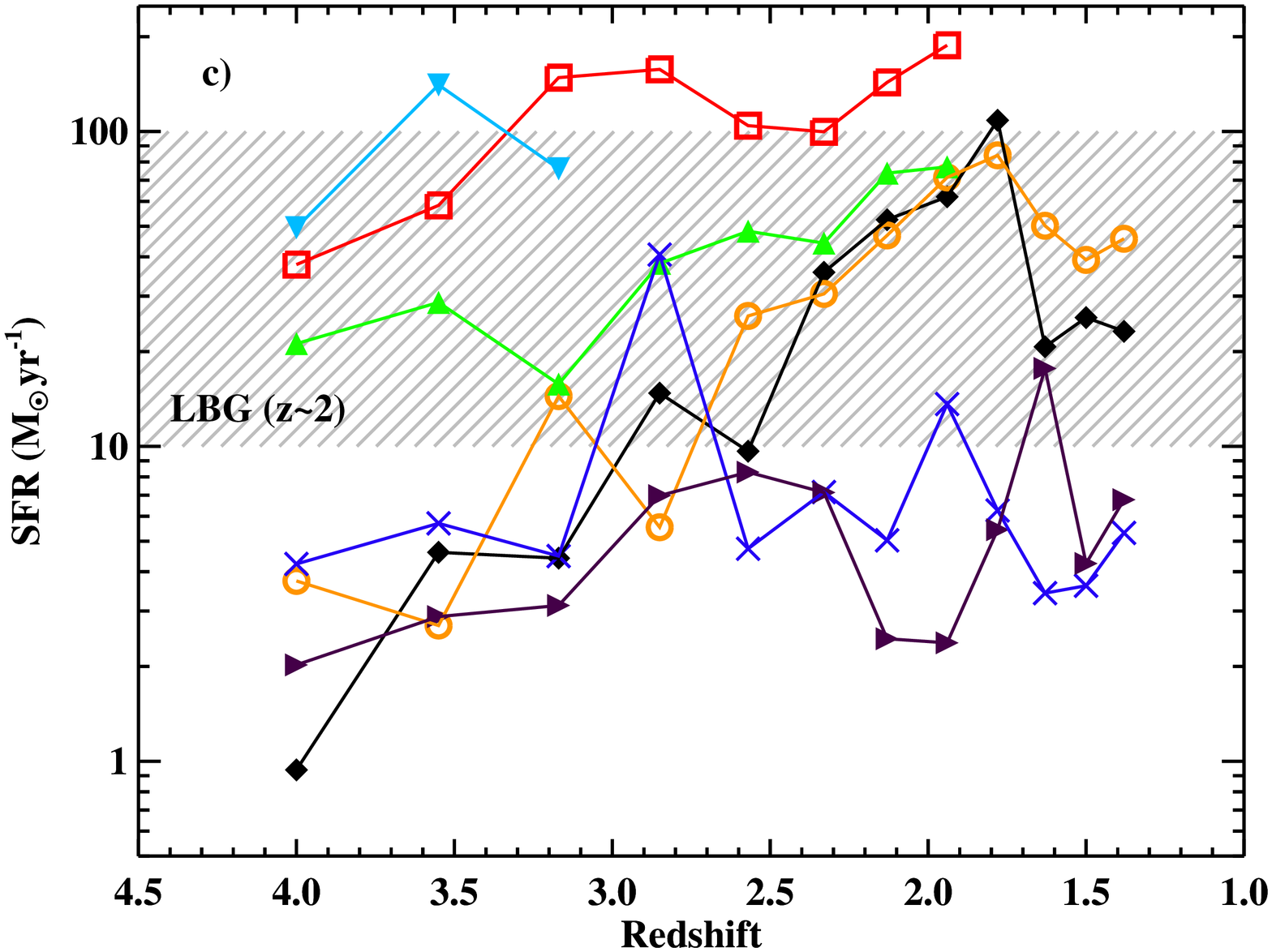}&
  \includegraphics[scale=0.3,angle=90]{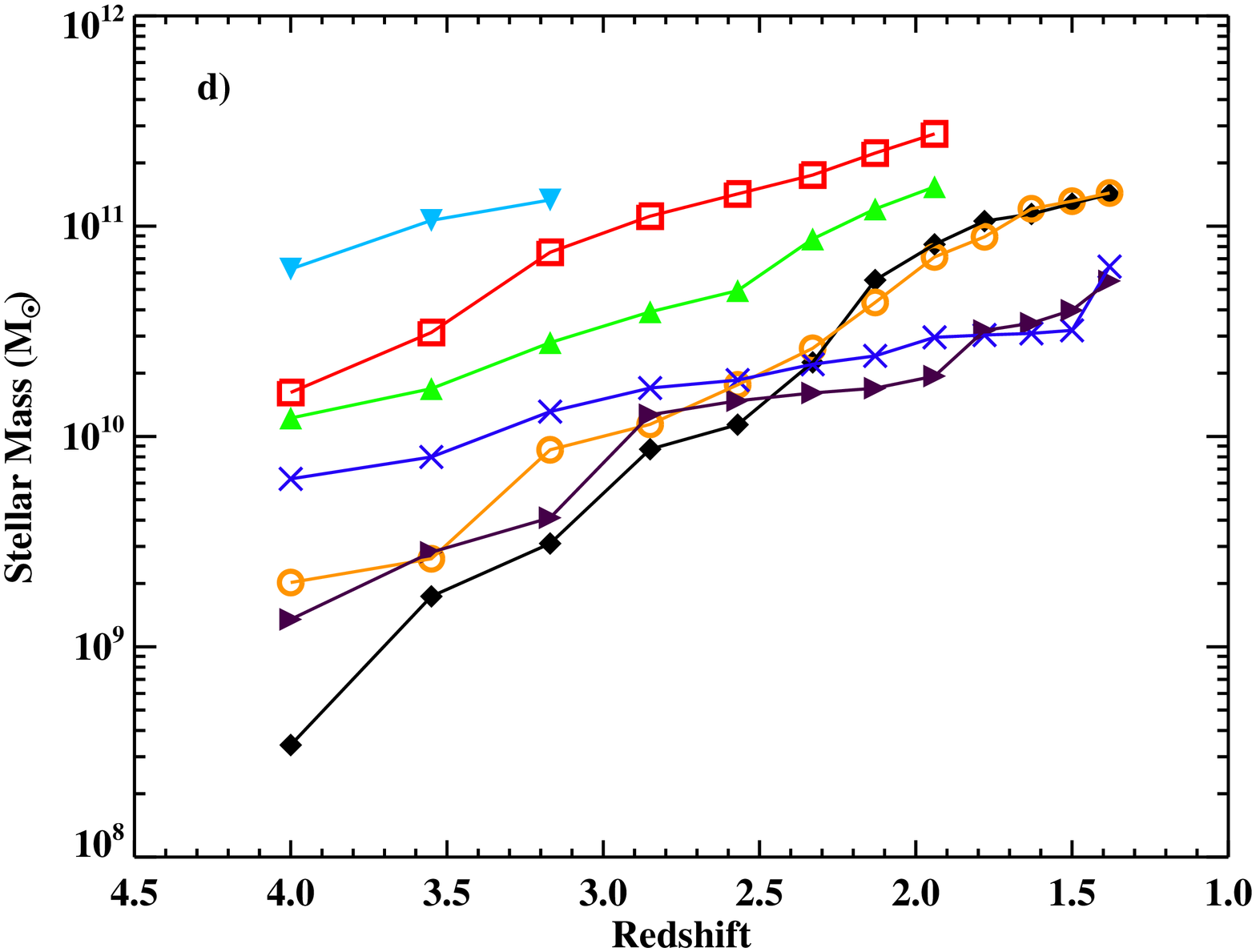}\\
  \includegraphics[scale=0.3,angle=90]{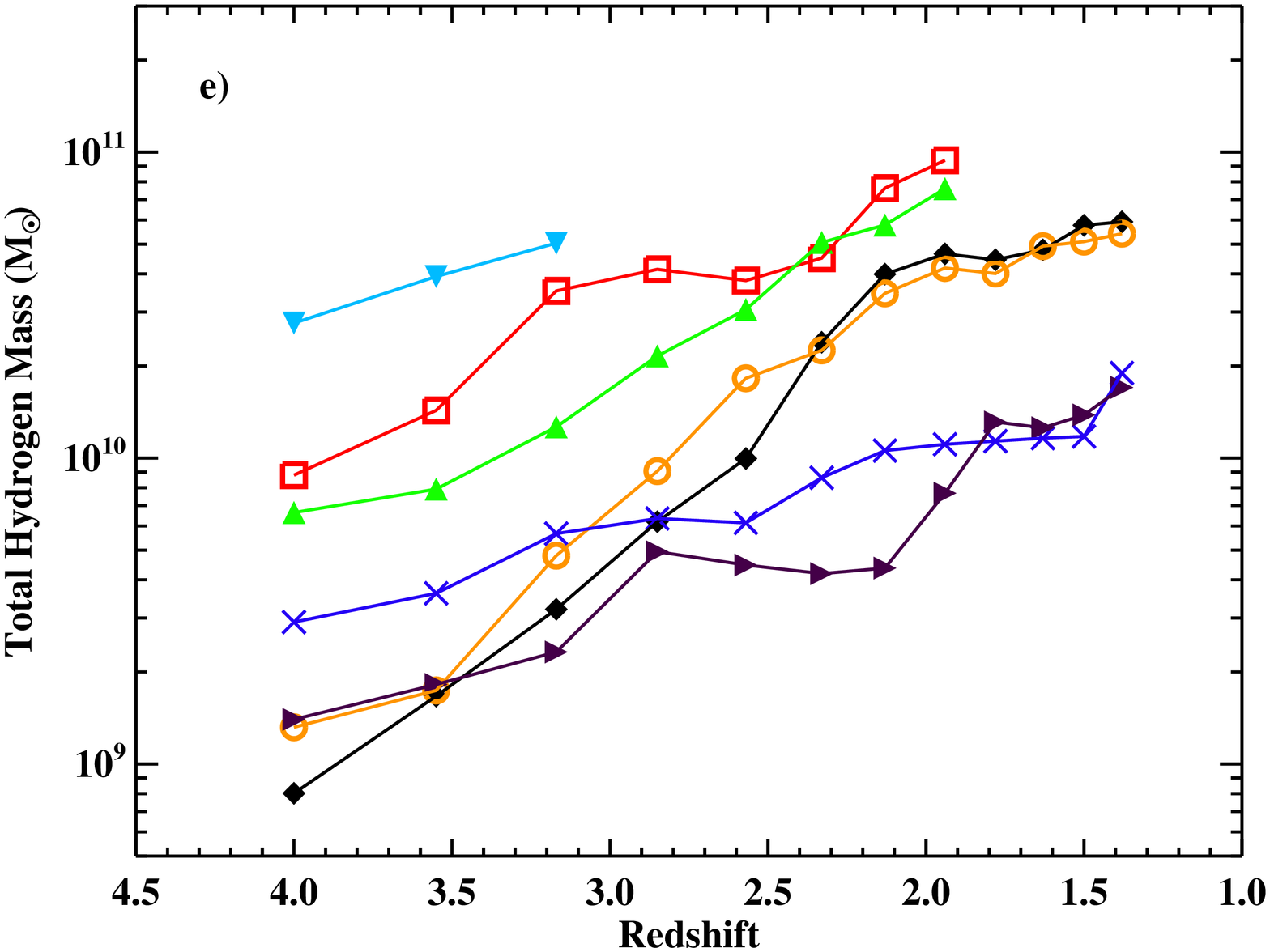}&
  \includegraphics[scale=0.3,angle=90]{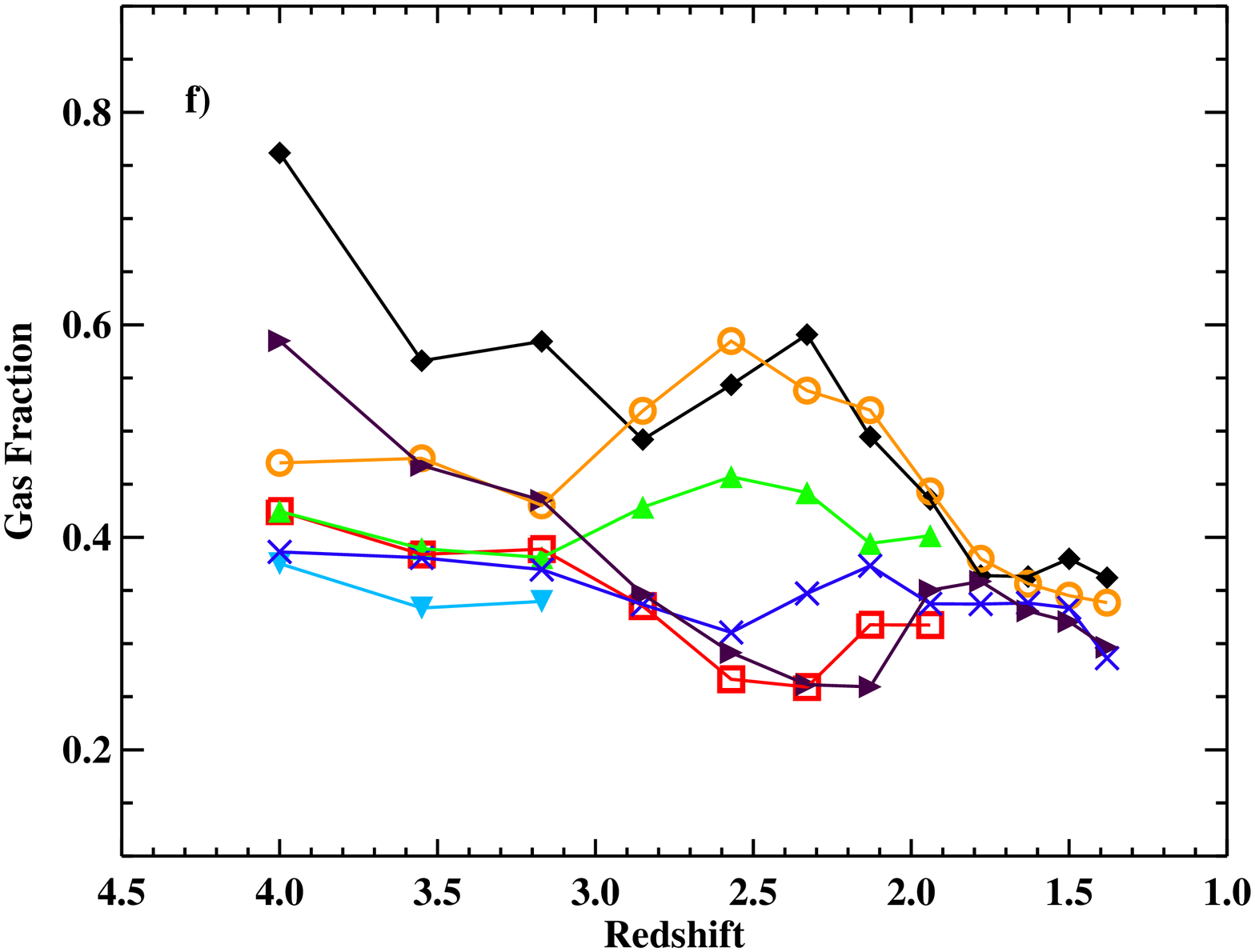}
\end{tabular}
\caption{Evolution of the galaxy properties within the virial radius.
Each galaxy is identified by a different line connecting different symbols.
a) Virial radii. b) Dark matter mass, where  
the dotted line marks the characteristic halo mass, and the horizontal dashed line is 
the critical mass above which a stable shock can develop 
\citep{dek06}. c) Star formation rate, where the shaded area 
represents the typical SFR of Lyman break galaxies at $z\sim2$ \citep{red10}
to highlight the general overlap. d) Stellar masses.
e) Hydrogen masses. f) Gas fraction, defined as the ratio of gas mass to total baryon mass.}
\label{fig:galprop}
\end{figure*}

\section{The simulations}\label{sec:sim}

\subsection{Simulation method}

A more detailed discussion of the simulations used in this work and comparisons 
with properties observed in high-redshift galaxies can be found in \citet{cev10}. Here, 
we briefly summarize the numerical procedures adopted and we review some of the uncertainties 
associated to these simulated galaxies that are relevant to this analysis.

The simulations have been run with the adaptive mash refinement (AMR) hydro-gravitational 
code ART \citep{kra97,kra03}. The code incorporates many of the relevant physical processes 
for galaxy formation, including gas cooling and photoionization heating, star 
formation, metal enrichment and stellar feedback \citep{cev09,cev10}. 
Cooling rates were tabulated for a given gas density, temperature, 
metallicity, and UV background (UVB) based on \textsc{cloudy} \citep{fer98}
and  assuming cooling at the center of a one kpc thick cloud, 
illuminated by a uniform radiation field. Photoheating 
is self-consistently modeled together with radiative cooling, through the 
same \textsc{cloudy} tables, assuming  a uniform UVB \citep{haa96}.
Self-shielding of dense, galactic neutral hydrogen 
from the UVB is approximated by suppressing the UVB intensity to the prereionization 
value ($5.9\times 10^{-26}$ erg s$^{-1}$ cm$^{-2}$ Hz$^{-1}$ sr$^{-1}$) above
gas densities $n_{\rm H}=0.1$ cm$^{-3}$. This threshold is consistent with 
the results from radiative transfer calculations (see Appendix \ref{rtcomp}).

Stars form according to a stochastic model that is roughly consistent with 
the \citet{ken98} law in cells where the gas temperature is below 10$^4$K and the 
density is above the threshold $n_{\rm H}=1$ cm$^{-3}$. 
The code implements a feedback model, in which the combined energy from stellar 
winds and supernova explosions is released at a constant heating rate over 40 Myr, 
the typical age of the lightest star that explodes as a type II supernova (\SNII). 
Energy injection by type Ia supernovae (\SNIa) is also included. The heating rate from 
\SNIa\ assumes an exponentially declining rate with a maximum at 1 Gyr.

The ISM is enriched by metals from \SNII\ and \SNIa. Metals are assumed to be 
released from each stellar particle by \SNII\ at a constant rate for 40 Myr since 
its birth, assuming a \citet{mil79} initial mass function (IMF) and matching the yields of \citet{woo95}. 
The metal ejection by \SNIa\ assumes the same exponentially declining rate. 
The code treats metal advection self-consistently and metals can diffuse and pollute the 
medium around the central disks. In these simulations, the dark matter 
particle mass is $5.5 \times 10^5$ \msun, the minimum 
star particle mass is $10^4$ \msun, the smallest cell physical size is always 
between 35 and 70 pc.

Each dark matter halo has been selected from a cosmological dark-matter-only simulation and 
resimulated with the addition of baryons, using a zoom-in technique that follows the evolution 
of the selected lagrangian region of $\sim1$ comoving Mpc$^3$ 
\citep[for details on the initial conditions see][]{cev09}. All the simulated galaxies are centrals to 
their halos and the analysis has been performed on a 
box of four times the virial radius (\rvir), centered at all times on the main halo.
Note that the selected galaxies do not reside in a sizable group or cluster
at $z=0$  because of the difficulties in simulating companion galaxies.

Although these simulations reproduce some of the main properties observed 
in high-redshift galaxies, like the Tully-Fisher and the mass-SFR relations \citep[see][]{cev10}, 
the use of approximated prescriptions in lieu of the complex interplay of 
physical processes that operate on scales smaller than our resolution limits the 
predictive power of these 
models and of simulations in general. 
Particularly relevant for our work is the uncertainty
associated with stellar feedback, implemented according to the prescription of \citet{cev09}. 
In these simulations, outflows eject hot gas and metals at high velocity 
outside the central disks into the IGM and CGM \citep[see][]{cev09}. The 
total outflow mass flux reaches, in some cases, about one-third of the inflow flux.
However, these winds may not be as strong as some observations suggest \citep[e.g.][]{ste10}
and AGN feedback is missing. As a consequence, a fraction of the gas inside the disks 
might actually be ejected by stronger outflows, altering the covering fraction of neutral 
gas and the metallicity distribution in the CGM \citep[see e.g.][]{fau11}. 
Unless the low density winds drastically perturb the shape and 
kinematics of the gas that inflows in dense narrow streams, these models capture the properties 
of the cold streams, but caution is advised in generalizing our results.

\subsection{Sample properties}

For each galaxy, the total dark matter and instantaneous stellar mass
are computed summing all the particles within the virial radius. 
Similarly, we obtain the total star formation rate (SFR) within the virial radius by averaging the 
masses of all the stars formed in the last 60 Myr. 
This time interval is arbitrarily chosen to be large enough to avoid fluctuations 
due to the stochastic star formation rate implemented in the simulation,
but the average SFR does not depend on this assumption.
The total hydrogen masses are computed from the hydrogen number density in each cell 
of the AMR grid as $m_{\rm H}=m_{\rm p}~n_{\rm H}~l^3$, with $m_{\rm p}$ the proton mass 
and $l$ the cell size. 
No correction for heavier elements, including helium, is applied
throughout this work, except when we quote a 
gas fraction where we multiply the hydrogen mass by 1.38 to correct for heavier elements.
Finally, we derive virial masses by combining together the baryonic and non-baryonic masses.

The sample is composed of seven massive galaxies that exhibit dark matter and baryonic 
properties roughly consistent with the observed population of Lyman break galaxies (LBGs). 
Galaxy properties as a function of time\footnote{Galaxies MW2, MW4, and MW5 were not available 
at low redshifts when this sample was assembled.} are displayed in Figure \ref{fig:galprop}
and listed in Table \ref{galprop} and Table \ref{gasprop} in Appendix \ref{app:table}. 
Panel a) and b) show the redshift evolution of the virial radius and dark matter mass. 
These galaxies span roughly more than a decade in mass.
As seen in panel b), they are all well above the characteristic Press-Schechter halo mass (M$_*$)
in the relevant mass range, more so at higher redshifts.
Galaxy MW1, and to some extent MW3, exhibit a faster overall growth rate than the 
other galaxies. The interval in mass and redshifts analyzed here is typical for cold gas accretion;
only at later times ($z<2$) halos approach the critical mass 
(horizontal dashed line) above which a stable shock develops
and cold accretion  becomes less efficient \citep{dek06,ker09b}.

The evolution of the SFR within the virial radius is presented in panel c).
While there is a tendency to have higher star formation rates in higher mass halos, 
the stochastic nature of the simulated SFR and different accretion histories 
produce a less ordered star formation history (SFH). The star formation in 
the sample brackets the typical values found in LBGs
\citep[shaded area for $z\sim2$;][]{red10}, with MW8 and MW9 usually below the
commonly observed range. The evolution of the stellar mass is shown in panel d) and follows the 
general halo growth. Finally, the gas properties are displayed in panels e) and f) where we show the
time evolution of the hydrogen mass and of the gas fraction, here defined as the ratio 
of gas mass to total baryonic mass. The simulated galaxies have a gas fraction 
comparable to the estimates from observed molecular gas at $z\sim 1-2$ \citep{tac10,dad10}.
The baryonic fraction inside the virial radius is at all times 
below the universal value $f_{\rm b}=0.165$, approaching $0.25-0.3 f_{\rm b}$ 
when only gas and stars are considered inside the central disks. 
MW2 is an exception, having a particularly high baryon fraction at all redshifts, 
probably reflecting an extraordinary high gas accretion rate \citep{cev10}.

\subsection{Satellite identification}

Being part of a larger cosmological simulation, each box contains also satellites
whose positions and sizes are used during the analysis 
to disentangle gas that resides in bound structures from gas in the streams.
For this reason, we identify the satellites using the \textsc{amiga} halo finder 
\citep{gil04,kno09}. This code locates the centers of the halos and iteratively 
computes their radii so that they define an overdensity $\Delta_{\rm vir}$ 
over the background density $\rho_{\rm b}$. 
For small child halos embedded within larger parent halos, 
this radius is defined instead as the distance at which the density profile
reaches a minimum as one moves away from the halo center. 
We adopt the internally computed $\Delta_{\rm vir}$ from \textsc{amiga}
as a function of redshift. We consider particles with velocities greater than 
the escape velocity to be unbound and we select only satellites with more than 182 particles.
This sets a lower limit to the satellite halo mass of $\sim 10^{8}$ \msun.

While \textsc{amiga} outputs several parameters for each of the identified halos,
in the following we compute masses and SFRs considering all the particles 
that are enclosed in the defining radius, purely based on positions.
Although not identical to the mass computed internally by \textsc{amiga}, these masses  are well 
correlated with the ones generated by the code. This simple definition allows 
us to compute self consistently the \HI\ masses and SFRs.
Due to this approximation, our satellite masses can be uncertain up to a factor of two.
Child halos contribute to the parent halo masses.


\begin{figure*}
\begin{tabular}{c c}
\includegraphics[scale=0.32,angle=90]{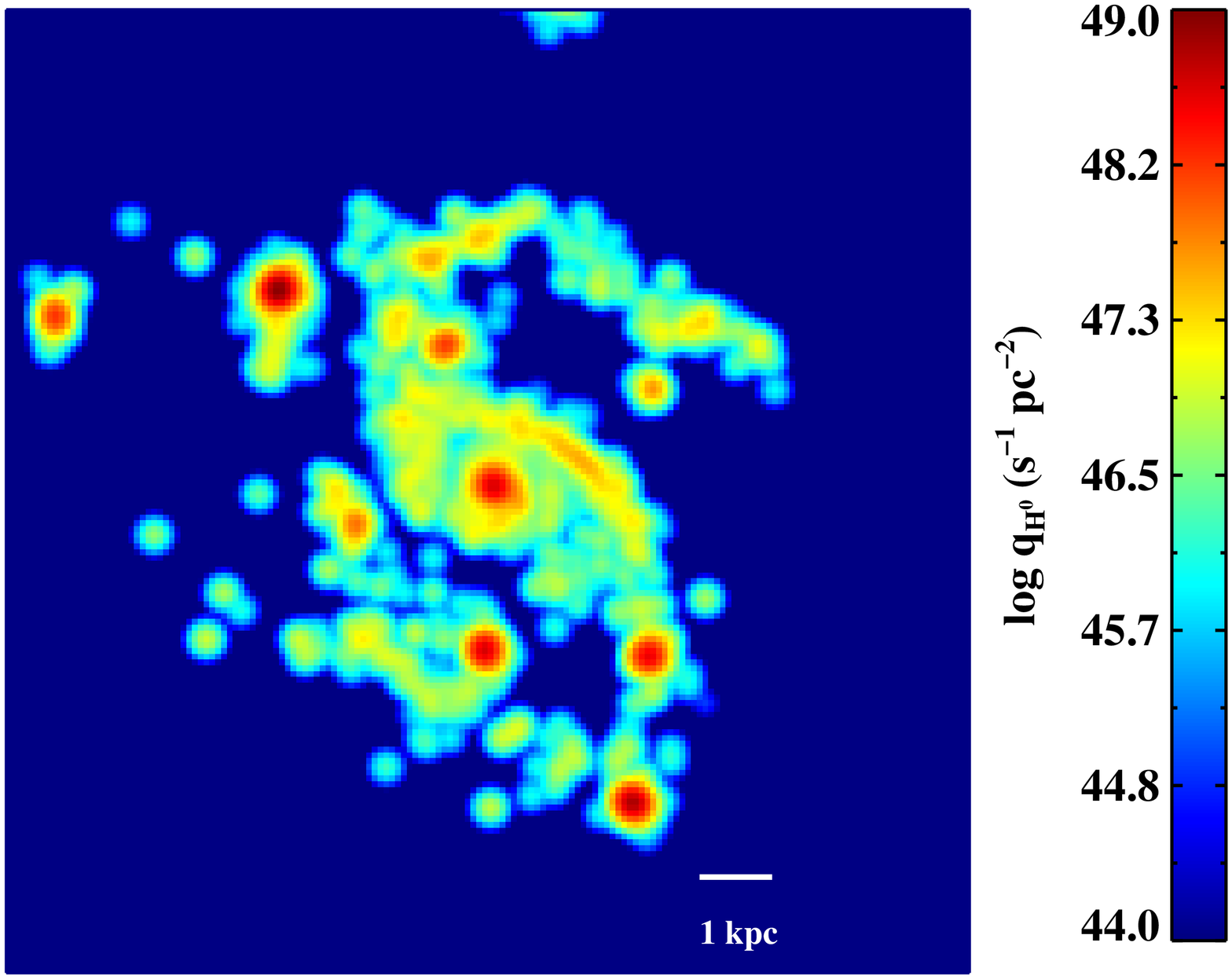}&
\includegraphics[scale=0.32,angle=90]{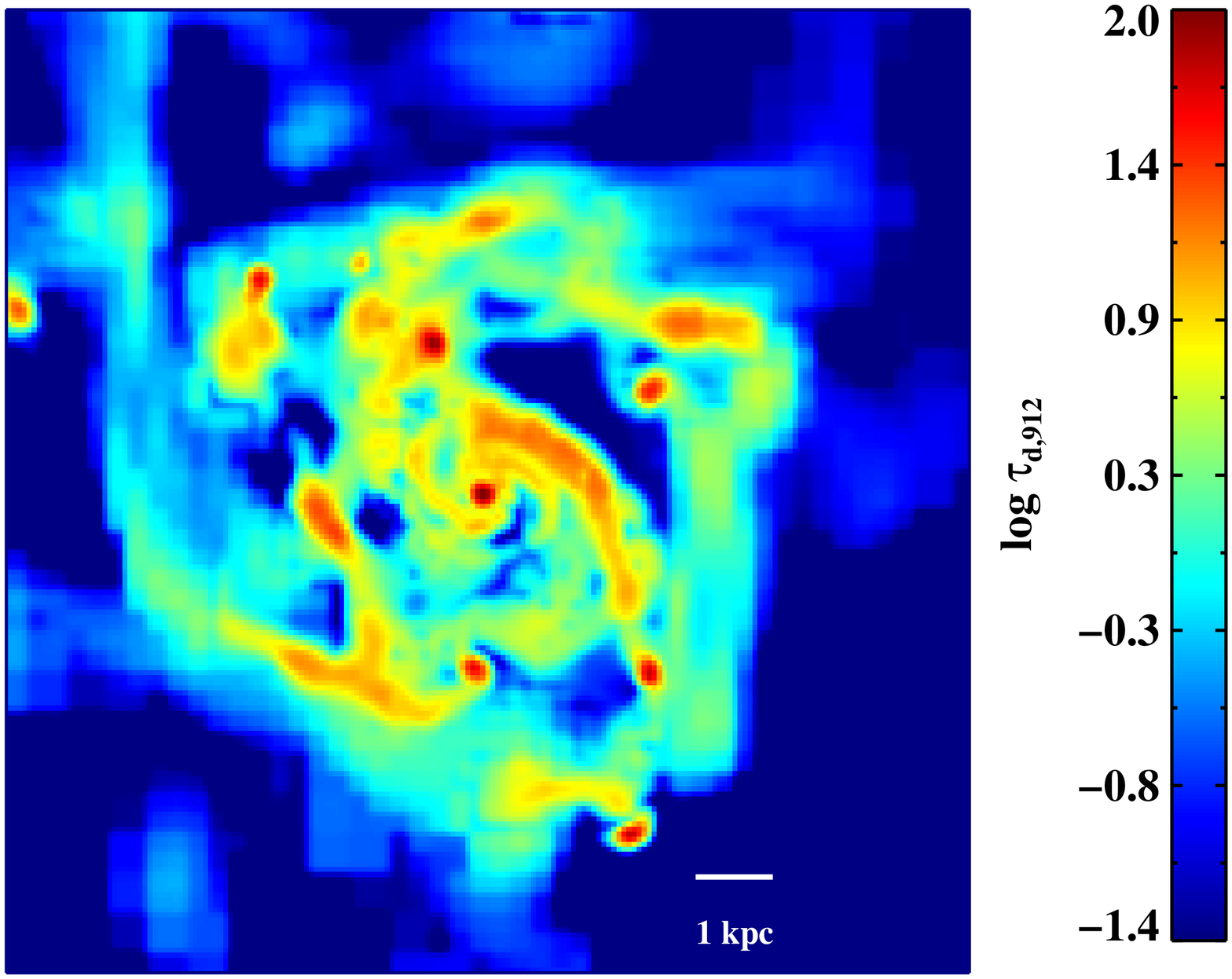}\\
\end{tabular}
\caption{Ionizing radiation and dust in the central disk of MW3 (face on).
Left panel: surface density of ionizing radiation smoothed with a Gaussian kernel 
of 240 pc, showing giant clumps with enhanced star formation.
Right panel: integrated optical depth of dust at 912\AA\ along the entire depth 
of the box (282 kpc). Star formation occurs primarily in 
overdense and enriched regions where the dust optical depth is high.}
\label{mw3_qhodust}
\end{figure*}

\section{Atomic hydrogen neutral fraction}\label{sec:rt}

\subsection{The radiative transfer post processing}\label{sec:rtpost}

A reliable estimate of the ionization state of the gas is essential for any study 
of absorption line systems. Ideally, the ionization state of the gas should be 
coupled to the hydrodynamic calculations, but due to the numerical complexity of this problem,  
codes that solve the (approximated) radiative transfer (RT) equation \citep[e.g.][]{gne01,abe02} 
together with the hydrodynamic equations has been developed and used only recently 
\citep[e.g.][]{raz06,pet09,wis10} and they remain 
expensive for high resolution simulations.
Currently, RT post-processing is the practical way to compute the ionization state of the gas
in our high-redshift simulations. The post processing approximation is justified by the fact that the 
radiation-diffusion timescale is short compared to the evolution timescale.

The neutral fraction $x_{\rm HI}$ of the atomic hydrogen in each AMR cell
is determined using a Monte Carlo RT code that includes both collisional ionization 
and photoionization due to Lyman continuum radiation 
from both the cosmological  UVB and local stellar sources, together 
with absorption and scattering by dust grains. This calculation improves upon results  
based on simple ray-tracing schemes. In fact, 
Monte Carlo transport does not introduce preferred 
directions in the photon path and takes into account the scattering of 
ionizing photons. Thus, a diffuse component of the UV radiation field is naturally included
(about 32\% of the time, photons scatter off of hydrogen atoms 
rather than being absorbed). Further, radiation from stars and dust absorption is modeled
locally for each stellar particle in order to reproduce the anisotropy of the radiation field
within the central disks and satellites galaxies.
Additional details on the numerical procedures and validating tests 
can be found in a companion paper (Kasen et al., in prep.). Here, we briefly summarize the 
models adopted for the source and sink of ionizing radiation.

\subsubsection{Collisional ionization}

At temperatures above $10^4$ K, hydrogen atoms are ionized by electron collisions. 
Neglecting cosmic rays that can affect the neutral fraction even at lower temperatures, 
electron collisions are the only factor that regulates the ionization state of the gas
in the absence of external radiation fields or in regions that are optically thick
to UV radiation. In our simplest model, we assume collisional ionization at equilibrium (CIE)
under the simplifying assumption of a dust free gas 
without external radiation \citep{gna07}. 
Under this approximation, gas below $T=10^4$ K is fully neutral, while gas
above $T>2\times 10^4$ becomes highly ionized. 
In models that include also UV radiation, collisional ionization is coupled to photoionization.
Since our RT calculation does not include photo-heating, the final temperature and the resulting
ionization by collisions are likely to be underestimated
in partially shielded regions. This problem is alleviated by the fact that 
the hydrodynamic code already includes a treatment for photo-heating
\footnote{Further discussion on this issue can be found in Appendix \ref{rtcomp}.}.

\subsubsection{UV radiation from cosmological background}

The UV radiation from unresolved background sources is the next ingredient 
we add to the RT calculation. Since the UVB spectrum does not vary 
much with frequency over the relevant interval for hydrogen ionization, we assume a constant 
mean intensity $J_{912}=5\times 10^{-22}$ erg~s$^{-1}$~cm$^{-2}$~Hz$^{-1}$~sr$^{-1}$ \citep{haa96}
over the energy range $13.6~\rm eV - 54.4~eV$, corresponding to the  
\HI\ ionization threshold and the cutoff at the \HeII\ threshold, respectively.
Further, the modest variation as a function of redshift in the interval $z\sim 2-4$, within the 
uncertainties of the UVB models \citep[e.g.][]{fau08,dal09}, justifies the use of a fixed value 
with time. Our analysis is sensitive to the specific assumptions of the UVB, 
but it will become evident from the discussion that a decrease in $J_{912}$ from 
$z=3$ to $z=4$ would not significantly affect (and even reinforce) our results.

\begin{figure*}
\begin{tabular}{c c}
\includegraphics[angle=90,scale=0.32]{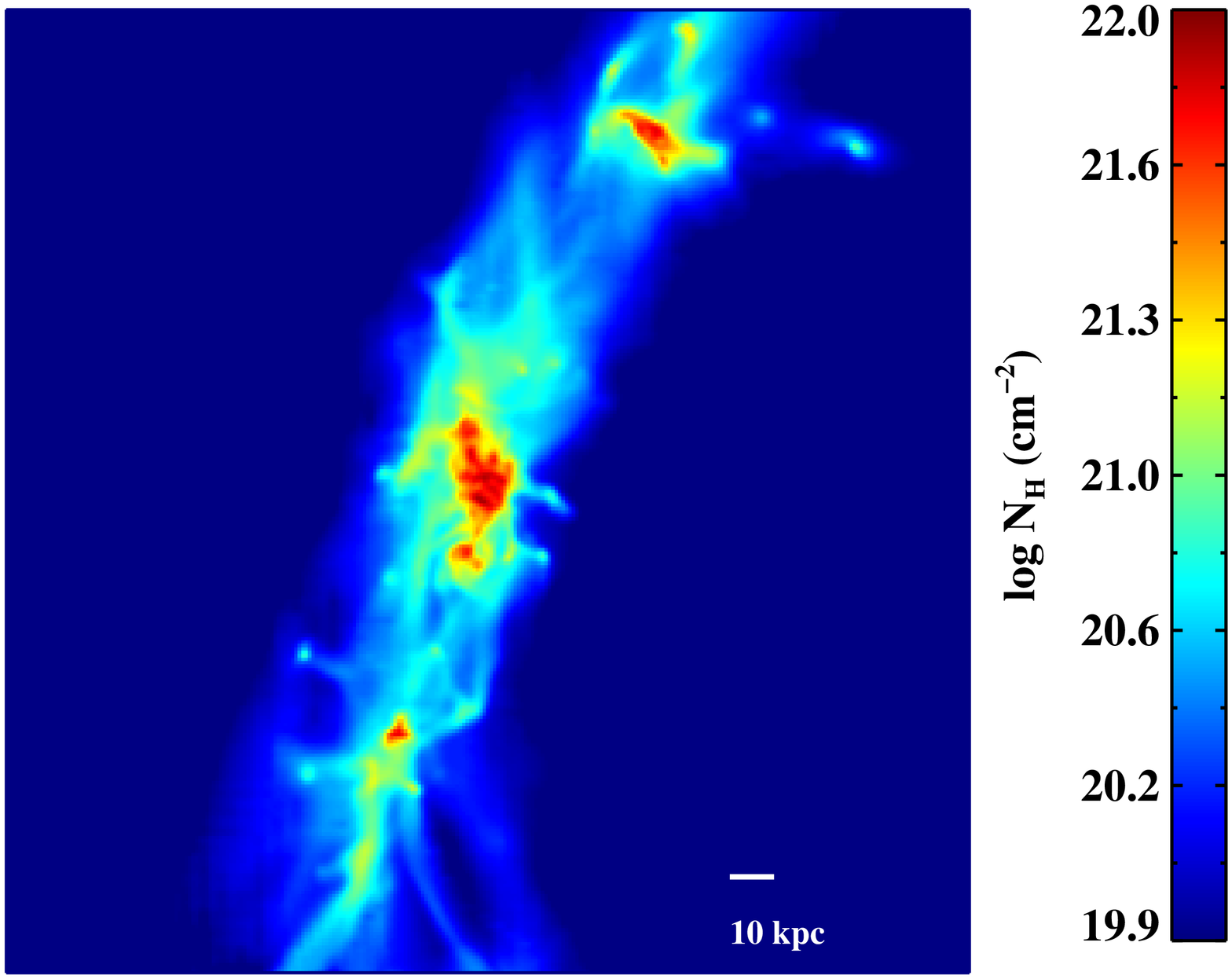}&
\includegraphics[angle=90,scale=0.32]{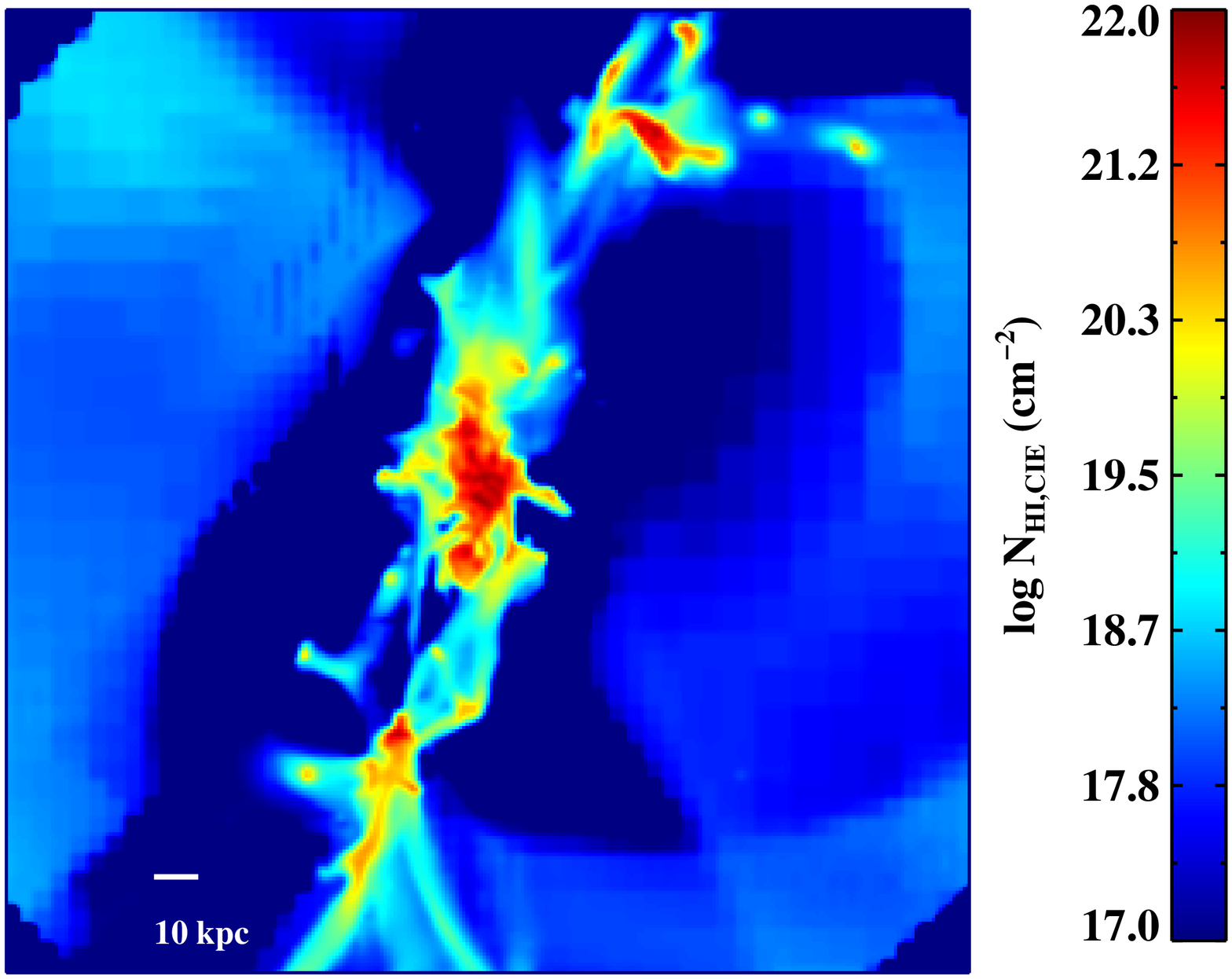}\\
\includegraphics[angle=90,scale=0.32]{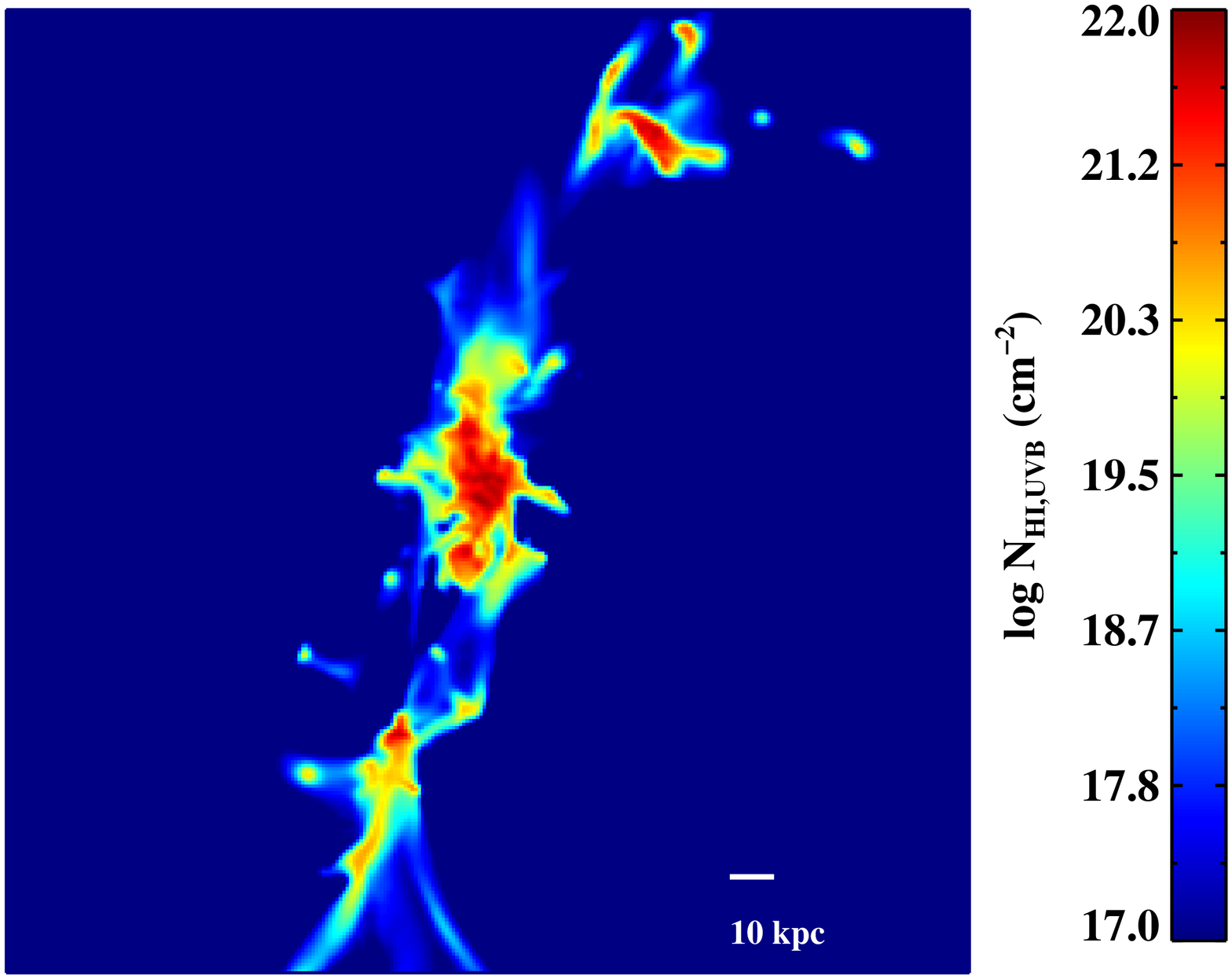}&
\includegraphics[angle=90,scale=0.32]{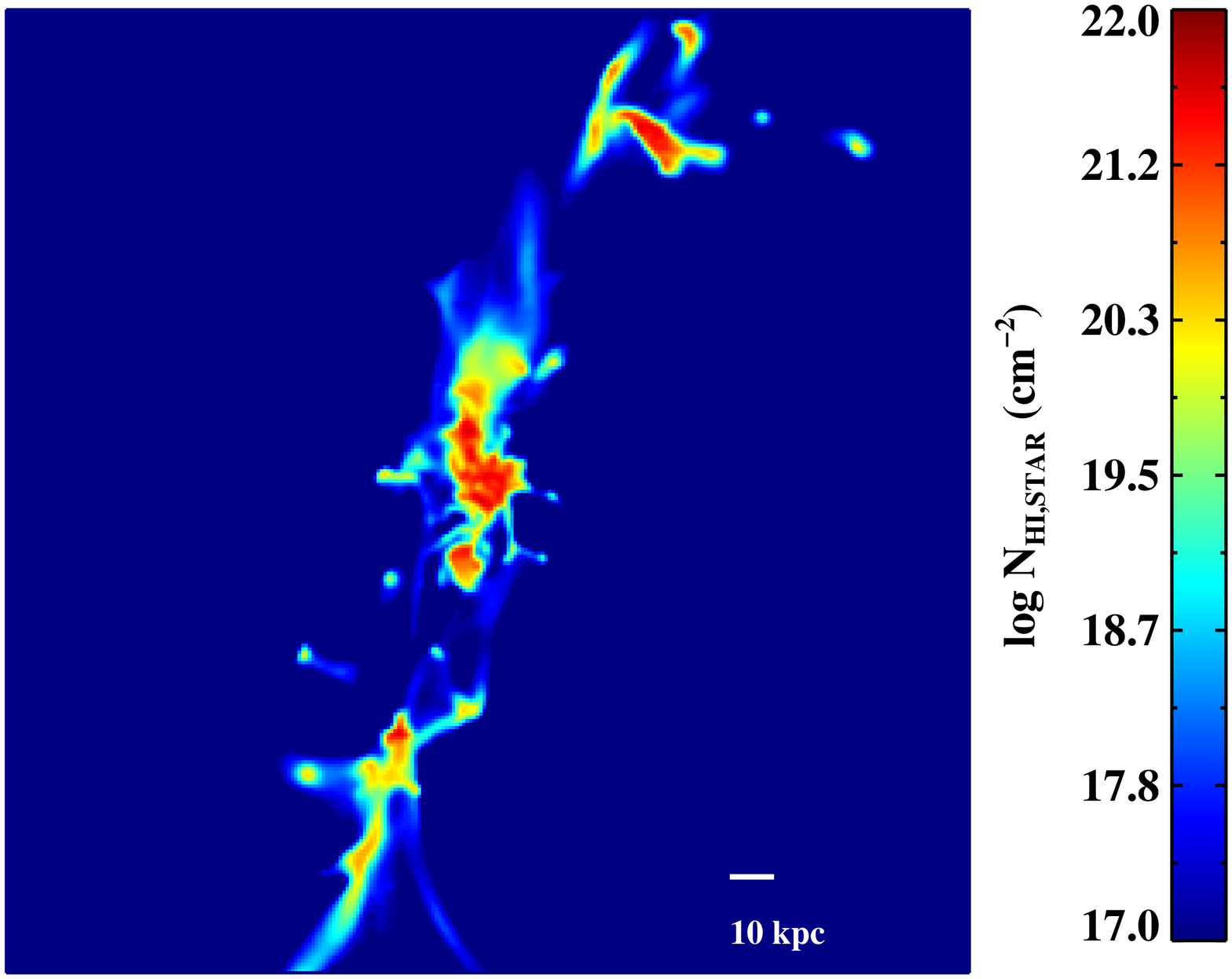}\\
\end{tabular}
\caption{Hydrogen column density for MW3 at $z=2.3$. Top left: \NH. Top right: \NHI\ from 
the CIE model. Bottom left: \NHI\ from the UVB model. Bottom right: \NHI\ from the STAR model. Most of the 
gas that resides in the streams is ionized by electron collisions and the UVB, while photons from 
newly born stars affect the high column density inside the central and satellite galaxies
and their immediate surroundings.}
\label{mw3_ref_rt}
\end{figure*}

\subsubsection{UV radiation from local sources}

An additional and important source of local ionizing radiation
is the Lyman continuum from newly born stars.
In our model, we estimate the rate of ionizing photons ($Q_{\rm H^0}$) emitted 
by a young star using stellar population synthesis models.
$Q_{\rm H^0}$ is a function of the metallicity, IMF, 
SFH, and age of the stellar populations. 
Among those, the most dramatic dependence is on the SFH and age, since the
ionizing radiation from a single burst can vary up to 
several orders of magnitude in a short time interval ($1-100$ Myr),
when the most massive stars that contribute to the Lyman continuum 
luminosity leave the main sequence.

As a first approximation, we estimate the total number of ionizing photons 
from a constant mean SFR from the entire simulated box 
(i.e. from both the central galaxy and satellites) using the proportionality
$Q_{\rm H^0} (s^{-1})=\alpha~SFR (\rm M_\odot /yr^{-1})$.  We compute the normalization $\log\alpha=53.168$ 
using \textsc{starburts99} \citep[][]{lei99}, with a Kroupa IMF \citep{kro01} above 
0.08 M$_\odot$, subsolar metallicity $Z=0.004$, and stellar models from \citet{lej01}. 
Stellar particles have different degrees of enrichment and the
mean metallicity ranges from $Z=0.004$ to $Z=0.013$ in our sample, but 
the variation of $Q_{\rm H^0}$ with metallicity is at most a factor of two across 
the range $Z=0.001-0.04$ ($\log\alpha=53.227-52.935$), and even less ($\sim$ 0.1 dex) 
with the choice of the available stellar models.
We redistribute the total $Q_{\rm H^0}$ among stellar particles which are younger 
than 20 Myr, in a fraction that 
is proportional to the stellar mass $M_{i,\rm star}$
\begin{equation}
Q_{i, \rm H^0}= Q_{\rm H^0}\frac{M_{i,\rm star}(t<20~{\rm Myr})}{\sum_{i} M_{i,\rm star}(t<20~{\rm Myr})}\:.
\end{equation}

Beside the intrinsic uncertainties in the stellar models and metallicity, other 
reasons make the photoionization rate from stars difficult to estimate. 
For example, our finite resolution affects the ability to properly resolve the clumpy ISM
at scales below $35-70$ pc. 
This enhances the leakage of Lyman continuum photons due to 
the absence of high density structures around stellar particles, but at the same time 
the escape fraction of ionizing photons is decreased by the lack of porosity \citep{woo05}.
A proper estimate of the error associated with the leakage of UV photons is difficult, 
but most of the ionizing radiation from stellar sources is absorbed in the proximity of the 
emitting regions at high hydrogen volume density. In these models, the escape fraction at the virial 
radius is between $5-10$ \%, varying with redshift and from galaxy to galaxy. As consequence, only 
a small fraction of the escaping radiation ionizes gas near the central disks with 
$N_{\rm HI}\sim 10^{18}-10^{20}$ \cmm\ (see next section).
Therefore, the associated uncertainties on the neutral gas and the cumulative cross 
section of optically thick hydrogen beyond 
\rvir\ with $N_{\rm HI} \gtrsim 10^{17}$ \cmm\ are expected to be minor.

The left panel of Figure \ref{mw3_qhodust} shows the ionizing radiation rate 
surface density (convolved with a Gaussian kernel of 240 pc only for visualization purposes), 
for the central disk of MW3 at $z=2.3$ and in a nearly face-on view. 
Most of the ionizing radiation is produced in clumps and 
in spiral density wakes. While the SFR in the entire box approaches 50 \sfr, the central disk 
accounts for only half of the total ionizing radiation (with SFR $\sim 27$ \sfr). 
About 50\% of the remaining photons come from a large satellite which is forming stars at 
$\sim 10$ \sfr. This example highlights the fact that photons from young stars that escape the 
surrounding halos can introduce low level anisotropies in the UV radiation field.

\subsubsection{Dust opacity}\label{sec:dust}

Dust is an important sink of Lyman continuum photons, particularly relevant 
for radiation from local sources. 
Using the metallicity in the gas phase from \SNII\ and \SNIa, the dust volume density
is given by
\begin{equation}\label{eq:dust}
\rho_{\rm d} = f_{\rm d}~\mu~m_{\rm p}~(Z_{\rm SN~Ia}+Z_{\rm SN~II})(n_{\rm HI}+0.01~n_{\rm HII})\:,
\end{equation}
where $f_{\rm d}=0.4$ is an estimate of the fraction of metals locked in dust \citep{dwe98} and $\mu=1.245$ 
is the mean particle weight, including helium. To mimic grain destruction, we suppress 
dust formation in ionized regions to only 1\% of what is found in the neutral phase. 
This fraction is the most uncertain quantity in the dust model \citep[see][section 7]{lau09}.

From the dust density we compute the dust optical depth at 912\AA, i.e. the hydrogen ionization potential, 
in each AMR cell as 
$\tau=\alpha~l$, with $\alpha=n_{\rm d}~(\sigma_{\rm s}+\sigma_{\rm a})=\kappa~\rho_{\rm d}~(1-A)^{-1}$
and $l$ the cell size. Here, $\sigma_{\rm s}$ and $\sigma_{\rm a}$ are the scattering and absorption 
cross sections, $A$ is the albedo, and $\kappa$ is the frequency dependent dust absorptive opacity. 
For the above quantities, we assume $\kappa=9.37\times 10^4$ cm$^{2}$ g$^{-1}$ and 
$A=\sigma_{\rm s}/\sigma_{\rm d}=0.248$ 
\citep{dra03}. During the RT calculations, at the relevant UV wavelengths, a linear fit 
to the \citet{lia01} data is used: $\kappa(\lambda)=9.25\times10^4-91.25\times(912 \AA-\lambda)$
and $A(\lambda)=0.24+0.00028(912\AA-\lambda)$.
A map of the dust optical depth for MW3 at $z=2.3$ in a nearly face-on view is in the right 
panel of Figure \ref{mw3_qhodust}. Here the projected dust optical depth is computed along 
a path of 282 kpc, the size of the entire box.

\subsection{Results of the radiative transfer calculation}\label{sec:mod}

For each galaxy, we run three different RT models, gradually including 
additional physical processes. In the first model (hereafter CIE model), 
we derive the neutral fraction assuming CIE, without any source of radiation. 
In the second calculation (UVB model), we include the UVB together with dust and 
collisional ionization. Finally, in our third model (STAR model) we add ionizing radiation
from local sources to the UVB model.
Figure \ref{mw3_ref_rt} presents an output from these calculations. In the top left panel, 
we show the projected \NH\ column density for MW3 at $z=2.3$. 
High column density gas is accreting onto the central galaxy through large radial streams,
with gas overdensities associated with satellites (two incoming galaxies along the streams and 
two closer in, near the central disk). In the other three panels, we display the \NHI\ column density from 
the different RT models. 

This figure captures the basic differences that arise from the different physical processes
included in the RT calculation. Part of the gas within the filaments has a temperature above 
$\sim 10^4$ K, and collisional ionization alone (top right panel) lowers the neutral column density 
by more than one order of magnitude. A comparison of the \HI\ map for the CIE and UVB models 
(bottom left) clearly shows that the CIE approximation largely overestimates 
the neutral fraction and that photoionization from the UV background affects most of the low density
gas in the streams. Indeed, the filaments are highly ionized, with patches of self-shielded 
neutral gas that surround the main galaxy and the satellites. Cold streams are not entirely self-shielded. 
Finally, the inclusion of local sources mostly affects the high 
column density gas (where stars form) and their immediate surroundings, where the column density 
is caused to drop below $N_{\rm HI}\sim 10^{20}$ \cmm. 
The low escape fraction from the galaxy disks (below 10\% at the virial radius)
implies a minor effect on the gas in the streams beyond \rvir\ without any appreciable 
difference compared to the UVB model for column densities below $N_{\rm HI}\sim 10^{18}$ \cmm. 

A more quantitative comparison between the different models is presented in Appendix \ref{rtcomp}.
There, we discuss the typical volume density for self-shielding, and the effect of local sources 
on the column density and mass of neutral hydrogen. We also provide a crude fitting formula 
to the UVB model useful to improve the CIE approximation.

\begin{figure}
\includegraphics[angle=90,scale=0.32]{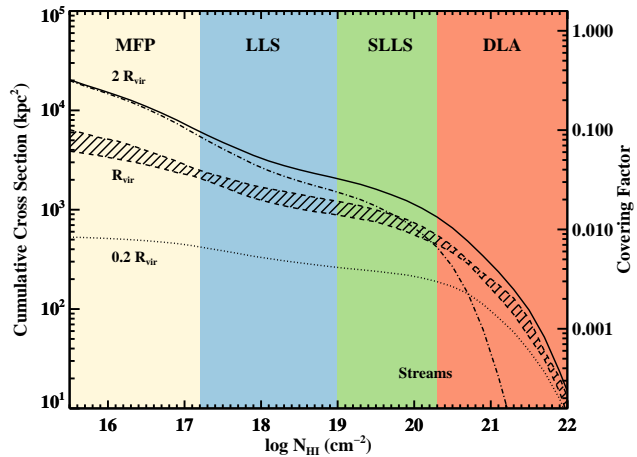}
\caption{Cumulative cross section for MW3 at $z \sim 2.3$ (STAR model)
averaged along three orthogonal directions. The different curves are for
a cylinder of radius 2\rvir (solid line),
a cylinder of radius 0.2\rvir\ (dotted line) and for the streams only 
within 2\rvir\ (dash dotted line).
The hatched region between dashed lines highlights the 
dispersion along the three projection axis for a cylinder of radius 1\rvir. On the right
axis, the physical area is translated into a fractional covering factor, normalized to the area within 2\rvir. 
Colors represent intervals of column density for DLAs, SLLSs, LLSs and MFP gas. 
Gas within galaxies contributes to half of the DLA cross section, 
while LLSs and MFP gas are mainly associated with cold streams. Only $\sim 10\%$ of the projected area 
is covered by optically thick gas and $\sim 1\%$ of the cross section is covered by 
primarily neutral gas at $N_{\rm HI} \gtrsim 10^{20} \, \rm cm^{-2}$).}
\label{fig:cfmw3}
\end{figure}

\section{The projected area of neutral gas}\label{sec:crsec}

In the previous section, we have shown that the streams are only partially shielded
and that both the UVB and local sources are necessary ingredients to study the 
neutral and ionized gas distribution across a large range of column density. 
In this section, we quantify the cross section 
of neutral gas in and around massive galaxies and the evolution of the covering factor.
In the remaining of our analysis, we consider the STAR model as our fiducial RT calculation, 
providing from time-to-time a comparison with results from the UVB model.  

\subsection{Neutral hydrogen cross section}

We derive the cross section that these simulations offers in projection as a function of 
the \HI\ column density by collapsing the AMR 3D grid onto a 2D regular grid with 
cell size equal to the size of the finest level in the AMR structure. 
We measure in each galaxy the covered area in \NHI\ bins of 0.2 dex, averaged 
in three orthogonal directions along the axes of the AMR grid. 
An example for MW3 at $z=2.3$ (STAR model\footnote{We highlight how the three 
RT calculations affect the cross section at various column densities in Figure \ref{xhi_all}.}) 
is in Figure \ref{fig:cfmw3}. For relative comparisons, we introduce 
in the right hand side axis the covering factor, defined as the cross section normalized 
to the area inside a circle of radius $2R_{\rm vir}$.
The solid line represents the cross section measured in the entire box, while the dot-dashed line highlights the 
contribution of the cold streams alone, obtained by masking all the gas cells inside one quarter of the 
virial radius of the main halos and satellites. Further, to isolate the contribution of the 
central galaxy, we display the cross section within two concentric cylinders of sizes 
$0.2R_{\rm vir}$ (dotted line) and $R_{\rm vir}$ (dashed lines). 
In the latter case,  we display the minimum and maximum cross section among the 
three projection axis with a hatched region. Obviously, different orientations have different cross section 
distributions, but the dispersion along the mean appears not to exceed the amplitude of the 
features that are visible in the distribution. The radius $0.25R_{\rm vir}$ is well suited to
separate streams and galaxies and it encompasses the interphase region where the flowing gas is reconnecting
with the central disk. However, the fraction of neutral gas in the column 
density interval $10^{19}-10^{20}$ \cmm\
that is associated with the streams depends on the location of this boundary.

We mark in this figure four \NHI\ 
intervals that define the ionization state of the atomic hydrogen and 
that are associated with different classifications of observed ALSs.
This is done to highlight at what intervals of column densities cold streams are predicted to
dominate the neutral gas cross section.
Specifically, at $N_{\rm HI} \geq 2\times 10^{20}$ \cmm,
the hydrogen is neutral, highly optically thick ($\tau \gtrsim 1000$), 
and gives rise to damped Lyman-$\alpha$ absorbers (DLAs). 
The interval $10^{19}~{\rm cm^{-2}} \leq N_{\rm HI} < 2\times 10^{20}$ \cmm\ 
marks the transition between fully neutral 
to ionized hydrogen and defines the  super Lyman limit systems (SLLSs) 
or sub-damped Lyman-$\alpha$ absorbers. 
The interval $1.6\times 10^{17}~{\rm cm^{-2}}  \leq N_{\rm HI} < 10^{19}$ \cmm\ defines Lyman limit systems
(LLSs) which are highly ionized, but retain enough neutral hydrogen to be 
optically thick ($1 \lesssim \tau \lesssim 60$). Finally, below $N_{\rm HI} = 1.6\times 10^{17}$ \cmm, 
the hydrogen becomes optically 
thin ($\tau \lesssim 1$) and it is highly ionized.  However, this gas is estimated 
to contribute to the mean free path (MFP) of Lyman photons
in the Universe at $z>3$ \citep{pro09b,pro10}. In this work, we consider the interval  
$3.2\times 10^{15}~{\rm cm^{-2}} \leq N_{\rm HI} < 1.6\times 10^{17}$ \cmm, 
which we dub as ``mean free path'' gas.

\begin{figure}
\includegraphics[angle=90,scale=0.32]{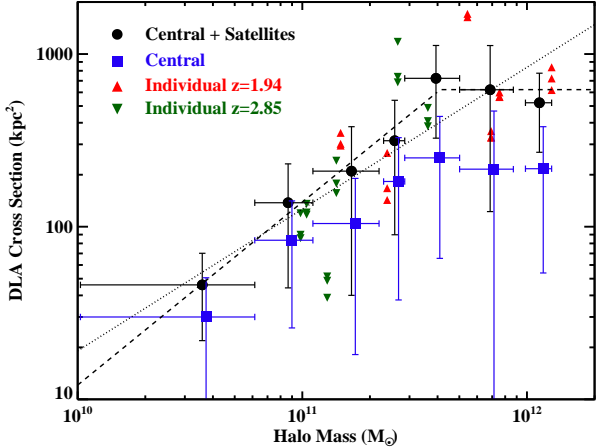}
\caption{Average DLA cross section within \rvir\ as a function of halo mass for 
the central galaxies only (blue squares) and central galaxies plus satellites (black circles).
All the galaxies in our sample, at all sampled redshifts, are included. 
The vertical error bar indicates the standard deviation 
about the mean, while the horizontal error bar indicates the interval of halo mass
used in the average. Also shown, individual galaxies at $z\sim2.8$ (green downward triangle) 
and $z\sim1.9$ (red upward triangle). A linear regression (dashed line) shows that the 
total cross section is nearly 
proportional to the halo mass below $4\times 10^{11}$ \msun, while it is roughly constant at higher masses. 
A shallower dependence is found over the entire available mass range (dotted line), consistent 
with DLAs originating above a fixed surface density threshold for a self-similar gas density
distribution. Central galaxies typically account for less than 50\% of the DLA mean cross section.}
\label{fig:dlaxsec}
\end{figure}

The cross section distribution within \rvir\ steeply rises from the high end towards 
$N_{\rm HI}\sim 10^{20}$ \cmm\
and then flattens between  $N_{\rm HI}\sim 10^{18}-10^{20}$ \cmm\, where gas is ionized.
At lower column density there is a mild increase.
Gas within 2\rvir\ exhibits a similar shape in its cross section, but the separation
between the two distributions increase systematically moving towards low column densities. 
From the example provided in Figure \ref{fig:cfmw3} we find that at 2\rvir,
$\sim 10\%$ of the projected area is covered by optically thick gas and $\sim 1\%$ 
is covered by predominantly neutral gas. Note that despite the lower covering factor, this gas may 
still contribute to the integrated opacity of the ionizing radiation
(see below). About $30\%$ of the area 
is occupied by gas with $N_{\rm HI} \gtrsim 10^{15}$ \cmm. Most of the MFP gas and of the LLSs are 
associated with the streams alone, and half of the area that is covered by SLLSs 
resides outside the virial halo of the massive central galaxy, with streams accounting for 
two third of the covering factor. A significant fraction of the neutral gas is located within 
halos with half of the DLA cross section inside the streams.
Above $N_{\rm HI}\sim 10^{21}$ \cmm, the fraction of DLAs in the streams drops below 20\%
and becomes negligible at even higher column densities.    
Only 20\% of the DLA cross section is found within 0.2\rvir, where it is
 likely associated with the central disk,
indicating that a significant fraction of DLAs can be found within clumps and satellites
inside the virial radius of the main halo, but outside the central disk.

That a non-negligible fraction of the DLA cross section is found within the virial radius
but outside the central disks appears to be a typical property of these simulations, 
as shown in Figure \ref{fig:dlaxsec}.
In each snapshot (STAR model), we identify an independent structure as 
an ensemble of contiguous cells above $N_{\rm HI}=2\times 10^{20}$
\cmm.  We define the central galaxy to be
the closest structure to the center of the main halo and we consider
satellites or clumps in the streams to be all of the remaining groups of contiguous
cells. The average cross section as a function of the halo mass is shown with blue squares for 
the central galaxies and black circles for the central galaxies and satellites. The amplitude
of the halo mass intervals, chosen to have an equal number of objects,  
is indicated by the horizontal error bar, while the vertical error bar is for the standard 
deviation along the mean. Values are listed in Table \ref{tab:dlacrss}.

In the entire sample, the central galaxy contributes 
$\sim 35\%-65\%$ of the DLA cross section, with galaxies residing in more massive halos 
showing smaller fractions. We conclude that in these simulations 
the satellites of massive halos and clumps that reside in the 
streams are at least as important to the DLA cross-section of massive dark matter halos as the
central galaxy \citep[see also][]{mal01,raz06}. Note that the number of objects included 
in this study is limited and the cross-sections listed here can be affected by sample variance.
Also, the cross section of DLAs is particularly sensitive to the disk sizes that are 
notoriously difficult to correctly reproduce in simulations, although our simulations 
show extended disks with $3-5$ kpc radii at high-redshift. \citep{cev10}.  
Further, these simulations have been selected to study cold gas accretion 
in massive galaxies and are not representative of the entire population of galaxies giving rise to ALSs.
Nevertheless, it is reassuring that the mean DLA cross section within \rvir\ ranges between 
$50-800$ kpc$^2$ for halos of $10^{10}-10^{11}$ \msun, in good agreement with the 
cross sections found at similar masses in the larger cosmological simulation by \citet{pon08} 
and \citet{tes09}. Our predictions are towards the lower end of the size distribution 
from \citet{cen10}, who finds a larger area as a consequence of strong feedback. This difference
emphasis how various feedback prescriptions affect the final cross sections.
 
The total area is nearly proportional to the halo mass below $4\times10^{11}$ \msun\ 
(slope $\gamma=1.06$ and normalization $\beta=3.16\times 10^{10}$ for a log-log linear regression; 
dashed line in Figure \ref{fig:dlaxsec}) and roughly constant at higher masses.
Across the whole available mass range, an approximate fit to the points in Figure \ref{fig:dlaxsec}
is given by the mass dependence of the area within the virial radius at a fixed 
time: $A_{\rm vir} \propto R_{\rm vir}^2 \propto M_{\rm vir}^{2/3}$.
This is consistent with DLAs originating from regions where the surface density is above a fixed threshold, 
provided that the distribution of densities within the virial radius (or 2\rvir) is 
self-similar among the halos.

Note that the relation found in Figure \ref{fig:dlaxsec} is based on only seven independent halos at 
all times and redshift is partially degenerate with the halo mass, 
since the most massive halos are found only at late times. It is therefore useful to 
inspect values for single galaxies, superimposed in this figure at $z\sim2.8$ and $z\sim1.9$
in three projections. While a mild evolution with redshift is visible (see also next section), 
the mean cross sections previously derived appear to be a fair representation
of this sample as a whole, more so considering the large intrinsic scatter 
from galaxy to galaxy.

\begin{figure*}
\begin{tabular}{c c c}
\includegraphics[angle=0,scale=0.25]{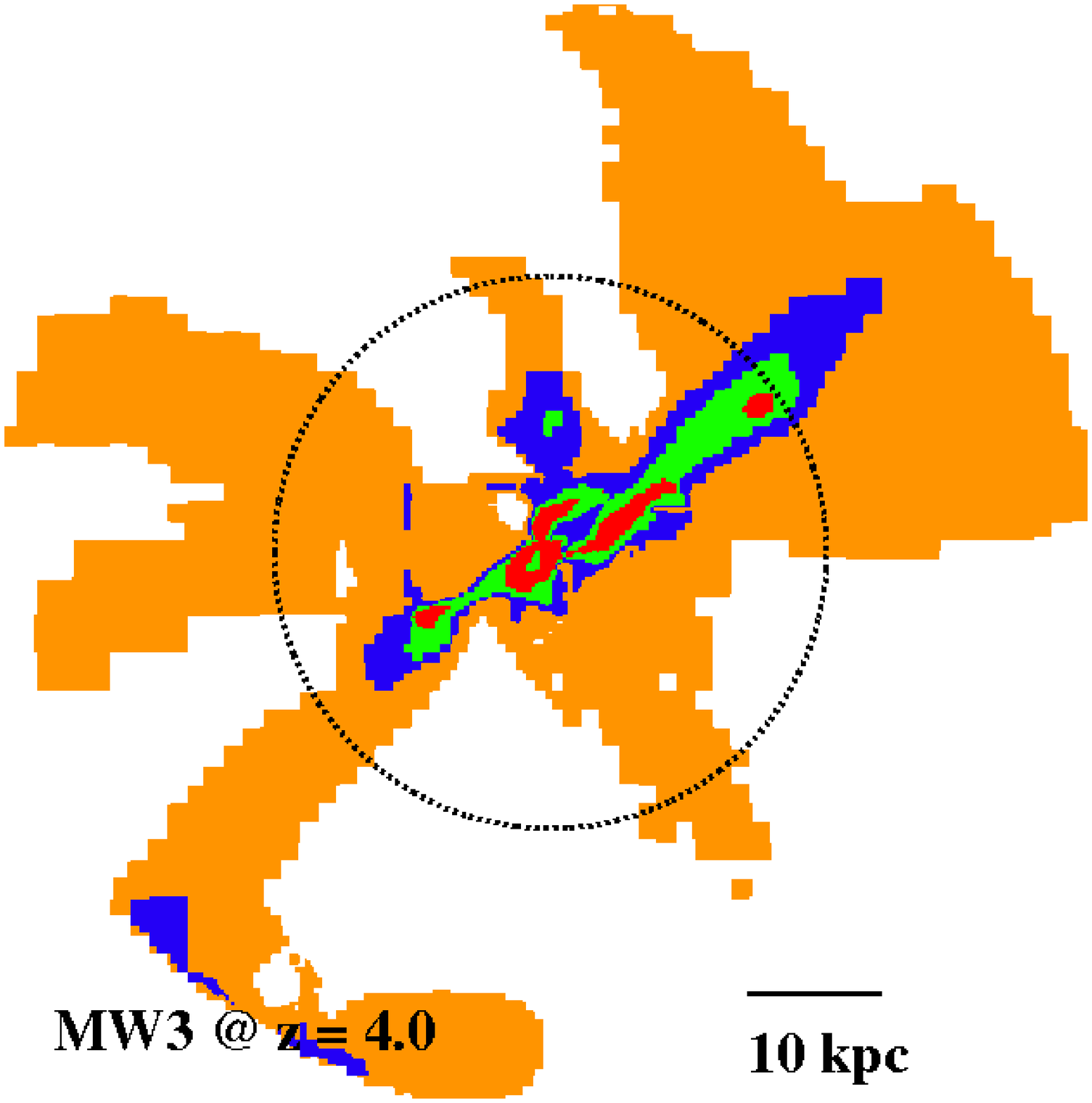}&
\includegraphics[angle=0,scale=0.25]{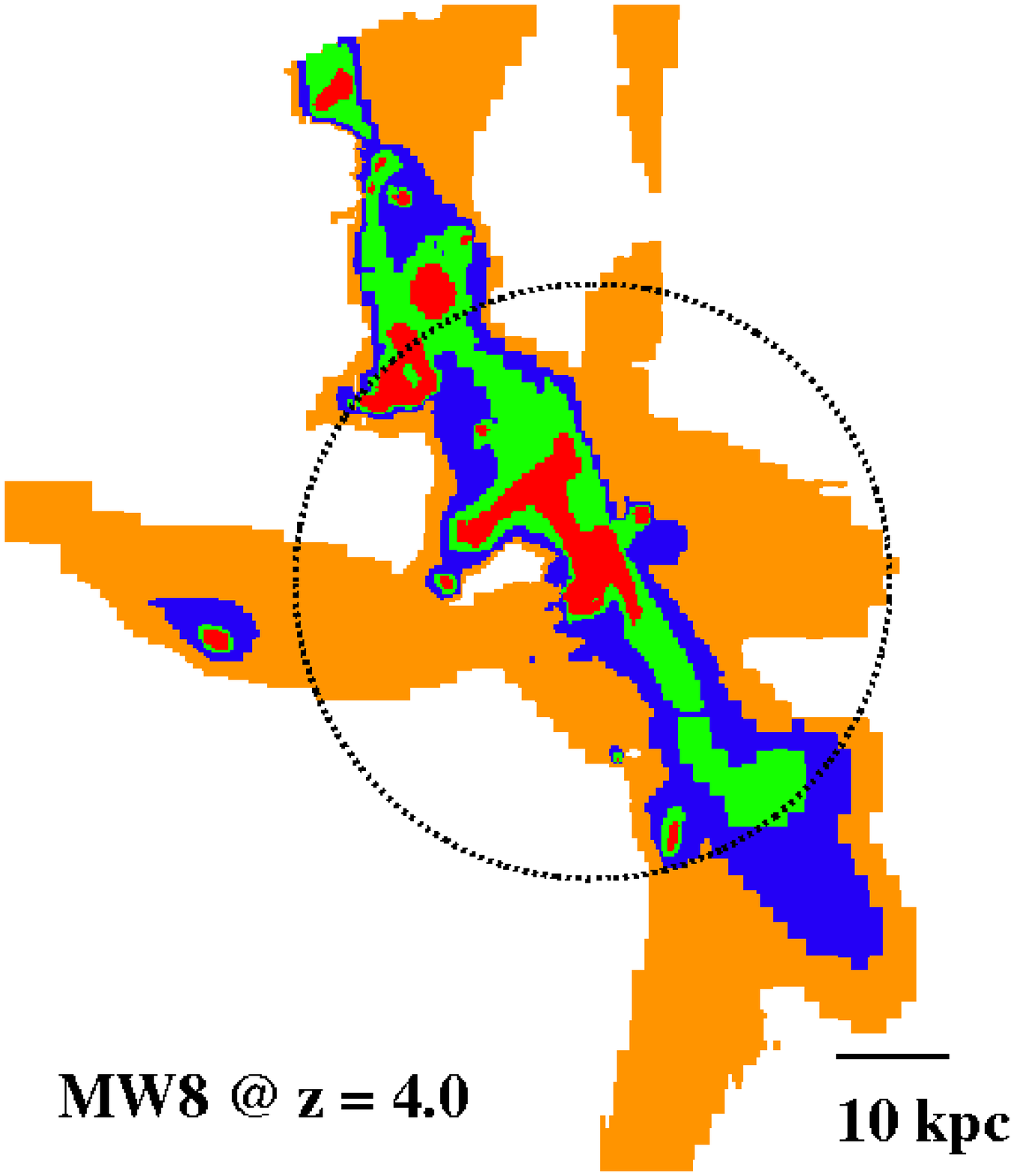}&
\includegraphics[angle=0,scale=0.25]{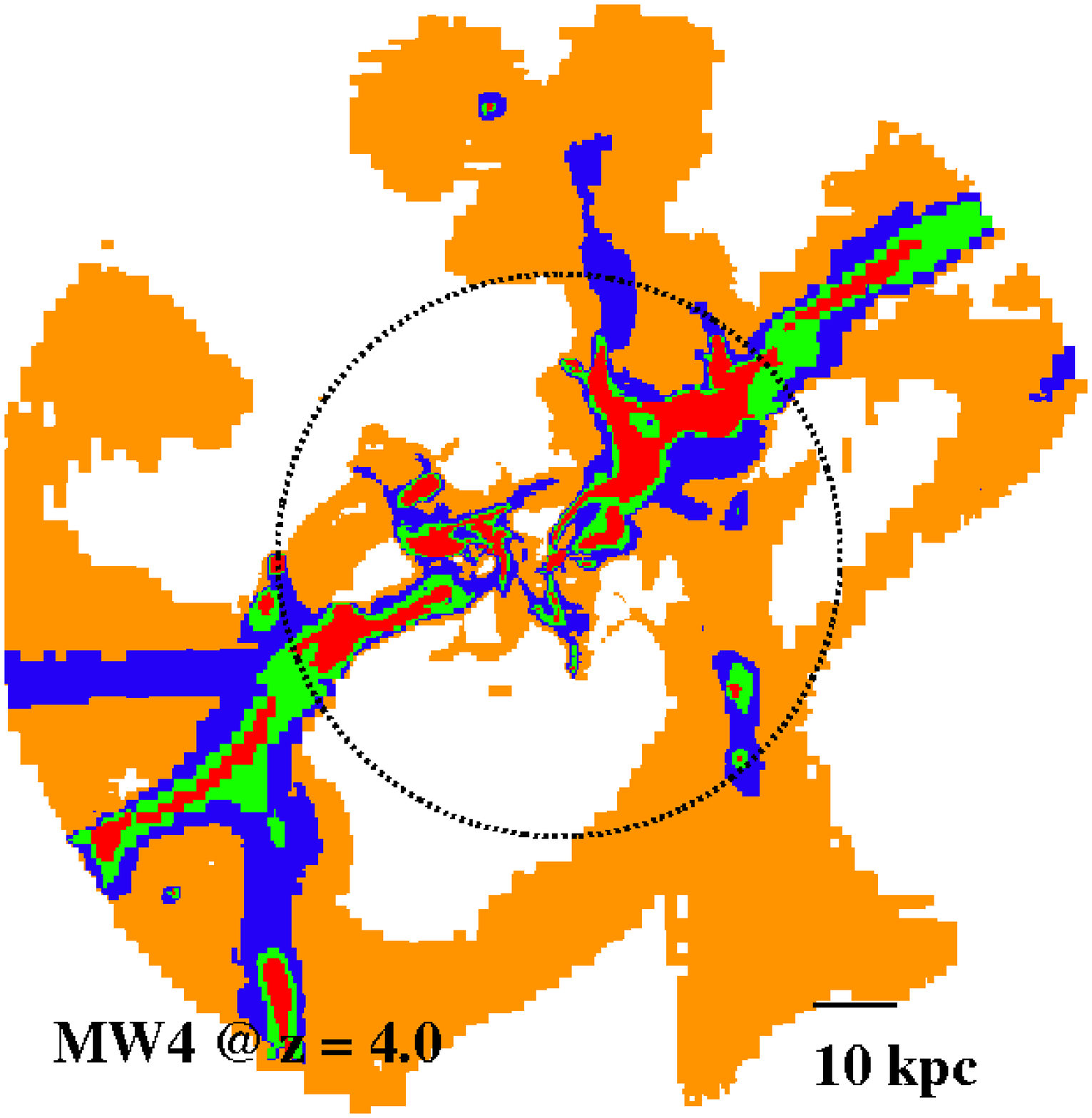}\\
\includegraphics[angle=0,scale=0.25]{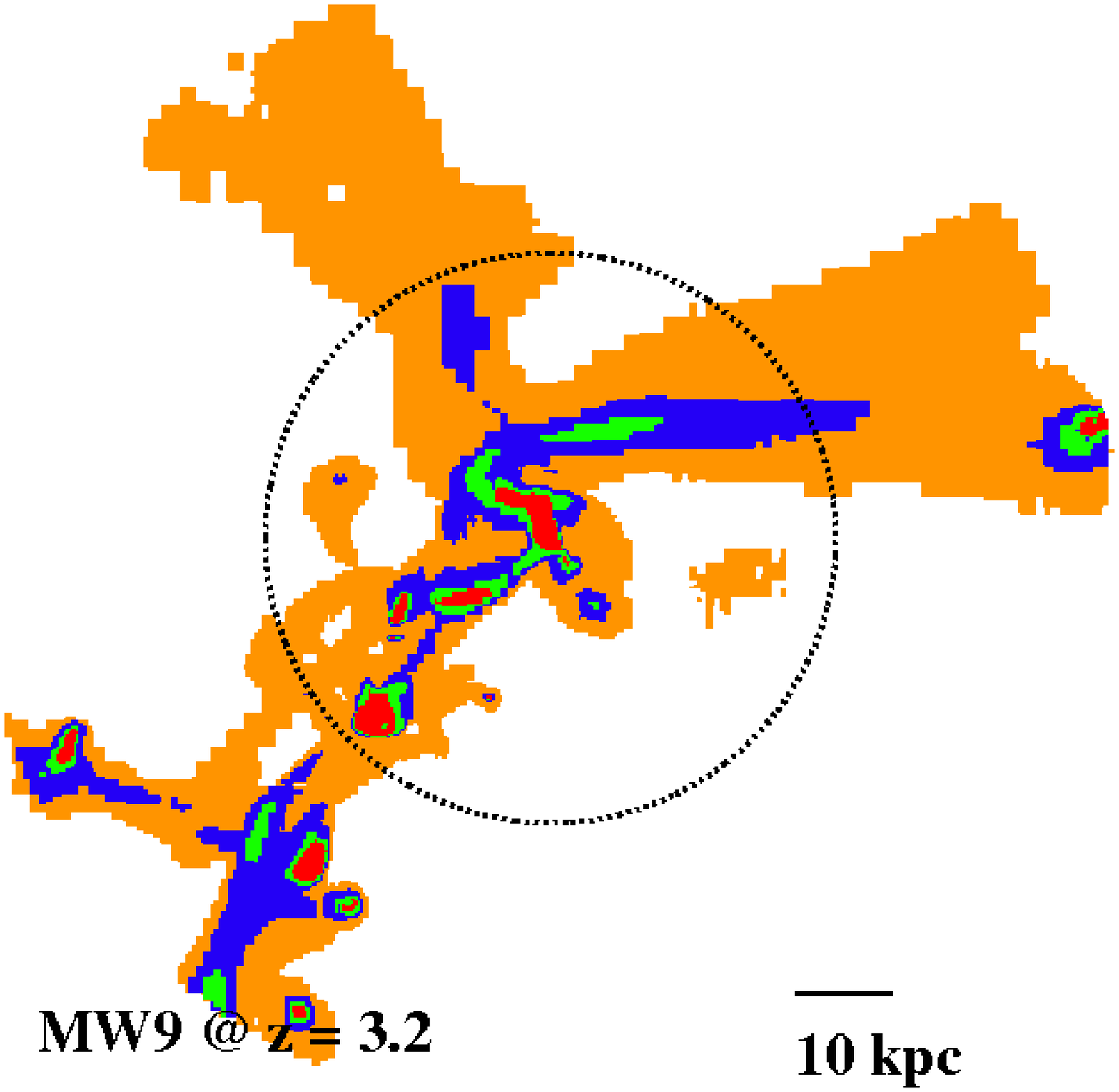}&
\includegraphics[angle=0,scale=0.25]{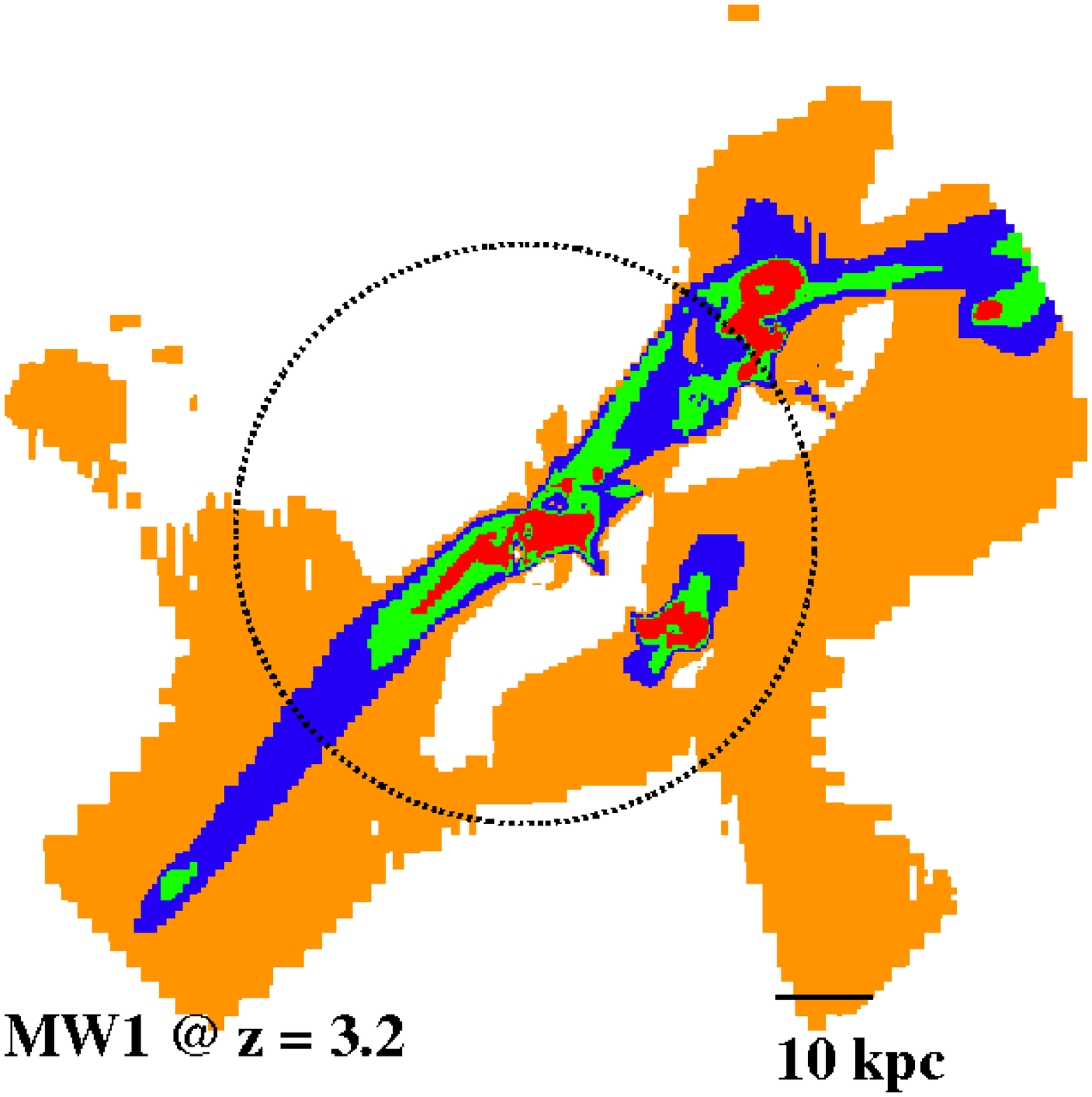}&
\includegraphics[angle=0,scale=0.25]{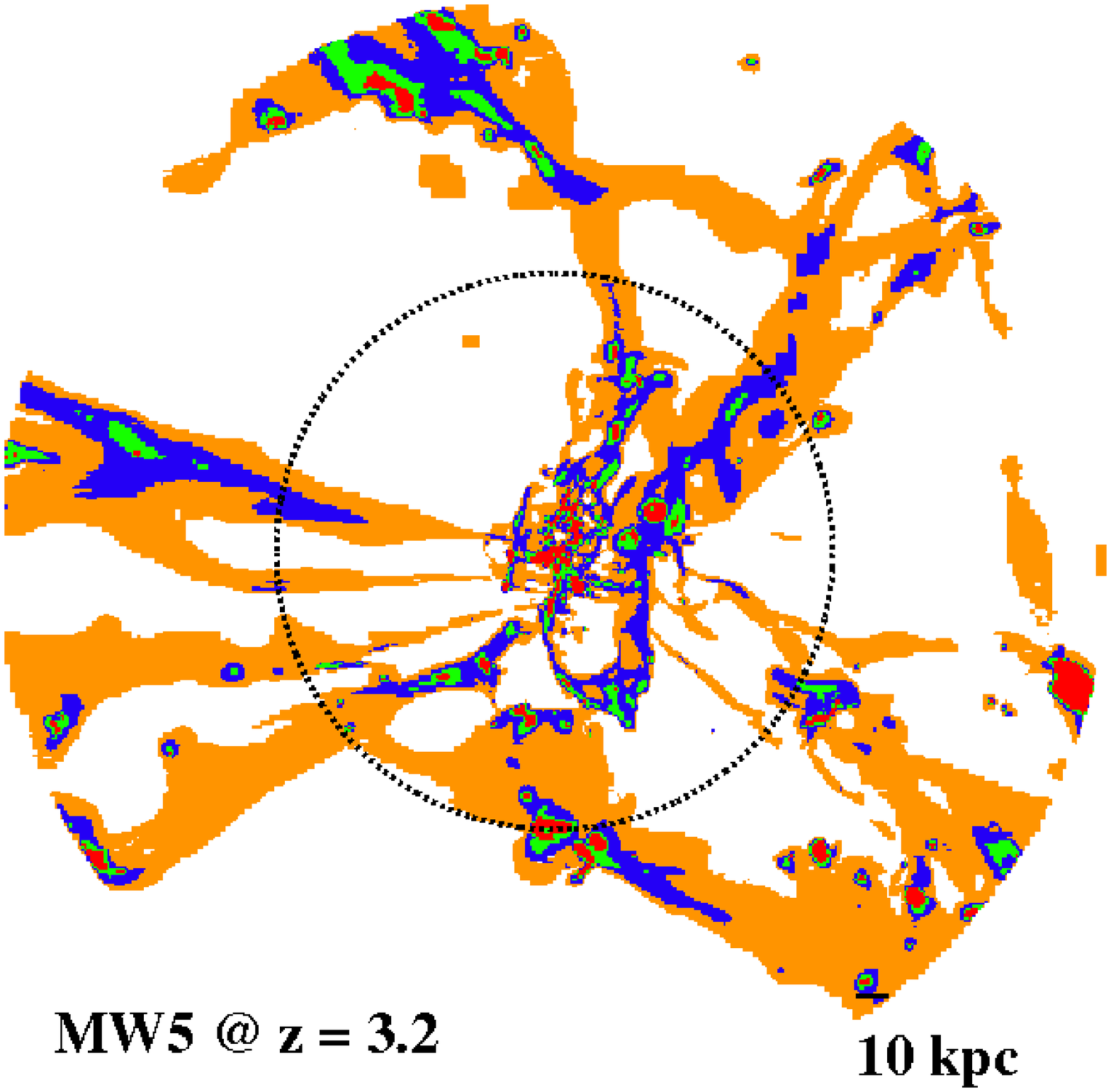}\\
\includegraphics[angle=0,scale=0.25]{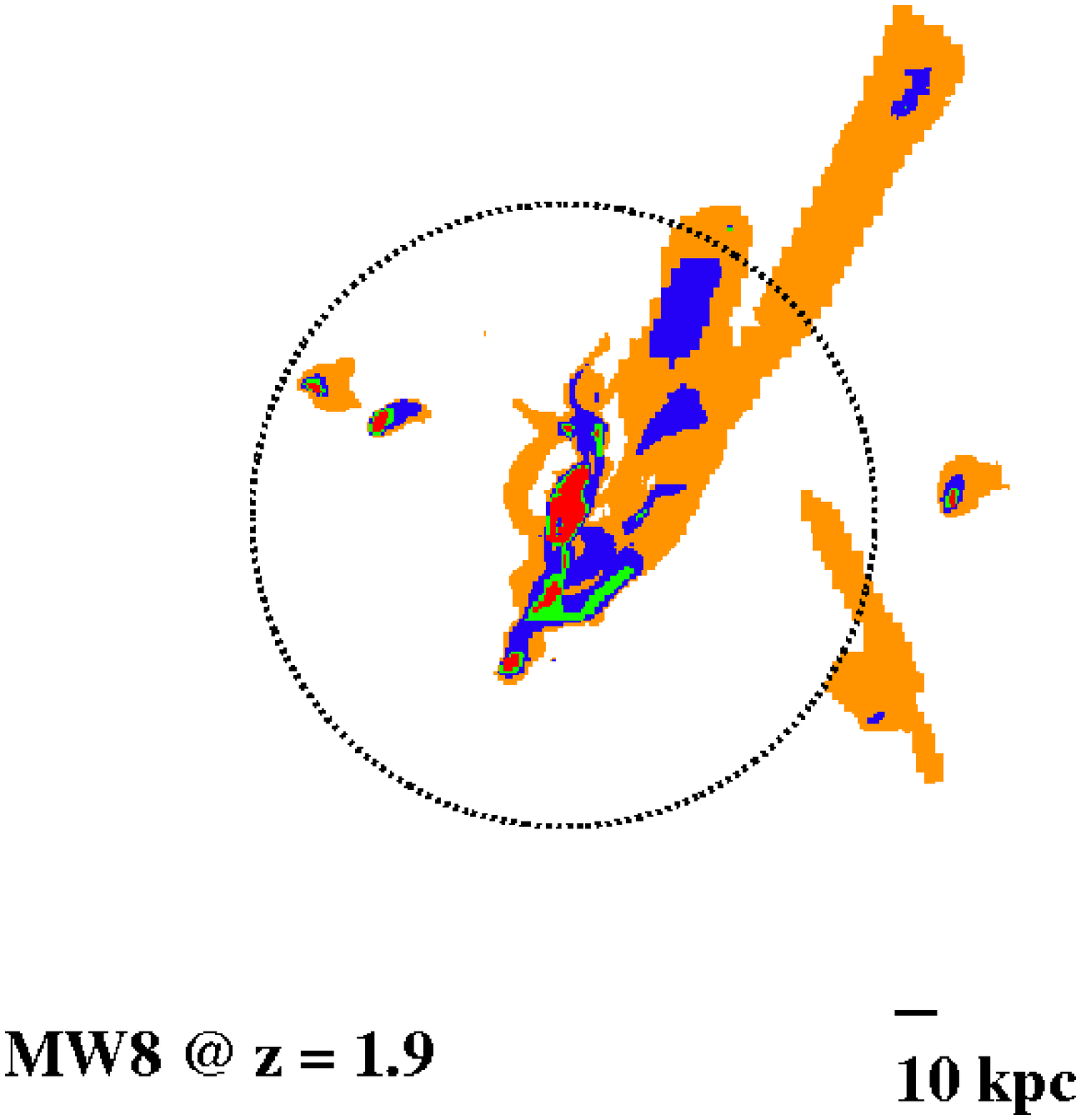}&
\includegraphics[angle=0,scale=0.25]{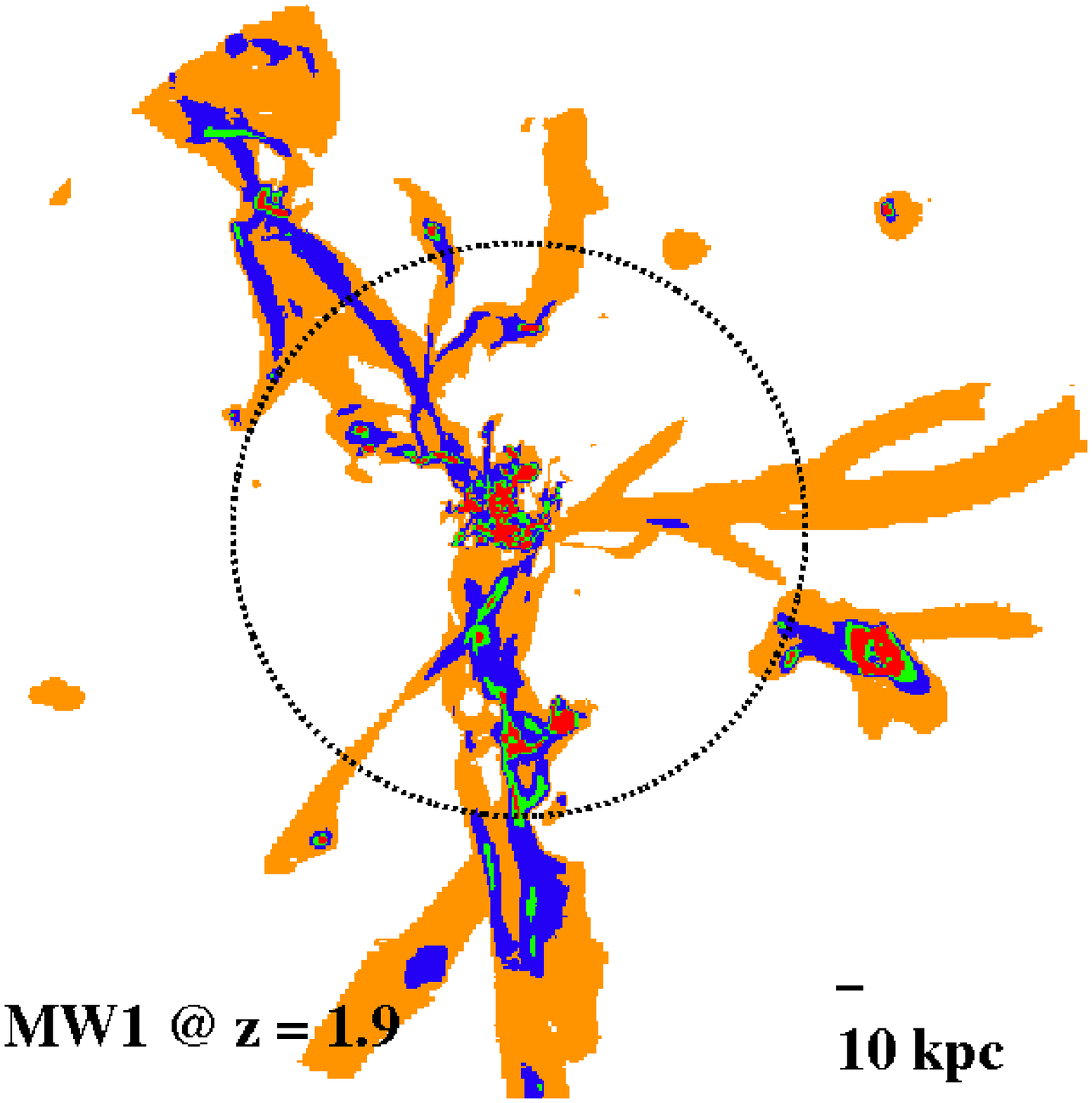}&
\includegraphics[angle=0,scale=0.25]{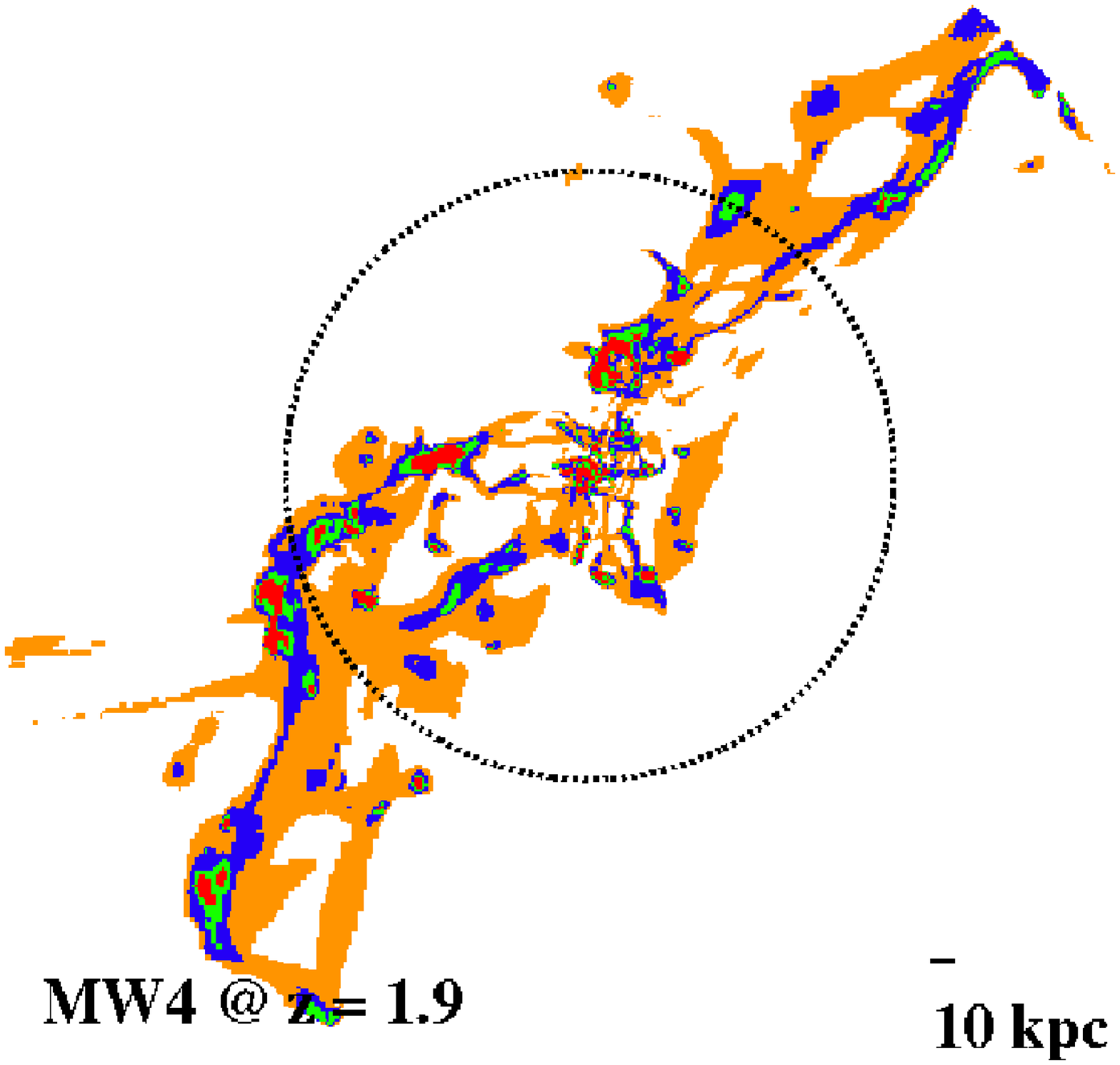}\\
\includegraphics[angle=0,scale=0.25]{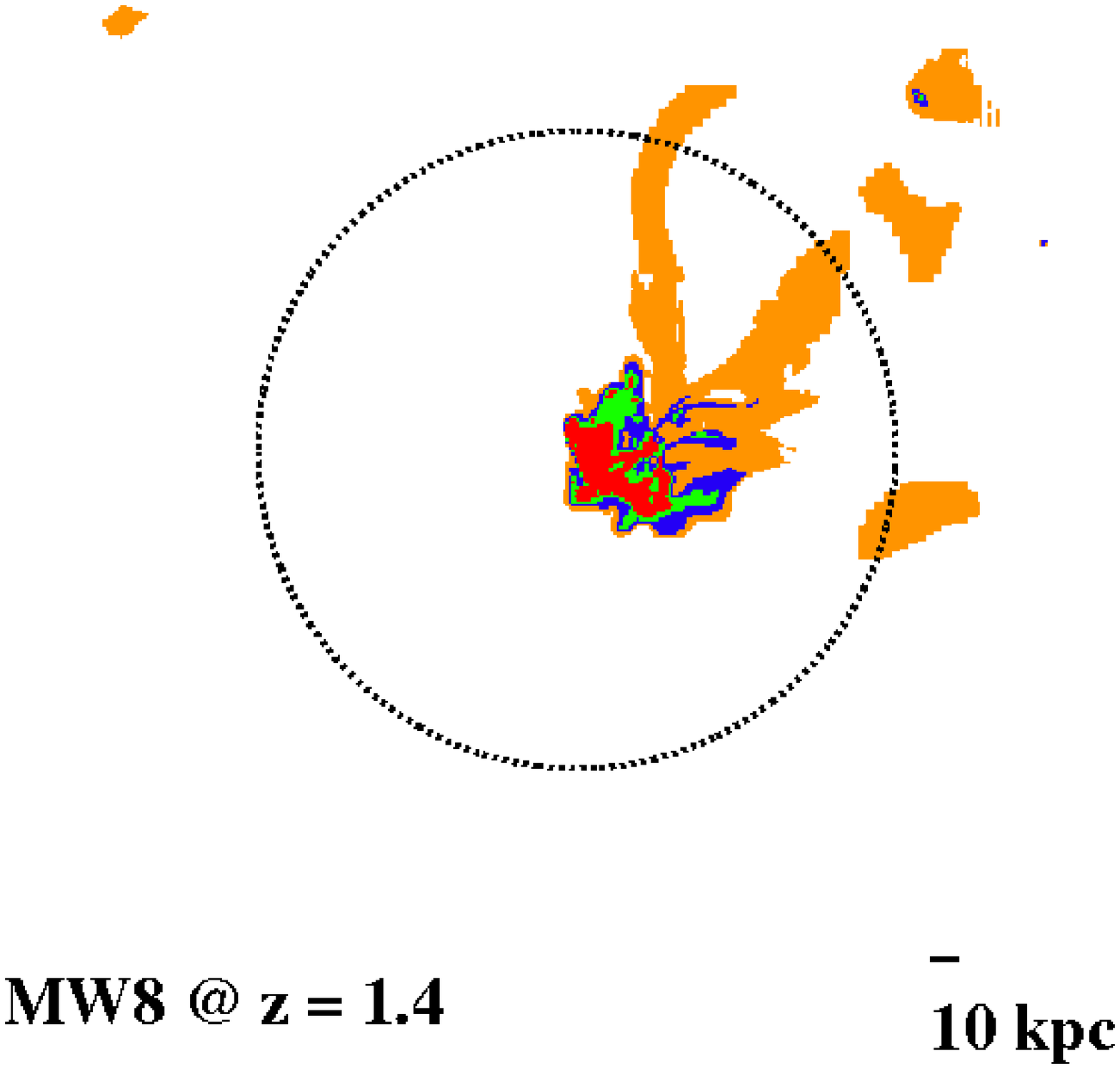}&
\includegraphics[angle=0,scale=0.25]{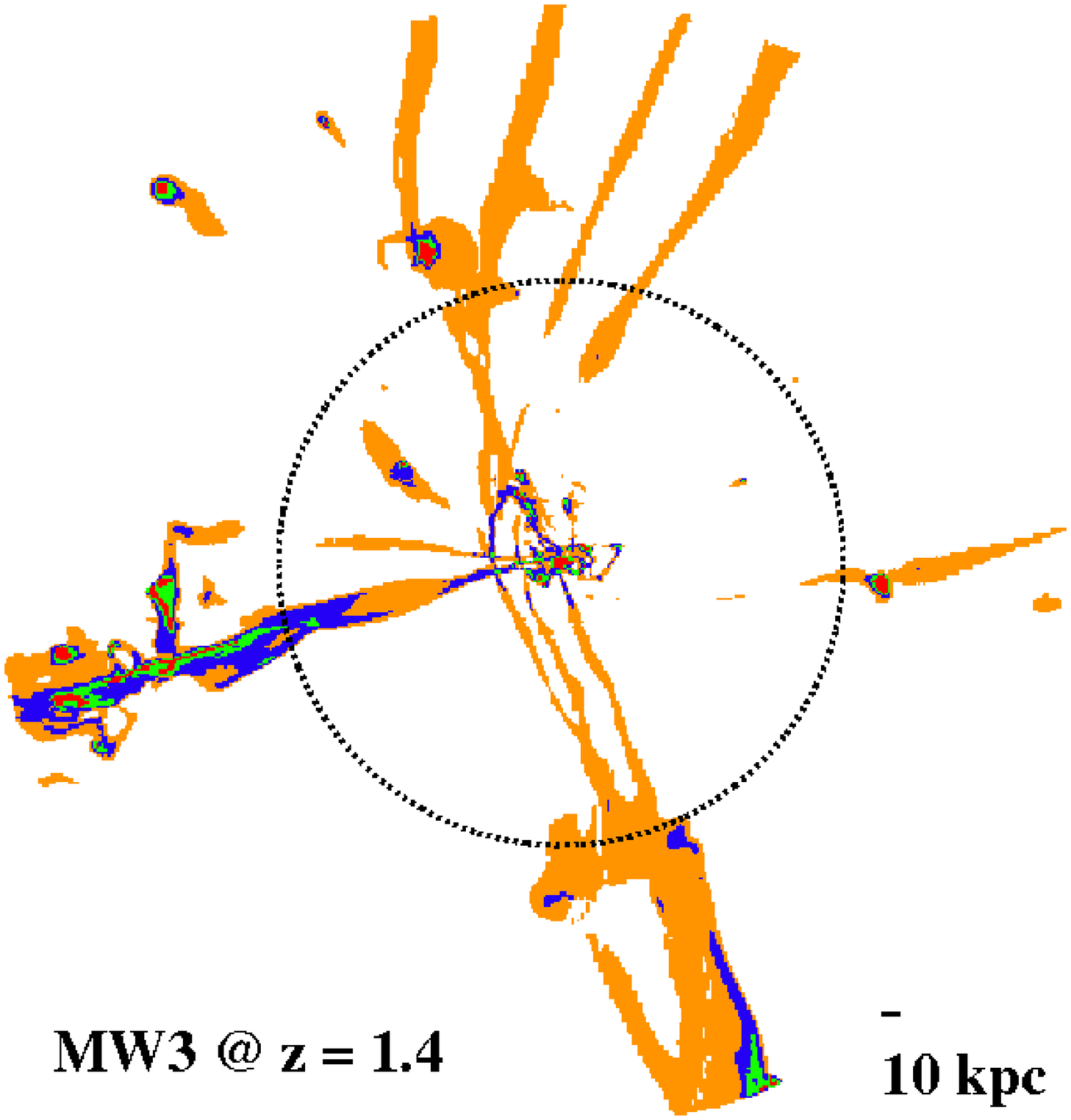}&
\includegraphics[scale=0.22]{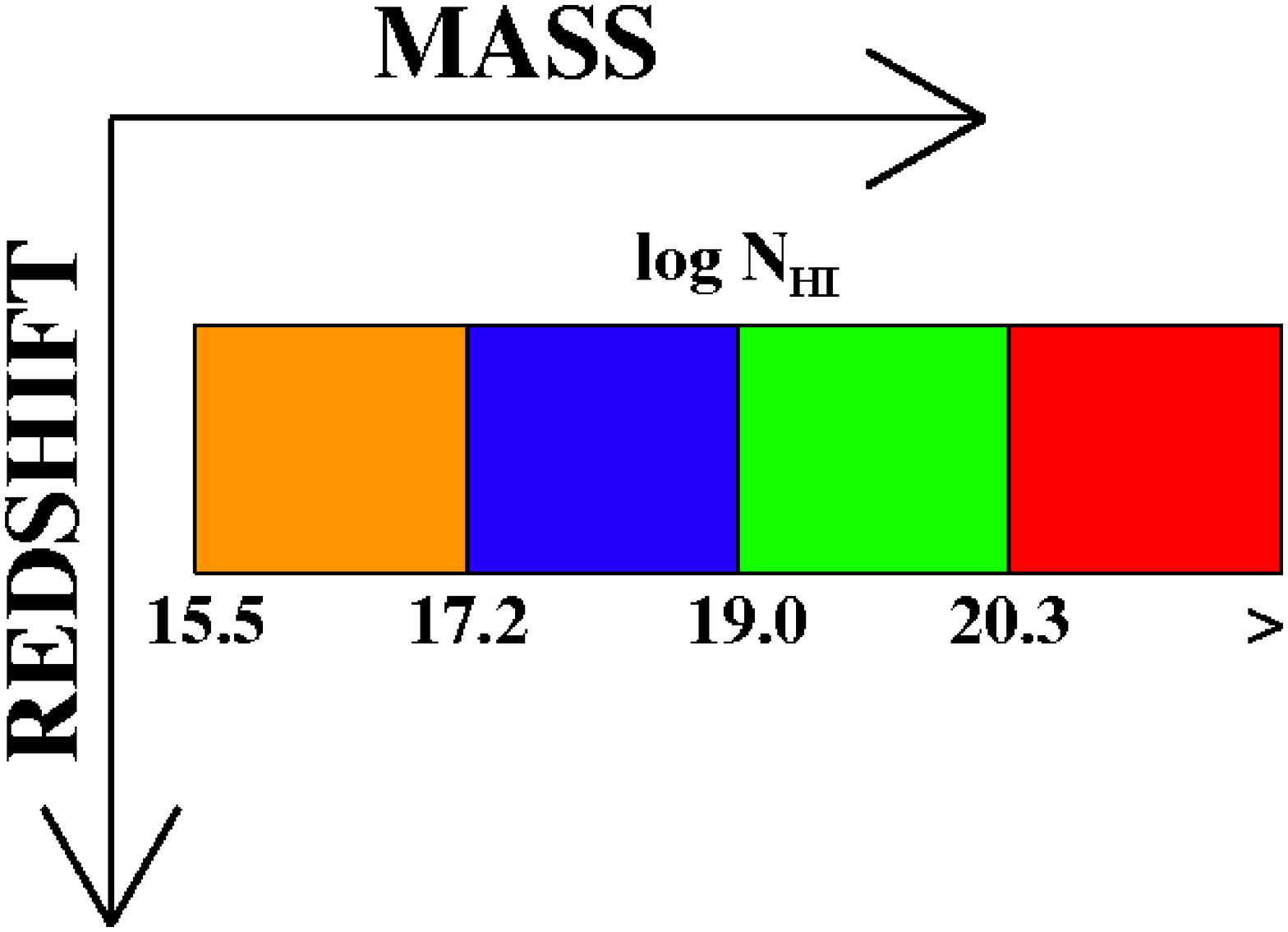}\\
\end{tabular}
\caption{Gallery of projected \HI\ column density  (STAR model) in selected galaxies and redshifts. 
Four intervals of column density are marked with different colors (DLAs in red, SLLSs in green, 
LLSs in blue and MFP gas in orange, similar to Figure \ref{fig:cfmw3}). 
Redshift is decreasing from top to bottom, and virial mass is increasing at constant 
redshift from left to right. The dotted circles mark the virial radius. 
Cold streams are ``patchy'', with neutral pockets of gas embedded in 
a more widespread ionized medium. A difference is seen between $z > 3$ and $z < 2$, and a marginal mass 
dependence is hinted, but the latter could be an artifact of numerical resolution.}
\label{fig:galccd}
\end{figure*}

\begin{figure*}
\includegraphics[angle=90,scale=0.52]{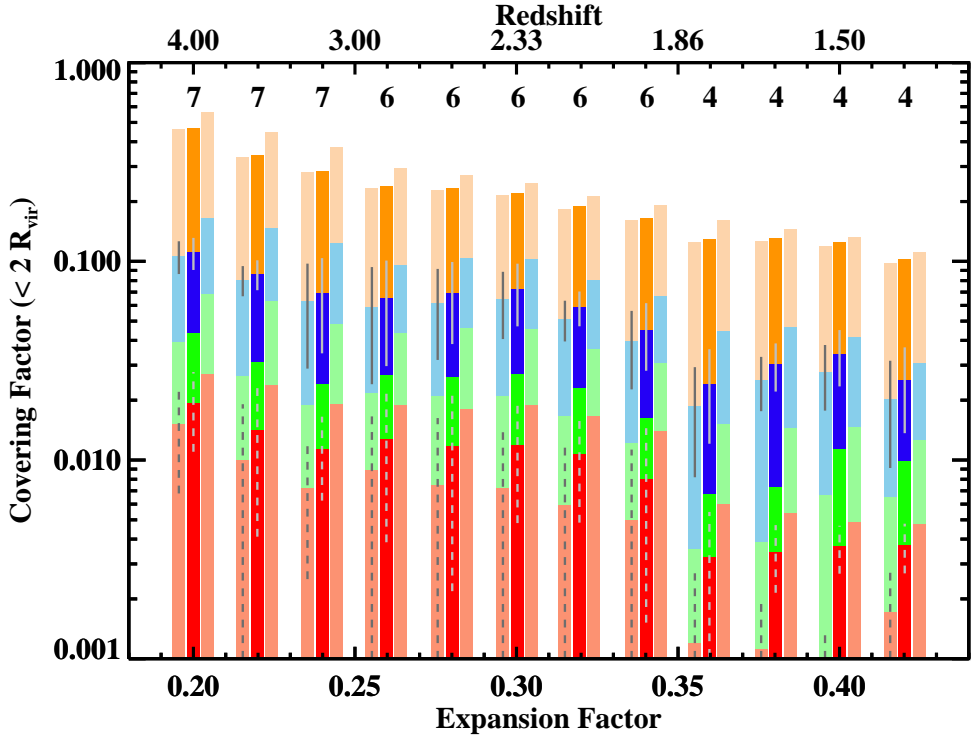}
\caption{Time evolution of the cumulative covering factor within 2\rvir.  
DLAs are in red, SLLSs in green, LLSs in blue, 
and MFP gas in orange. The central columns are for the galaxies and streams in the STAR model, 
the left columns are for the streams alone in the STAR model and the right columns are 
for the UVB model, galaxies and streams. The numbers above the bars indicate 
the number of galaxies in each bin of redshift. Gray dashed and solid lines indicate the 
standard deviation for the mean covering factor of DLAs and LLSs respectively. 
The covering factor slowly decreases with redshift at all column densities, reflecting 
the decline of cosmological density.}
\label{fig:cfred}
\end{figure*}

\begin{figure*}
\includegraphics[angle=90,scale=0.52]{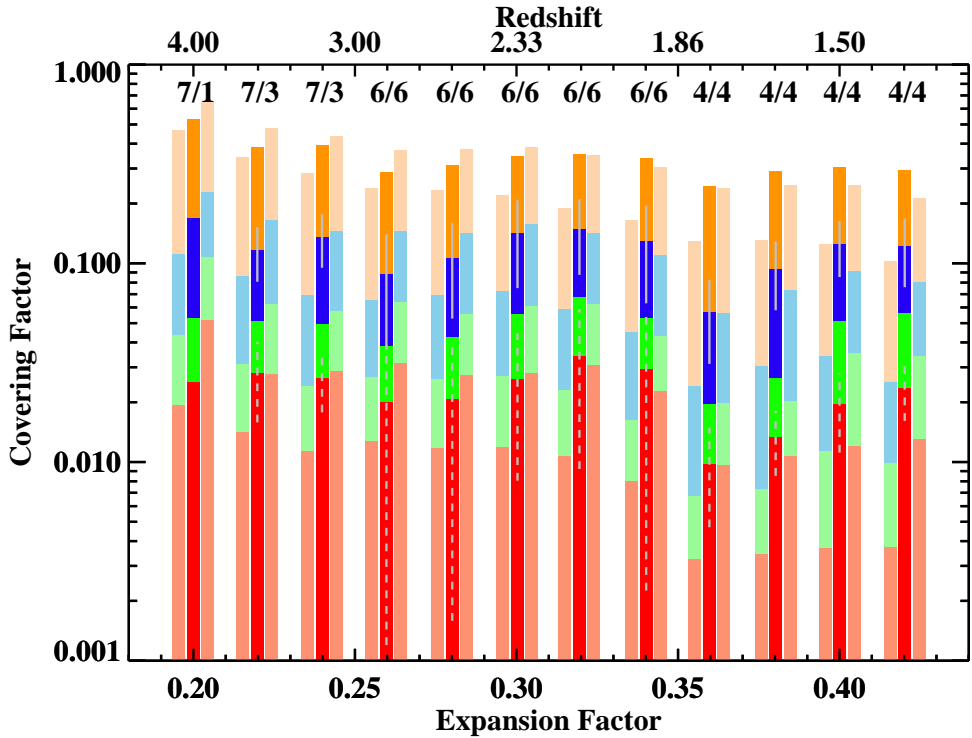}
\caption{Same as Figure \ref{fig:cfred}, comparing  covering factors that are 
measured in different areas.  In the center, the covering factor is 
measured within a fixed aperture of 10 arcsec in radius. 
Shown for comparisons are the covering factor defined within \rvir\ and 2\rvir, on the right and left, 
respectively.  The number of galaxies included in each bin is shown above 
the columns (variable/fixed apertures) and the standard deviation around the mean is 
highlighted with solid and dashed lines. 
The variation with time is much weaker when the aperture is fixed, 
reflecting the decrease of covering area with distance from the halo center.}
\label{fig:diffcf}
\end{figure*}

\subsection{Redshift evolution of the covering factor}\label{statan}

A qualitative description of the covering factor evolution within 2\rvir\ is provided in 
Figure \ref{fig:galccd}, where we show a gallery of projected \HI\ column densities (STAR model) 
for a selected number of galaxies. Redshift decreases from the top to the 
bottom and mass at fixed redshift increases from the left to the right. 
DLAs are marked in red, SLLSs in green, LLSs in blue and MFP gas is in orange. 
In each panel, the galaxy center and virial radius are indicated by a dotted circle. 
Gas that surrounds the galaxies is non-uniform and the cold streams are
``patchy''  with neutral regions that are embedded in a widespread 
ionized medium. 

An evolution in redshift appears from this figure.
The gas coverage is larger at early times than at late times.
Evolution in the gas concentration is evident for DLAs.  At highest redshifts, 
there are large regions of neutral gas within the streams.  These become more 
confined to the central disks and satellites at later times.
A trend in mass is also visible, although less pronounced, with larger gas coverage in 
lower mass halos. Also, the topology of the streams seems to evolve with time and mass.
In low mass systems and at early times, wide streams penetrate 
inside the virial radius to deliver cold gas to the forming galaxies. At lower 
redshifts and in more massive systems, the streams break into a series of narrow 
filaments \citep[see][]{ker09,fau10b,stw10}. In our simulations, numerical resolution 
could be responsible for the increased smoothness in the smallest galaxies at the 
highest redshifts and higher resolution is required to confirm whether this trend is real.

\begin{figure}
\includegraphics[angle=90,scale=0.32]{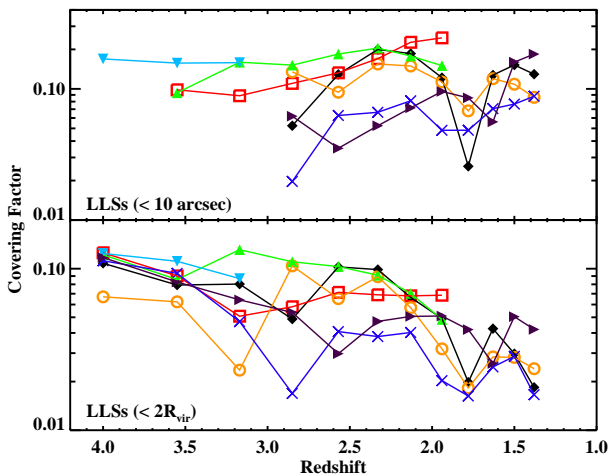}
\caption{Redshift evolution of the covering factor for LLSs 
in individual galaxies (color coded as in Figure \ref{fig:galprop}) 
within 2\rvir\ (bottom panel) and for a fixed aperture of 10 arcsec (top panel).  
Trends similar to those inferred from Figure \ref{fig:cfred} 
and \ref{fig:diffcf} are visible when inspecting individual galaxies, regardless to their mass.}
\label{fig:allp}
\end{figure}

A more quantitative analysis of the redshift dependence of the covering factor is 
given in Figure \ref{fig:cfred}. Each bar, represents the mean values for DLAs (red), 
SLLSs (green), LLSs (blue) and MFP gas (orange) at one redshift and within 2\rvir. 
The central column is for the galaxies and streams in the STAR model, 
the left column is for the streams alone in the STAR model and the right column is 
for the UVB model (streams and galaxies). Values for DLAs and LLSs in individual galaxies 
are listed in Table \ref{tab:crsc}. All covering factors are given as cumulative, i.e.
integrated from the lowest column density to infinity (numerically approximated to 
$N_{\rm HI}=10^{25}$ \cmm). The numbers at the top of the bars 
indicate the number of galaxies included in each redshift bin. The standard deviation 
of the sample is shown for DLAs and LLSs with gray dashed or solid lines
to highlight the wide scatter from galaxy to galaxy, partially associated 
to the mass (see Figure \ref{fig:allp}).

Figure \ref{fig:cfred} confirms the impression that the covering factor within 2\rvir\
is slowly declining with time. At the highest redshifts, $30-50\%$ of the area is covered by MFP gas, 
$7-10\%$ is covered by LLSs and $1-2\%$ is occupied by DLAs \citep[compare with ][]{fau10b}. 
By $z\sim2$, the covering factor becomes $1\%$ for DLAs, $5\%$ for LLSs and
$\sim 20\%$ for MFP gas. In the range $z=2-1.3$, the covering factors are 
$\sim 10\%$, $3-4\%$, and $0.3-0.4\%$ respectively. 
Note that the already limited statistics are dominated by lower mass 
galaxies below $z\sim 1.9$, while the highest redshift bins may be more prone to resolution effects.

In the redshift interval $2<z<3$, where the sample is more homogeneous, the 
time evolution of the covering factor can be parametrized by $\sim a^{-d}$, with
$a$ the expansion factor. This trend can be explained by 
the universal expansion for a self-similar density distribution of neutral hydrogen. 
In these simulations, the cumulative \HI\ column density 
PDF can be described by $\propto N_{\rm HI}^{-c}$, with $c \sim 0.2$ for $N_{\rm HI} \lesssim 10^{20}$ \cmm\
and $c\sim 1$ at higher column densities. In our sample, the viral mass and the virial  radius 
grow with time as $\sim a^{3}$ and $\sim a^{2}$, and 
the virial column density decreases as $\sim a^{-1}$. For a self-similar density distribution
of neutral hydrogen, it follows that $c \sim d$, in agreement with the evolution in Figure \ref{fig:cfred}.

Comparing next the UVB model (right bars) with the 
STAR model, a systematic offset towards higher values is apparent 
from the former. The inclusion of local sources does not drastically alter 
the covering factor (a factor of two for DLAs and LLSs), but it is clear that 
ionizing radiation from stars has appreciable effects especially at intermediate column 
densities and it is required for a consistent comparison with observations \citep{sch06}.
The redshift evolution is still present in the UVB model, reassuring us that it is not 
driven by local sources. Furthermore, this trend cannot be artificially induced by our 
choice of a constant UVB, since the intensity of the UVB decreases towards the 
highest redshifts here considered.

In Figure \ref{fig:cfred}, we finally compare the covering factor from the streams and galaxies
with the streams alone.  Overall, the covering factor of highly optically thick gas 
in the streams is decreasing faster with time than that of galaxies. 
For the MFP gas, this effect is absent, with streams responsible for the totality of the covering factor
at all times. Streams contribute to more than 80\% of the LLSs covering factor over the full redshift
interval. Instead, a rapid evolution is seen for SLLSs and
DLAs. Streams are responsible for as much as 70\% of the covering factor at the highest redshifts, 
but less then $30\%$ of the covering factor at later times as neutral gas accumulates 
in proximity of the central galaxies an satellites.

Being a normalized quantity, the covering factor within 2\rvir\ is suitable for comparisons 
among different galaxies and redshifts. However, it does not provide direct information on the 
physical cross section and its evolution. Further, observations probe either the gas 
cross section or the covering factor within a given projected distance from a source.
In the latter case, the most natural quantity to compute is the covering factor within a 
fixed angular distance from the halo center. In Figure \ref{fig:diffcf}, we show the 
mean covering factor within \rvir\ (right columns), 2\rvir\ (left columns) and within 
a circular aperture of 10 arcsec in radius (central column). Again, the number 
at the top indicates the number of galaxies included in each bin
with the second number for the fixed aperture. The variance about the mean for LLSs and 
DLAs is shown with a solid and dashed line, respectively. At the redshifts examined here, 
an angular distance of  10 arcsec corresponds to $70-85$ proper kpc, equivalent to 
$\sim 3-5$ times the virial radius at the highest redshifts and to 
$\sim 0.5-1$ times the virial radius at the
lowest redshifts. The snapshots in which the extraction box exceeds the simulated region are not included.
Further, for a consistent comparison,  we consider a path length through the grid of twice the virial radius
in all three cases. 

The covering factor measured within \rvir\ is higher than the one inside 2\rvir, particularly
for DLAs. This is trivially expected for a gas distribution with 
declining column density. Considering now a fixed angular distance, the covering factor exhibits a much weaker
evolution with a modest increase in the interval $2<z<3$, where the sample is more homogeneous.
As the universe expands, the mean column density within 2\rvir\ drops and the fraction of 
the area within 2\rvir\ subtended by gas above a 
fixed column density threshold decreases accordingly. 
However, the virial radius is increasing with respect to the size of the extraction aperture,
resulting in a nearly constant or slightly increasing covering factor.

Within $\sim 100/h~\rm kpc$, the covering factor of 
absorbers in the transverse direction of QSO host galaxies
is significantly higher than our estimates, being
unity for $N_{\rm HI}>1.6 \times 10^{17}$ \cmm\ \citep{hen06}
and 30\% for $N_{\rm HI}>10^{19}$ \cmm\ \citep{hen07}.  
This discrepancy may originate in the environment around QSOs,
more massive than the galaxies in our sample. Further, these simulations do not resolve 
overdensities on very small scales and consequently the cross section distribution 
suffers from an additional uncertainty. For example, a population of neutral clouds 
of $\lesssim 100$ pc in size that are pressure confined by the hot halo gas 
\citep[see e.g.][]{dek08,pro09c,bir10} may be missed in these simulations and
the cross section of optically thick gas would be underestimated. 
Similarly, shells or bubbles associated with radiative shocks on small scales may not be 
properly resolved \citep[cfr.][]{sim06}.

The weak redshift evolution seen in Figure \ref{fig:cfred} and Figure \ref{fig:diffcf} may be
the result of having combined galaxies at different mass. To explicitly test this hypothesis, in Figure 
\ref{fig:allp} we display the LLS covering factor within 2\rvir\ (bottom panel) and within 10 arcsec
(top panel) for individual galaxies. Although with significant scatter from object to object, the covering factor 
within 2\rvir\ appears to be roughly constant or slowly declining regardless to the galaxy mass.
Similarly, the covering factor within a 10 arcsec aperture is slowly increasing at all masses.

\section{Predicted absorption line statistics}\label{alsingal}

In the previous section we discussed the redshift evolution of the covering factor in 
streams and galaxies, where we found that only a small fraction of the virial area 
is covered by optically thick gas. In this section, we wish to translate
the cross sections and covering factors previously measured into more direct observables
that can be used in comparison to real data to uncover signatures of cold gas accretion. 
We start by introducing some of the formalism adopted in typical measurements of ALSs. 
Readers familiar with this notation can skip the next sub-section
and continue from Section \ref{sec:fnx}, where we derive the 
column density distribution function for cold-stream fed galaxies 
and discuss their visibility in absorption.

\subsection{Common formalism}

To probe any form of gas that lies along the line of sight, observers collect large spectroscopic samples 
of QSOs that serve as randomly distributed background sources \citep[e.g.][]{pro05,ome07}. 
In this experiment, there is no {\it a priori} knowledge of the foreground system \citep[see however][]{raf10}
and one can directly measure the probability to intersect gas along a random sightline. This quantity translates
to a sky covering factor that is the percentage of total observed 
area covered by gas.

One can define the incidence  
\begin{equation}
\ell (z) = \frac{N_{\rm abs}}{\Delta z}
\end{equation}
as  the number of systems $N_{\rm abs}$ detected across the total searched redshift
$\Delta z$. This quantity is related to the probability of 
intersecting a system along the line of sight at redshift $z$.
In analogy with the idealized experiment of a body moving in a sea of particles 
with number density $n$ and cross section $\sigma$, the following equality holds
\begin{equation}
\ell (z) dz = n_{\rm abs} \sigma_{\rm abs} dl_{\rm c}
\end{equation}
with $n_{\rm abs}$ the comoving number density of absorbers,  $\sigma_{\rm abs}$ their
comoving cross section and  $dl_{\rm c}$ the infinitesimal comoving path length.

Within a given cosmology $H(z)=H_0[\Omega_\Lambda+(1+z)^3\Omega_m]^{1/2}$
and   $dl_{\rm c}=c/H(z)~dz$.
Replacing z by an ``absorption length'' $X$
\begin{equation}
dX=\frac{H_0}{H(z)}(1+z)^2dz
\end{equation}
and using the identity $\ell(z)dz=\ell(X)dX$, the incidence can be rewritten as 
the number of systems per absorption length \citep{bah69}
\begin{equation}\label{eq:lofx}
\ell (X) = \frac{c}{H_0} n_{\rm abs} \phi_{\rm abs} 
\end{equation}
where the cross section $\phi_{\rm abs}$ now represents the physical area of the absorber.

Equation (\ref{eq:lofx}) is the zeroth moment of the column density distribution function \fnx\
that characterizes the number of ALSs 
per absorption length as a function of the \HI\ column density. 
From the definition of \lox, it follows that
\begin{equation}\label{eq:fnx}
f(N_{\rm HI},X) = \frac{c}{H_0} n_{\rm abs}(N_{\rm HI},X) A_{\rm abs}(N_{\rm HI},X)\:, 
\end{equation}
with $ A_{\rm abs}$ the physical area per unit \NHI.
Equation (\ref{eq:fnx}) reveals that absorption line statistics directly probe the shape and 
the evolution of the physical cross section times the number density of absorbers 
rather than the covering factor of gas around these systems.

\begin{figure}
\includegraphics[angle=90,scale=0.32]{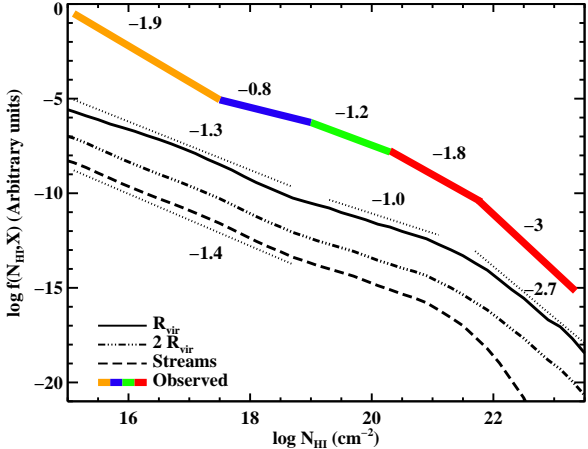}
\caption{Comparison of the shape of the simulated and observed column density function \fnx.
The normalizations of the different curve are arbitrary, to highlight the shapes only.
The simulated \fnx\  from the STAR model at $z\sim3.5$ is shown for gas inside 
the virial radius (solid line), for gas within 2\rvir (dash-dotted line), 
and with for the streams only (dashed line). The observed \fnx\ at $z \sim 3.7$ from 
\citet{pro10} is shown as a colored curve. Power law regressions to the simulation results 
in intervals of column density are shown as dotted lines.
The numbers denote the power law indexes.
Simulations of massive galaxies at high resolution reproduce the observed cutoff at high column density,
but predict an \fnx\ that is shallower than observed around $N_{\rm HI}\sim10^{20}$ \cmm\ and for SLLSs.
Cold streams dominate the \fnx\ below $N_{\rm HI} \sim 10^{18}$ \cmm,
but a larger contribution from the intergalactic medium is required at lower 
\NHI\ to satisfy the constraints imposed by the observed mean free path of ionizing
radiation.}
\label{fig:fnx_shape}
\end{figure}

\subsection{The column density distribution function}\label{sec:fnx}

To understand the nature of the gas probed in absorption, it is common practice 
to reproduce the \fnx\ in its shape and normalization from simulations
of large volumes, required for statistical significance. 
Here we take a different approach where we derive an \fnx\ that, by construction, 
is due to gas associated only with massive galaxies and the streams with
merging galaxies that feed them. Any similarity or discrepancy with the 
observed distribution function informs the features 
of \fnx\ that reflect the gas distribution in these particular class of systems
as predicted by these simulations. 
The reader should keep in mind that our sample and our \fnx\ are
not intended to be representative of the full ALS population.

Our model for \fnx\ follows the definition in Equation (\ref{eq:fnx}). 
The physical area per unit \NHI\ is constructed from the average cross section
distribution function over all the galaxies at a common redshift
\begin{equation}
\overline{A_{\rm gal}(N_{\rm HI},z)}=\sum_i^{N_{\rm gal}}
\frac{\phi_{\rm gal,i}(N_{\rm HI},z)}{N_{\rm gal}\Delta N_{\rm HI}}\:.
\end{equation}
We consider the virial radius to define the gas that is physically associated with a 
galaxy, but we construct also a second model considering the cross section within 2\rvir\
to stress the effects of the cold streams.
For the comoving number density of absorbers, we use the cumulative 
number density\footnote{The comoving number density is computed following the approximations
described in \citet[][Appendix A]{dek06}. Recent simulations \citep[e.g.][]{kly10} reveal 
deviations from these approximations, but they become relevant only above $z\sim4$.} 
of galaxies that are more massive than the least massive galaxy 
at a given redshift. In practice, we construct an average cross section that we attribute to 
all the halos of equal mass or more massive than the galaxies in our sample.
This procedure implies an extrapolation towards higher masses, since 
systems much above $10^{12}$ \msun\ are not represented in our sample.
Due to the scatter from galaxy to galaxy in the cross section and to its mass dependence, 
$\overline{A_{\rm gal}(N_{\rm HI},z)}$ and the final \fnx\ should be regarded as approximations.

In Figure \ref{fig:fnx_shape} we compare the functional form of 
the \fnx\  derived for the STAR model at $z\sim3.5$ within the virial radius (solid black line),
within twice the virial radius (dashed-dotted line), and in the streams alone (dashed line).
The observed distribution at $z\sim 3.7$ from  \citet{pro10} is superimposed with colors.
For DLAs and SLLSs (i.e. $\mNHI > 10^{19} \, \rm cm^{-2}$ where one
resolves the damping wings of \lya), \fnx\ is derived from observations of individual absorbers,
but at lower column densities observations provide only integrated constraints on the
distribution function. For this reason, the shape of \fnx\ in the LLSs and MFP region
is more uncertain. In Figure~\ref{fig:fnx_shape}, we report the preferred model from \citet{pro10}.

Although not a perfect match, these models reproduce some 
of the features of the observed \fnx. Three different intervals can be identified.
First, there is a steep decrease at high column density with an evident break 
around $N_{\rm HI}=3\times 10^{21}$ \cmm. From our model, we measure a power law 
$\sim -2.7$ for $\log N_{\rm HI} \gtrsim 21.7$ (in \cmm; dotted line), similar to the 
observed slope $\sim -3$. This cutoff naturally emerges from our 
high resolution simulations of massive 
disks, while it is hard to reproduce in cosmological simulations at
lower resolution \citep[e.g.][but see \citealt{alt10,mcq11}]{pon08,tes09,cen10}.
The inclusion of molecules would suppress even further the number of systems with \HI\ above 
$10^{22}$ \cmm (see Figure \ref{xhi_all}), bringing the model to even better agreement 
with observations. 

Moving towards lower column densities, the model exhibits a plateau with slope 
$\sim -1$ between $N_{\rm HI}=3 \times 10^{18}-10^{21}$ \cmm.  The observed \fnx\
distribution also flattens below $10^{20}$\cmm\ \citep{ome07,pro10}
which has been interpreted as the result of transitioning from
predominantly neutral gas to predominantly ionized \citep{zhe02,mcq11}. 
In fact, our model flattens at somewhat high $N_{\rm HI}$ and underpredicts the
observed incidence of SLLSs relative to DLAs. Provided that 
SLLSs arises from galaxies, perhaps this discrepancy would be resolved by including 
less massive galaxies than those modeled here. As previously noted for the cross section,
we might be missing a population of small clouds of neutral hydrogen due to limited resolution. 
Finally, below  $N_{\rm HI}=10^{18}$ \cmm\ the distribution 
function becomes moderately steeper with slope $\sim -1.3$.  This qualitatively follows
the observed distribution, but it is not sufficiently steep to reproduce
the integrated mean free path to ionizing radiation at $z \sim 4$
\citep{pro09b}.  We infer, therefore, that gas beyond the virial
radius must contribute significantly at these column densities \citep[see also][]{koh07}.
As already seen in Figure \ref{fig:cfmw3}, due to the stream contribution (dashed line)
the \fnx\ steepens mildly below $N_{\rm HI}\sim 10^{18}$ \cmm, but gas within 2\rvir\ is
not enough to reproduce the observed slope (dash dotted line). Contribution from the IGM at 
larger radii may be required.

\begin{figure*}
\includegraphics[angle=90,scale=0.5]{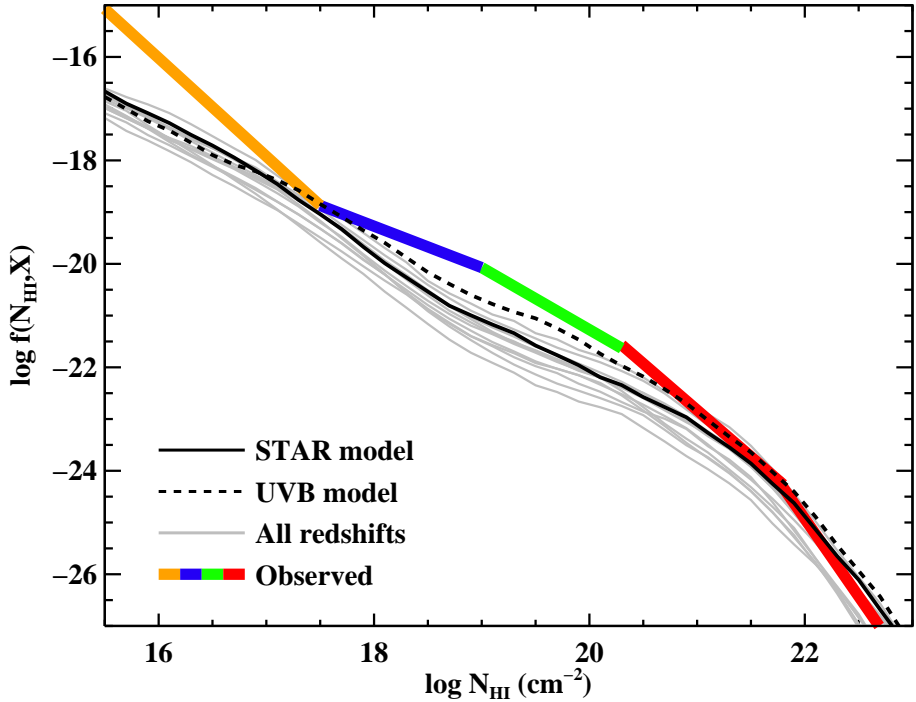}
\caption{Column density distribution function from observations at $z \sim 3.7$ 
\citep[colors;][]{pro10} and simulations for the STAR (solid line) and the UVB (dashed line) 
models at $z=3.5$ within the virial radius. Shown in gray are simulated \fnx\ for different redshifts
in the interval $z=4-1.4$. There is a satisfactory agreement between theory and observations
for DLAs and at $N_{\rm HI}\sim 10^{18}$ \cmm, 
while the two distributions diverge at lower column densities. 
SLLSs and high density LLSs are underrepresented by these models. 
The \fnx\ exhibits modest redshift evolution in shape, in agreement with observations.}
\label{fig:fnx_inv}
\end{figure*}

Figure \ref{fig:fnx_inv} compares again the observed \fnx\ with the model inside the virial 
radius, including at this time the absolute normalization. The observed \fnx\ has been 
corrected to match our adopted cosmological parameters. Values are listed in Table \ref{tab:fnx}. 
This figure reveals an overall satisfactory agreement 
between the observations and the models  for high column density DLAs and at 
$N_{\rm HI}\sim 10^{18}$ \cmm.  As already noted, the two distributions diverge entering 
the optically thin regime, suggesting a missing IGM contribution in the simulations. 
Also missing are SLLSs and high column density LLSs that are not found in large 
enough numbers within the halos of massive galaxies.  
The UVB model, superimposed as a dashed line, has a larger fraction of 
LLSs and SLLSs, but not enough to account for the missing systems. 
It is interesting to note that in these models 
SLLSs originate from gas at the outskirts of the disks and are ionized by the leaking UV photons from 
stars that are forming in the disk \citep{sch06}. While most of the \fnx\ is shaped by the UVB 
\citep[][]{raz06,nag10}, RT effects from local sources produce non-negligible effects. 
Albeit with all the uncertainties discussed, from this simple comparison we infer that massive disks
shape the \fnx\ at the highest column densities and cold streams contribute non negligibly to 
the LLS population, particularly between $N_{\rm HI}\sim 10^{17}-10^{18}$ \cmm.

Finally, in gray, we superimpose different \fnx\ for the redshift interval $z=4-1.4$.
These models are remarkably invariant with redshift
and this limited evolution in the \fnx\ is consistent with the observation 
that the column density distribution function of DLAs 
preserves its shape over 12 Gyr \citep{pro09}. This is also 
consistent with the present day distribution function from 21-cm observations in
\HI\ rich disks. Approaching the optically thin regime, a larger degree of evolution is observed 
\citep{pro10,rib11}, which is attributed to radiative transfer effects. 
This variation is not captured by our model, probably because massive galaxies are 
less sensitive to the effect of the UVB than the IGM and low mass galaxies which 
are needed in order to fully match the observed distribution. 

The limited evolution in the models presented in Figure \ref{fig:fnx_inv} can be qualitatively understood 
from visual inspection of Figure \ref{fig:galccd}. As the virial radius grows with time, 
the gas preserves its distribution with centrally concentrated neutral regions surrounded by SLLSs. 
In turn, this column density interval marks the transition to 
more filamentary structures that dominate the LLSs and MFP cross section.
More quantitatively, this non-evolution is related to the fact that 
the neutral hydrogen column density PDF is found to be nearly self-similar.
Therefore, in our simulations, the lack of evolution in \fnx\ is driven by 
the distribution of gas within each halo (i.e. the cross section) 
more than by the cosmological evolution (i.e. the number density of systems).

\subsection{Are cold streams visible in absorption?}\label{incidence}

By integrating the column density distribution function, 
one can compare predictions from the simulations with the 
observed incidence of absorbers.
Figure \ref{fig:fnx_inv} shows that the normalization of the simulated
\fnx\ is not too far from the observed one.
In these simulations, massive galaxies ($M_{\rm vir}\sim 10^{10}-10^{12}$ \msun) 
alone can account for $\sim 20\%-30\%$ of the observed DLA and LLS population. 
The inclusion of satellites and streams up to twice the virial radius further increases
this contribution \citep[cfr.][]{mal03}. Although uncertain up to a factor two, these numbers suggest that a
non-negligible fraction of observed ALSs originate in galaxies and streams as represented 
in these simulations.

This result is not at odds with the low covering factor for optically thick gas discussed in 
the previous sections. This is because the incidence of absorbers with background QSOs is sensitive to 
the physical cross section times the number density of halos. Given a large enough number density of 
halos, even galaxies with relatively small covering factor can offer enough total cross section to 
account for a significant fraction of the observed incidence. 
Most of the DLAs are estimated by theory to originate in the 
mass interval $10^{9}-10^{11}$ \msun\ \citep[e.g.][]{pon08,tes09}, with a peak around 
$\sim 5\times 10^{10}$ \msun. In models with strong feedback \citep[e.g.][]{nag07,cen10}, 
the peak shifts to even higher masses. It is not surprising that our sample that 
includes galaxies with $5\times 10^{10}-3\times 10^{11}$ \msun\ at $z\sim 3$ is within a factor of a
few of the observed incidence. 

Combining the observed incidence for DLAs and LLSs with the cross section 
from our simulations, we can infer the minimum halo mass required to fully match 
observations. At $z\sim3.5$ ($z\sim2.8$), 
the average cross section within 2\rvir\ is $\sim 304~\rm kpc^2$ ($\sim 530~\rm kpc^2$) 
for DLAs and $\sim 1548~\rm kpc^2$ ($\sim 2010~\rm kpc^2$) for LLSs.
The halo number density required to match the observed incidence of DLAs and LLSs
\citep{pro09,pro10,rib11} is $\sim 0.062~\rm Mpc^{-3}$ ($\sim 0.032~\rm Mpc^{-3}$), corresponding to a minimum 
halo mass of $1.6\times 10^{10}$ \msun\ ($3.6\times 10^{10}$ \msun). This is a factor 
1.4 (2.7) below the minimum mass included in this sample. 
This calculation is very crude since, for example, at lower 
masses the mean cross section is expected to decrease (e.g. Figure \ref{fig:dlaxsec}).
Further, these results are sensitive to the radius adopted for the cross section determination, 
here chosen to be 2\rvir, i.e. the size of the re-simulated box.  
Since ideally one would adopt the mean separation between halos, larger than 2\rvir\ at these masses,
our numbers may be even underestimated. With these limitations, our calculation suggests that only a small 
extrapolation towards lower masses is needed to fully account for the observed incidence of ALSs. 

In summary, despite their small covering factor in each halo, the cold streams
 should have already been detected in absorption in
large spectroscopic surveys. Since most of the cross section in cold streams 
is predicted to be at $N_{\rm HI}\lesssim 10^{19}$ \cmm (see Figure \ref{fig:cfmw3} and Figure \ref{fig:fnx_shape}), 
LLSs are the best candidates for the cold gas in the inflowing streams.
If there are other components of cold gas in the halo
that may not exist in proper abundances in our current simulations  
(e.g. high velocity cloud analogs or massive outflows), then 
other indicators are required to distinguish between them 
and the cold streams.

\section{Observable properties of the CGM}\label{alsingal2}

In a second class of experiments designed to probe cold gas in absorption, observers select 
an object or a class of objects of interest and use background sources to probe the gas 
distribution around the foreground systems. Using a background quasar (or galaxy) 
and knowing the angular separation between the probe and the object of interest, 
this experiment yields a direct measurement of the angular \HI\ distribution 
\citep[e.g.][]{chr07,mon09,fum10b} about the object of interest 
or of the gas covering factor and kinematics \citep[e.g.][]{ste10}. 
In the next section, we provide a summary of the formalism adopted
in the analysis of absorption lines as well as a more 
technical description on how line profiles are derived from these simulations.
Starting from Section \ref{sec:meta}, we discuss predictions for the 
metallicity and kinematics of absorption lines, providing a comparison with recent 
observations.

\subsection{Common formalism and numerical procedures}

\subsubsection{The equivalent width}

The normalized profile of an absorption line is characterized by the equivalent width
\begin{equation}
W(\lambda_0)=\frac{\lambda_0}{c} \int^{+\infty}_{-\infty} \left[1-\bar{I}(v)\right]dv\:,
\end{equation}
with $\bar I$ the normalized intensity and $\lambda_0$ the rest frame wavelength at the line center.
The equivalent width provides a direct measurement of the light absorbed along the line of sight.
Through the radiative transfer equation, $W(\lambda_0)$ can be related to the gas optical depth
that in turn depends on the gas column density and kinematics and, via the oscillator strength,  
on the quantum mechanical properties of the transition. 

Three different regimes are typically identified in the so-called curve of growth, that describes the 
variation of the equivalent width with both column density and oscillator strength.
At very high column densities, in the damped regime, all the light at the line center is absorbed 
and the equivalent width is a solely function of the column density.
At intermediate column densities, on the saturated portion of the curve of growth, 
the line center is again fully black, but the equivalent width is most 
sensitive to the line of sight component of the gas velocity that regulates 
the line width. Finally, at low column densities, in the linear regime, the equivalent width 
becomes proportional to the column density in the observed direction and the absorption profile is
no longer zero at the line-center. Averaging along different sightlines, one can further probe 
the mean covering factor around a class of sources by looking at the fraction of transmitted 
light close to the line center.


\begin{figure}
\begin{tabular}{c}
\includegraphics[angle=90,scale=0.4]{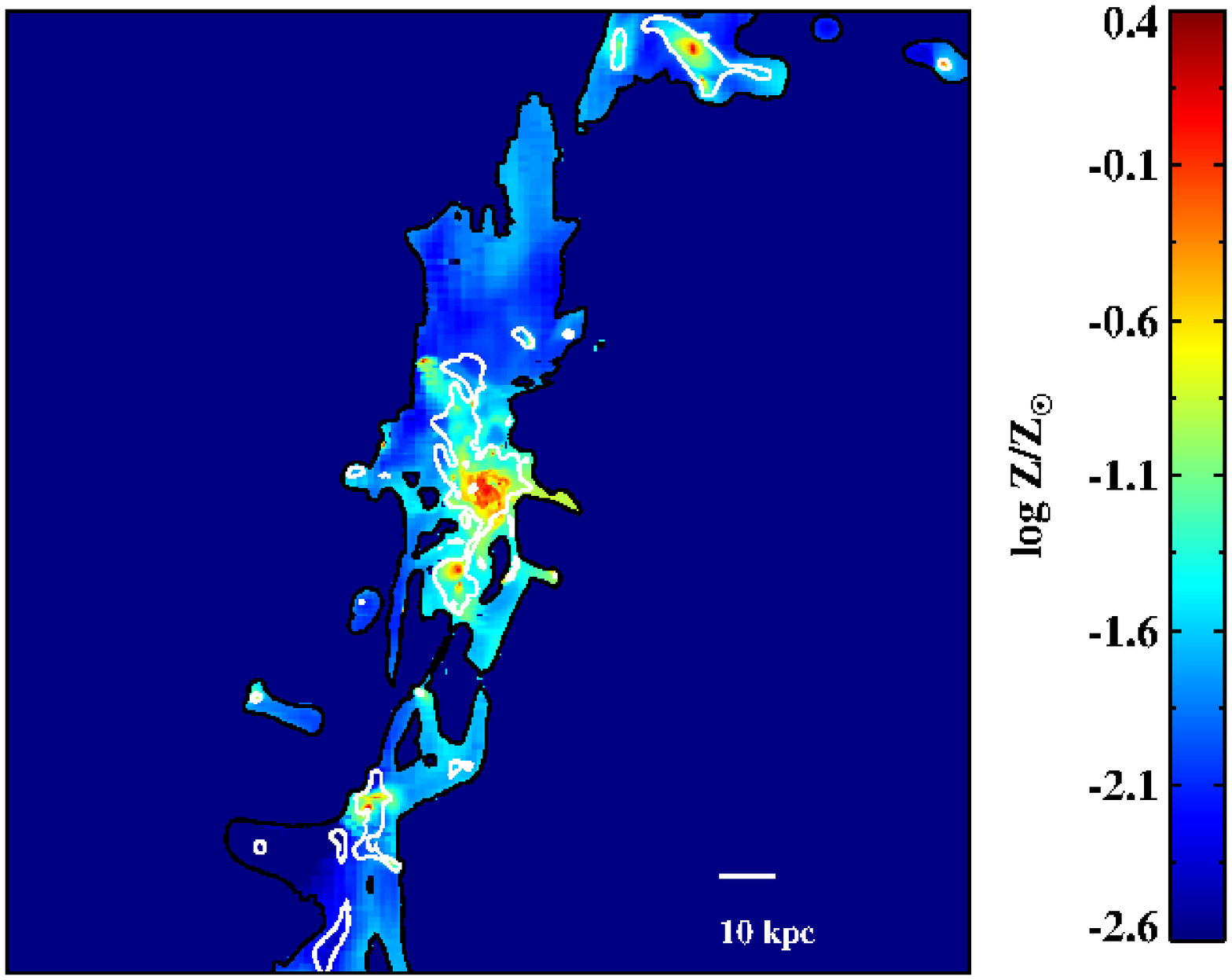}\\
\includegraphics[angle=90,scale=0.4]{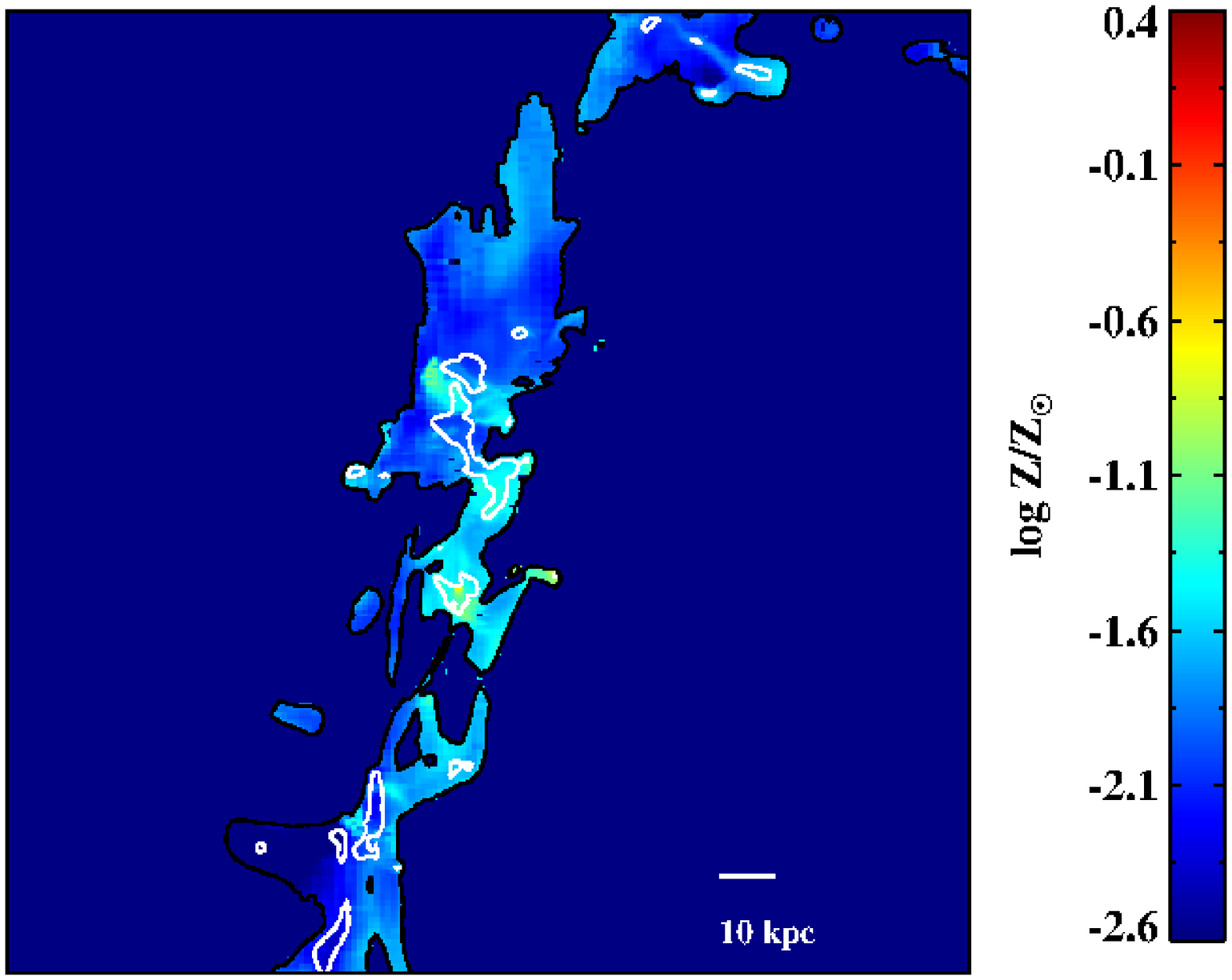}
\end{tabular}
\caption{Metallicity in the optically thick gas of MW3 at $z=2.3$, weighted by 
hydrogen column density. Top: galaxies and streams. Bottom: streams alone.
The black and white contours denote $\rm \log (N_{\rm HI}/cm^{-2})=17.2$ and 
$\rm \log (N_{\rm HI}/cm^{-2})=20.3$ respectively. 
A metallicity gradient is visible. Gas in and around galaxies is more enriched than
in the streams alone.}
\label{fig:metal}
\end{figure}

\subsubsection{Techniques for the simulated line profiles}

To reproduce the absorption line-profile for Ly$\alpha$ and \SiIIl1260, 
we compute the transmitted intensity $I_{\nu}$ of a background source with intensity 
$I_0$ in a random projected sightline $i$
\begin{equation}
I_{\nu,i}=I_0\exp(- \tau_{\nu,i})\:, 
\end{equation}
where the optical depth $\tau_{\nu,i}$ is given by 
\begin{equation}
\tau_{\nu,i}=\sum_{k|i} N_k s \phi_k(\nu)\:. 
\end{equation}
Here, the sum is over all the AMR cells aligned along the $i$-th line of sight, 
and $N_k$ is the column density in each cell. The frequency integrated absorption cross 
section $s$ is given by 
\begin{equation}
s=\frac{\pi e^2}{m_e c} f\:,
\end{equation}
with $f$ the oscillator strength, $e$ the electron charge, $m_e$ the electron mass
and $c$ the speed of light.  The frequency-dependent line-profile is 
\begin{equation}
\phi_k(\nu)=\frac{H_k(u,a)}{\Delta\nu_{D,k}\sqrt{\pi}}\:,
\end{equation}
with the Voigt function 
\begin{equation}\label{eq:voigh}
H(u,a)=\frac{a}{\pi}\int_{-\infty}^{+\infty} dy \frac{\exp(-y^2)}{(u-y)^2+a^2}\:.
\end{equation}
Here, $a=\gamma / (4\pi\Delta\nu_D)$, $u=(\nu-\nu_{l})/\Delta\nu_D$, $y=v/b$ and
 $\Delta\nu_D=(\nu_0 b)/c$. 
In the above equations,  $b^2=(2kT/m_X+\zeta^2)$ is the 
broadening parameter given by that gas temperature and element mass, 
$\gamma$ is the sum over the spontaneous emission coefficients, $\nu$ is the observed 
spectral frequency, and $\zeta=10\rm~km/s$ the turbulent velocity. We assume the central galaxy 
is at zero velocity (corresponding to the rest-frame frequency $\nu_0$) and we compute for each 
gas cell the line center frequency $\nu_l$ corrected for the gas velocity 
along the line of sight. We integrate Equation (\ref{eq:voigh}) using the formalism 
provided in \citet{zag07}. 

All the spectra are computed at a resolution of $4~\rm km/s$.
For \lya, we assume $f=0.4164$ and $\gamma=6.265\times 10^8$. For \SiIIl1260, 
we use  $f=1.007$, and $\gamma=2.533\times 10^9$ and we derive the \SiII\ density
from the neutral hydrogen density and metallicity, assuming $\rm
\log(Si/H)_\odot +12=7.51$ \citep{asp09}. To compensate for the different ionization 
potential of silicon and hydrogen, we boost the \SiII\ volume density by a factor of 10 in cells 
with $x_{\rm HI}<0.1$. The amplitude of this boost depends on the gas and radiation field
properties, but it will become clear from our analysis that a variation of several order 
of magnitudes in the \SiII\
column density is necessary to significantly alter our results. In this simple model, 
we also neglect \SiII\ recombination at low temperature and a lower collisional rate 
in shielded regions. These effects alter the \SiII\ column density, but only in regions of small 
covering factor with minor consequences on the average absorption lines. 
A complete analysis of metal lines requires the inclusion of collisional ionization and 
photoionization, which is beyond the scope of this paper.

\begin{figure}
\begin{tabular}{c c}
\includegraphics[angle=90,scale=0.32]{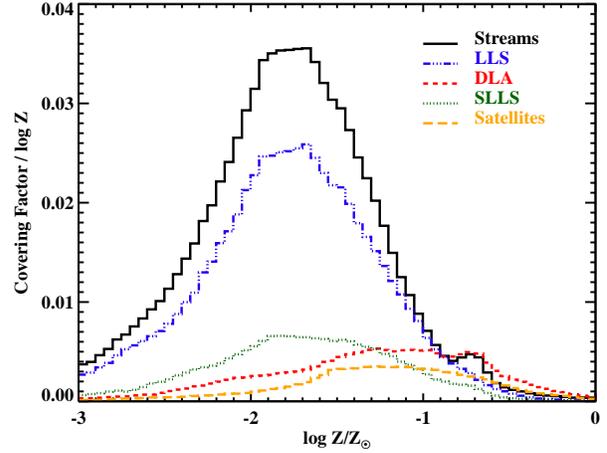}
\end{tabular}
\caption{Distribution of metallicity, column density weighted, in the entire sample
for DLAs (red dashed line), SLLSs (green dotted line) and LLSs 
(blue dot-dashed line). Also shown are the streams alone (black solid line) and satellites 
without centrals (orange long dashed line). 
The smooth stream component is more metal poor than the galaxies, 
with typical $Z\sim0.01 Z_\odot$. However,
gas in streams is not pristine, but already enriched by previous episodes of star formation.}
\label{fig:metalhist}
\end{figure}

\subsection{Metallicity}\label{sec:meta}

The metal enrichment of the gas may help distinguish
between gas that is accreting from the cosmic
web and the more enriched material that is located in proximity to a galaxy or is outflowing 
from regions with intense star formation. Production and diffusion of metals is known to be 
problematic in numerical simulations \citep{wad08}, but the AMR code used for our simulations 
should minimize problems of metal transport typical of SPH simulations \citep[but see][]{she10}. 
Caution is advised in generalizing these comparisons of metallicity between our simulations  
and observations because our current simulations are limited to supernova-driven
winds which do not reproduce outflows as massive as suggested by some observations
\citep[e.g.][]{ste10}, and also because the  specific IMF \citet{mil79}  is assumed in the simulations.

In Figure \ref{fig:metal}, the column density weighted metallicity in solar units 
\citep[$Z_\odot=0.0134$;][]{asp09} for a projection of MW3 at $z\sim 2.3$
is shown. On the top, we display metals in both streams and galaxies
for optically thick gas; on the bottom, we present the metallicity in the streams alone.  
Contours are for $N_{\rm HI}=1.6\times 10^{17}$ \cmm (black line) and  
$N_{\rm HI}=2\times 10^{20}$ \cmm (white line). A spatial gradient in metallicity is visible.
DLA gas inside the main galaxy and satellites is significantly enriched, with values close to 
$1/10$ solar \citep[in reasonable agreement with DLA observations;
e.g.][]{pro03}. The fraction of metals decreases moving far from the central disks and at lower column 
densities. The bulk of the LLSs is enriched at $1/10-1/100$ solar, with the
streams skewed at the lowest metallicity.

A similar metal content characterizes the entire sample at $z=1.3-4$, almost independently of 
redshift and mass. Figure \ref{fig:metalhist} shows the distribution of column density weighted
metallicity $\psi$ in DLAs (red dashed line), SLLSs (green dotted lines) and LLSs 
(blue dash dotted line). The metal distributions for streams alone 
(only for $N_{\rm HI}>1.6\times 10^{17}$ \cmm; solid line) and for satellites without centrals 
(orange long-dashed line) are superimposed.
This distribution has been normalized so that the integral over the metallicity gives the average 
covering factor in the entire sample for a fixed range of column density
\begin{equation}
\psi(\Delta N_{\rm HI},Z)= \frac{\phi(\Delta N_{\rm HI},\Delta Z)}{\Delta z ~\phi_{\rm tot}}\:. 
\end{equation}
Here, $\phi$ is the physical cross section.

From Figure \ref{fig:metalhist}, we see that most of the area that is covered by optically thick 
gas is occupied by metal poor cold filaments, with a distribution centered at $Z\sim10^{-2} Z_\odot$ 
and a spread of over $\sim 1$ dex. Cold streams are responsible for a metal poor population of 
LLSs, while optically thick absorbers that originates within 
0.25\rvir\ of centrals and satellites 
can be enriched above $\log Z/Z_\odot = -1$. DLAs are characterized by a broad 
distribution of metals centered at  $\log Z/Z_\odot \sim -1$ with a tail to lower metallicity.
Satellites alone have a distribution similar to the one of DLAs.
SLLSs originate from ionized material in the surroundings of the disks and have an equally 
broad distribution, centered around $\log Z/Z_\odot \sim -1.5$. 
Outflows more vigorous than the ones generated in our simulations are expected to 
produce even higher metallicity around the star-forming regions.
Since more primordial gas that replenishes galaxy 
disks with fresh fuel for star formation is predicted to have one order of magnitude less metals
than gas in galaxies, we conclude that metal poor LLSs in proximity to galaxies 
are the best candidates for cold gas in inflowing streams. The metallicity distribution 
in LLSs \citep[e.g.][]{proc10} may offer a crucial test to distinguish between 
cold inflowing streams and gas that already resides in the galaxies or is outflowing 
\citep[see also][]{gia11}.

\begin{figure}
\begin{tabular}{c}
\includegraphics[angle=90,scale=0.32]{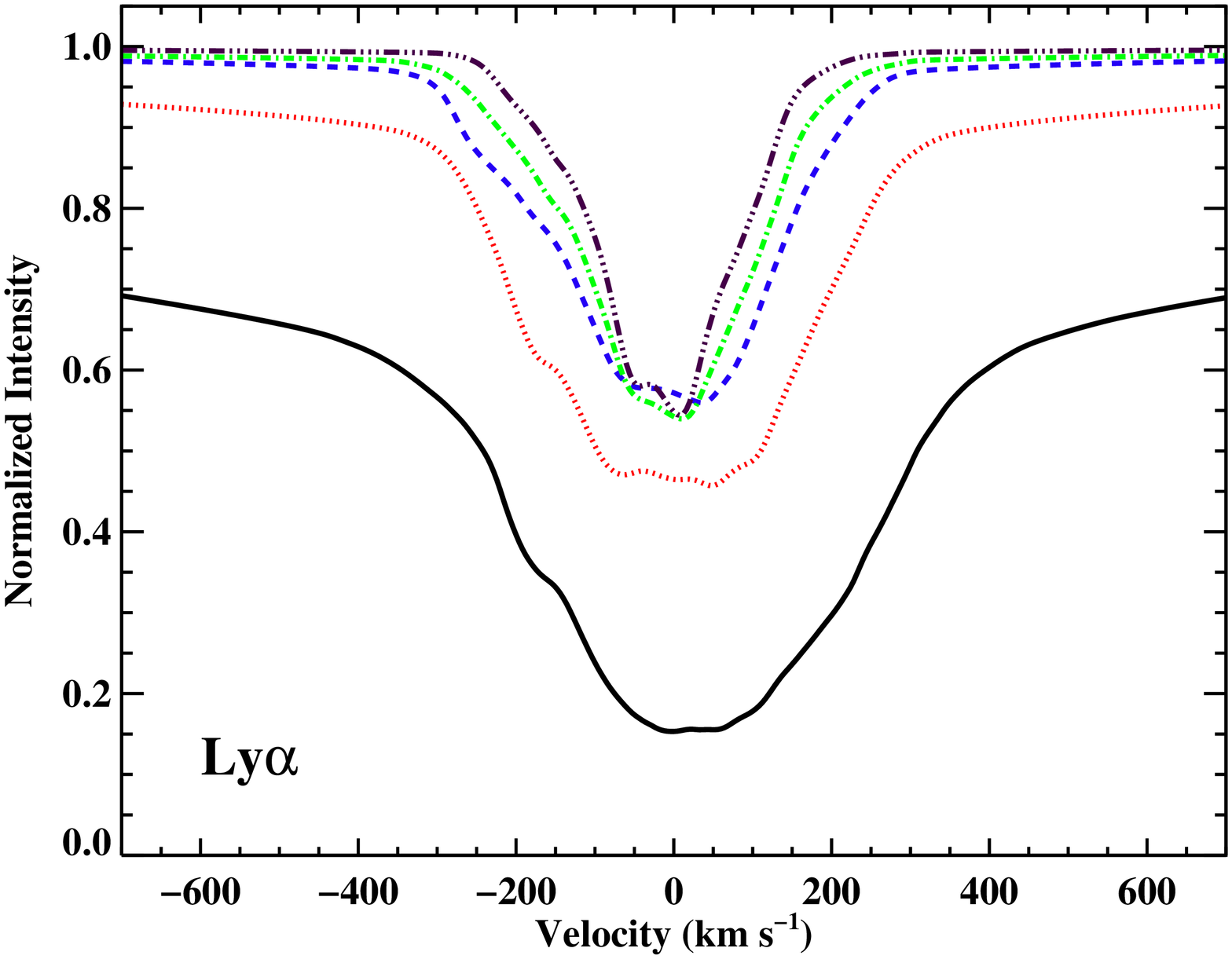}\\
\includegraphics[angle=90,scale=0.32]{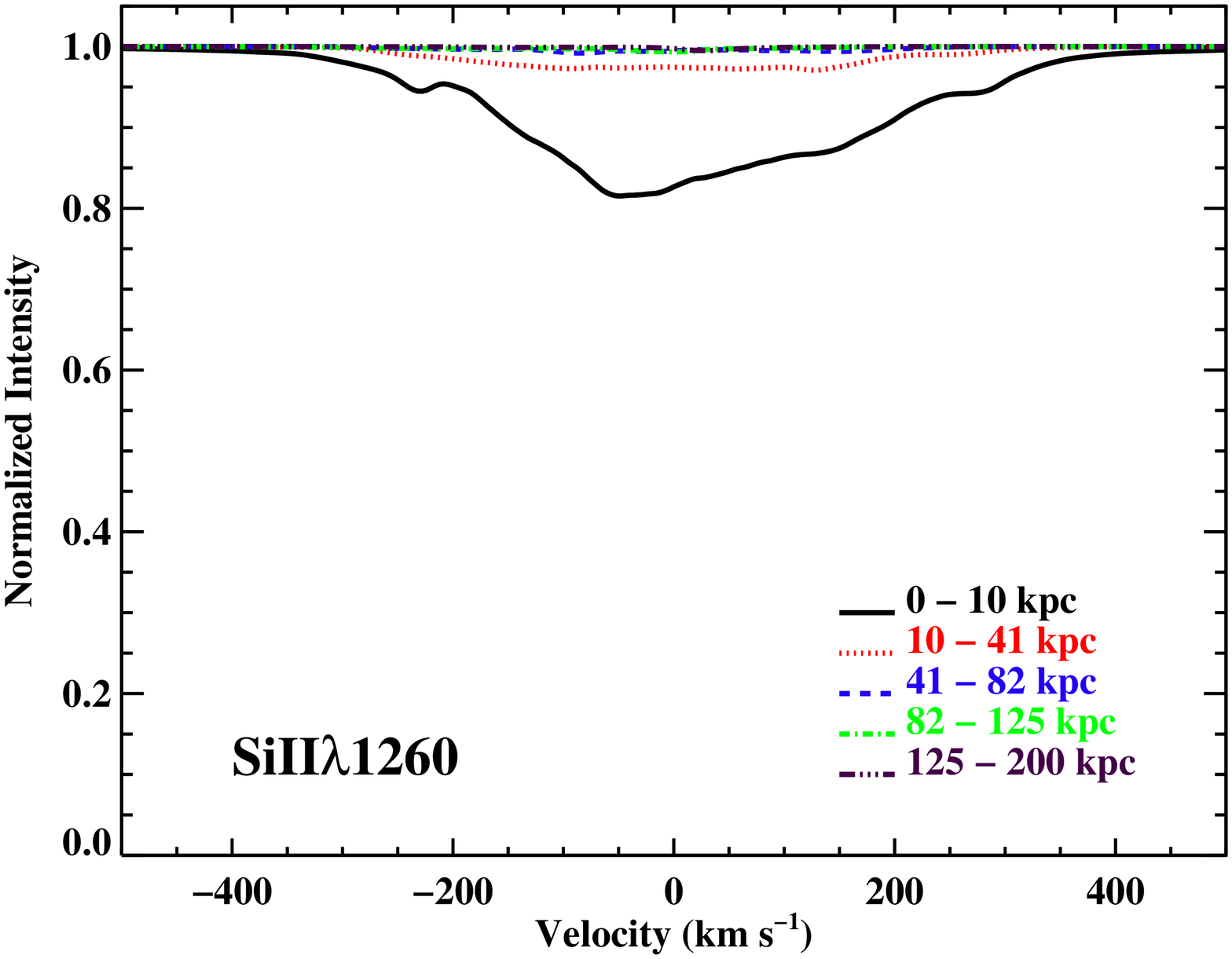}
\end{tabular}
\caption{Averaged \lya\ (top) and \SiIIl1260 (bottom) absorption lines 
for background sources at different 
impact parameters from the central galaxy. 
The \lya\ profiles are never saturated due to the low covering factor and are characterized
by a FWHM of $\sim 200-300~\rm km/s$ for $b>41~\rm kpc$. The low covering factor, 
low intrinsic Si/H abundance and low metallicity produce very weak average \SiIIl1260 absorption.}
\label{fig:spect}
\end{figure}

\subsection{Absorption line profiles}

The information on the metallicity and covering factor can be combined together 
with kinematics to formulate predictions on the strength and shape of absorption
line profiles at different projected distances from the galaxy centers. 
These models can then be compared with the population of ALSs observed 
in the CGM of Lyman break galaxies at $z\sim2-3$ \citep{ste10}. 
We perform this calculation by stacking several spectra generated
along multiple sightlines in a subset of our sample (MW1, MW2, MW3 and MW8 in two projections) 
at $z\sim2.3$ and $z\sim3.2$. The absorption profiles at $z\sim2.3$ for \lya\ 
(top panel) and \SiIIl1260 (bottom panel) are shown in Figure \ref{fig:spect}, where each spectrum  
is a composite of independent sightlines in five intervals of impact parameter, 
chosen to match the observations presented in \citet{ste10}.

Despite the fact that \lya\ saturates at low \HI\ column densities, none of the spectra 
are black at the line-center due to non-unity covering factor of neutral gas.
The transmitted normalized intensity at the line center ranges in the interval 
$0.5-0.6$, similar to what is observed. Damping wings are visible within 
the innermost $40~\rm kpc$, where most of the highly optically thick gas resides.  
The line profiles are characterized by a full width at half maximum (FWHM) velocity of 
$\sim 400-600~\rm km/s$ for $b<41~\rm kpc$, related to 
the kinematics of the central galaxies and the inner satellite relative motions,
as well as the bulk velocity of the incoming streams.
Conversely, at larger impact parameters ($b>41~\rm kpc$), a velocity 
of $200-300~\rm km/s$ FWHM is most likely associated with the inflowing gas 
that is streaming at $\sim 200~\rm km/s$ or more \citep{dek09}.
The line profiles are symmetric, despite the fact that the gas is infalling towards the central  
galaxies, indicating that there are no preferred directions in the velocity field
in a stack of multiple sightlines. Inspecting the metal lines, it is evident that the low
covering factor and the intrinsic low Si/H abundance,
together with the low metallicity of the streams, produce absorption lines that are 
weaker than the observed values at all impact parameters $b>10$\,kpc.

\begin{figure}
\begin{tabular}{c c}
\includegraphics[angle=90,scale=0.32]{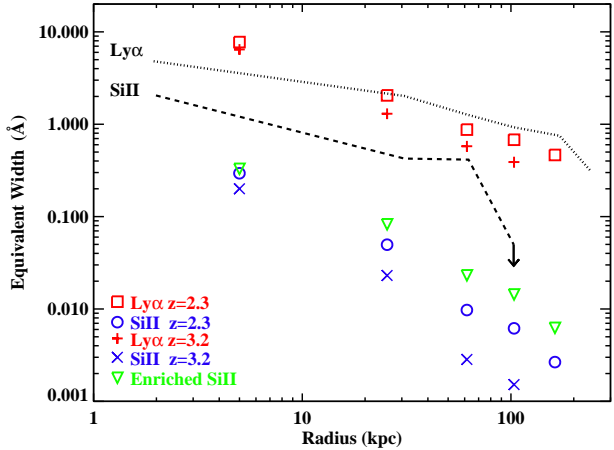}
\end{tabular}
\caption{Equivalent width profile
at $z\sim 2.3$ and $z\sim 3.2$ for \lya\ (red open squares and crosses) and \SiIIl1260\ 
(blue open circles and x's). The black dotted and dashed lines show the observed 
rest frame equivalent widths in LBGs at redshift $z\sim 2-3$ \citep{ste10}. 
Massive galaxies that are accreting gas through cold streams have a large enough covering 
factor and line-of-sight velocity to account for the observed strength of the \lya\ absorption. 
Conversely, the metal poor streams cannot reproduce the observed equivalent width in metal lines.
Boosting the degree of enrichment by introducing an artificial metallicity floor
at $0.1Z_\odot$ (green downward triangles) is not enough for matching 
the observations.}
\label{fig:ew}
\end{figure}

\begin{figure}
\includegraphics[angle=90,scale=0.32]{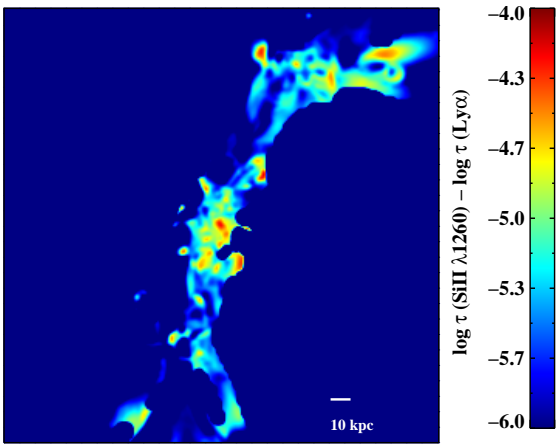}
\caption{Optical depth integrated over velocities in the interval $\pm 250~\rm km/s$ for \SiIIl1260 
relative to \lya\ for gas with $N_{\rm HI}>10^{13}$ \cmm. The low intrinsic Si/H abundance 
and metal content of cold streams suppress the metal line opacity by more than 
five orders of magnitude compared to \lya.}
\label{fig:taukin}
\end{figure}

In Figure \ref{fig:ew} we present a comparison between the line strength from simulations
and observations. For each radial bin, the rest frame equivalent width of \lya\
at $z\sim 2.3$ and $z\sim 3.2$ are shown with red open squares and crosses, respectively.
Similarly, we superimpose the equivalent width of \SiIIl1260\ 
at $z\sim 2.3$ (blue open circles) and $z\sim 3.2$ (blue x's).
Values are listed in Table \ref{tab:ew}.
The black dotted and dashed lines are for the observed rest frame equivalent width from 
spectra of LBGs at redshift $z\sim 2-3$ \citep{ste10}. 
As expected from direct inspection of the absorption profiles, massive galaxies that are 
accreting gas through cold streams have a large enough covering factor
and sufficiently complex kinematics in \HI\ gas to match 
observed \lya\ values. We emphasize that this is achieved through the
gas associated with satellites and streams alone; the supernova-driven 
winds in our current simulations are hot and not as strong as the outflows 
suggested by \citet{ste10}. While winds, commonly observed in star-forming galaxies, 
are essential ingredient for realistic models, the fact that the same observations
can be reproduced by two extreme scenarios stands as a reminder that 
evidences for outflowing/inflowing gas are subtle and more refined 
models are now required.

That these simulations fail to reproduce the 
strength of \SiII\ at all impact parameters $b<100$\,kpc
is not surprising since, due to the lack of strong outflows of cold gas, 
low-ion metals are tracing the distribution of hydrogen in the galaxies and streams.
We can consider whether, for example, the discrepancy for \SiIIl1260\ might be attributable 
to the low enrichment of the cold streams.
In Figure \ref{fig:taukin}, we show the ratio of the \SiII\ to \lya\ optical depth, 
integrated in the velocity interval $\pm 250~\rm km/s$. Only gas with $N_{\rm HI}>10^{13}$ \cmm\
is displayed. The low abundance of silicon relative to hydrogen at solar metallicity 
suppresses the metal line opacity by five orders of magnitude and the 
low metallicity of cold streams decreases the \SiII\ opacity even further, 
by a factor of $\sim 10-100$. For these reasons, metal lines that originate 
in cold streams are weak and likely to be undetected. A higher degree of enrichment is not 
enough to boost the equivalent width by one order of magnitude, 
as demonstrated  by a simple experiment \citep[see also][]{kim10}. 
If we artificially impose a floor in metallicity at $Z=0.1Z_\odot$
for all the gas with $N_{\rm HI} > 10^{15}$ \cmm, we obtain the equivalent widths 
shown by the green downward triangles in Figure \ref{fig:ew}. A much higher degree 
of enrichment increases the line strength only by a modest value.

We conclude that cold streams are unlikely to produce the large
equivalent widths of low-ion metal absorption around massive
galaxies. If the average equivalent widths reported by \cite{ste10}
are confirmed (e.g.\ with higher resolution observations), an
alternate source of opacity must be included in the simulations.  
This could include, for example, absorption by neighboring galaxies (not modeled here),
small clumps of cold gas embedded in a hotter medium, and/or galactic-scale outflows. 
However, in order to satisfy the constraints 
imposed by the \lya\ absorption, these phenomena cannot carry a very large 
column density of neutral hydrogen nor contribute to the kinematics apparent in the cold flows
unless they replace the streams altogether. 
This would fail to supply the gas necessary to produce the observed
high SFR plus a similar outflow rate.  Unless these SFRs and outflows are sustained
by a large amount of reservoir gas, the inflows and outflows must live together.
This is a natural consequence of the fact that the inflows are along narrow
dense streams, while the outflows are covering a large solid angle between the
streams. Additional work in simulations is needed to 
explore stronger wind models and to understand the 
interaction between outflowing and inflowing gas \citep[e.g.][]{raz09,pow10,van11}, before we can
reconcile both the \lya\ and metal absorption profiles with 
a combination of inflows and outflows.

\section{Summary and Conclusions}\label{sec:sumcon}

We investigated the characteristics of cold gas in cold-stream fed galaxies in absorption
using a sample of seven galaxies drawn at random for zoom-in cosmological hydro-AMR simulations with
high resolution ($35-70~\rm pc$), in the halo mass range 
$10^{10}-10^{12}$ \msun\ between $z\sim1.4-4$. 
We considered the contributions of the cold gas in the central galaxies
and of the inflowing cold gas in the two components of the streams that feed 
the high-redshift massive galaxies from the cosmic web --- the incoming small 
galaxies and the larger smooth component.
Our analysis focused on the mass dependence and time evolution of optically thick 
and thin gas cross section and covering factor, on the metallicity distribution 
in the streams and massive galaxies, and on the \lya\ and metal line kinematics. 
The simulations were compared with available observations of absorption line systems in the 
foreground of quasars and with kinematics of the circumgalactic medium 
in Lyman break galaxies to highlight what observables are most useful 
to detect cold streams in absorption. The limited sample size and the fact that 
very strong outflows are not included in these simulations make our predictions 
rather specific to the study of cold streams, and caution is advised in generalizing 
our findings to the full population of ALSs. Our results and findings can be summarized as follows.

\begin{itemize}
\item After post processing the simulations with a radiative transfer code that includes 
  dust, collisional ionization, and photoionization from the UV background and
  local stellar sources, 
  cold streams appear highly ionized and not entirely self-shielded structures.
  Pockets of high column density neutral hydrogen, mostly associated with the central galaxy 
  and instreaming galaxies, 
  are embedded in a ionized medium with $N_{\rm HI} \lesssim 10^{19}$ \cmm. 
  The densest regions are mostly affected by the UV radiation from local
  sources because the UV background is completely shielded in regions of
  $n_{\rm H}> 0.1~\rm cm^{-3}$.
  While the UV from stars is largely confined by hydrogen and dust absorption in 
  the star-forming regions, the non-zero escape fraction is responsible for
  the ionization of the gas in the immediate surroundings of disks 
  ($10^{18}\lesssim N_{\rm HI}/\rm cm^{-2} \lesssim 10^{20})$. 
  Since most of the mass is found at the highest 
  column densities, UV from stars decreases the total 
  neutral hydrogen mass in galaxies.

\item The smooth component of the cold streams dominates 
  the cross section of neutral hydrogen below 
  $N_{\rm HI}=10^{18}$ \cmm\ within 2\rvir. 
  At $z \leq 2$, less than half of the cross section in the range
  $N_{\rm HI}=10^{19}-10^{20}$ \cmm\ is due to the streams.
  At these low redshifts, most of the cross section for $N_{\rm HI}>10^{20}$ \cmm\ 
  is in galaxies, with the smooth component of the streams accounting 
  only for a small fraction of the total budget.
  However, a rapid evolution is seen with redshift, with the
  gas clumps in the streams responsible for 
  more than half of the DLA/SLLS covering factor at $z \sim 4$.  

\item  No more than 60\% of the DLA cross section is associated with the 
  central galaxies. Satellites and gas clumps in streams are important or even the 
  dominant contributors to the neutral gas cross section, especially in the most 
  massive galaxies.
  
\item  At all redshifts, the covering factor at 2\rvir\ is 
  $<40\%$ for $N_{\rm HI}>10^{15}$ \cmm, $<10\%$ for LLSs, and  $<1\%$ for DLAs. 
  The \HI\ gas covering factor within 2\rvir\ appears to decrease slowly 
  from $z\sim4$ to $z\sim1.5$, and the covering factor within a fixed aperture of 
  10 arcsec ($\sim 70-85~\rm kpc$) seems not evolving or slowly increasing with redshift.
  However, the small sample and the non-homogeneous sampling of the halo mass function 
  at all the redshifts prevent us from deriving robust scaling relations.   
  In these simulations, a certain degree of self-similarity is found in the column density probability 
  distribution function for neutral hydrogen and the time evolution of the different 
  covering factors appears to follow the universal expansion.
  
\item Cold streams mostly appear as LLSs and contribute significantly to the 
  column density distribution function in between $N_{\rm HI}=10^{17}-10^{18}$ \cmm.
  Massive disks, as described by these high resolution simulations, shape the 
  column density distribution function for very high column density DLAs. 
  These models, especially if molecules are included, can reproduce the observed 
  knee at $10^{21.5}$ \cmm\ in the \fnx.  Conversely, there is a deficiency of systems 
  between $10^{18.5}-10^{20}$ \cmm, likely 
  due to the missing low mass systems in our simulated sample and to finite numerical resolution which 
  does not capture a clumpy medium. The simulated \fnx\ is very shallow
  below $10^{17}$ \cmm, indicating that cold streams in proximity to massive galaxies are not the 
  only sources of opacity for Lyman limit photons in the Universe; an additional contribution 
  from the IGM is required. Finally, in agreement with observations, the simulated \fnx\ is 
  not evolving in shape with redshift, due to a nearly self-similar distribution of neutral hydrogen 
  in halos.

\item Despite the small covering factor, comparing the observed and
  simulated incidence of absorption line systems, it appears that
  massive galaxies and cold streams are responsible for at least
  $30\%$ of the absorbers detected in the foreground of QSOs. In fact,
  due to a large enough number density of galaxies in the universe,
  the cross section required to match the observed incidence is only a
  factor of a few above the mean values in this sample. A modest
  extrapolation towards lower halo masses can fully reproduce the
  observed population. To the extent that cold streams are indeed the
  common mode of accreting gas in massive galaxies at high redshift,
  as emerges from the cosmological simulations, these streams have already
  been detected in absorption line surveys, primarily as metal-poor LLSs. 
  
\item In our simulations, most of the cross section within 2\rvir\ 
  is occupied  by metal poor gas ($Z\sim 10^{-2} Z_\odot$),
  that has been partly  enriched by previous episodes of star formation. DLAs and
  SLLSs exhibit broad metallicity distributions, with a peak at
  $Z\sim 10^{-1} Z_\odot$ and $Z\sim 10^{-1.5} Z_\odot$,
  respectively. Metallicity is a valuable tool to disentangle gas 
  that is associated with galaxies or winds (with metallicity 
  distribution skewed above $Z\sim 10^{-1.5} Z_\odot$) from LLSs 
  in the cold streams (with the lowest level of enrichment). 

\item Because of the partial covering factor of neutral gas, 
  composites of \lya\ absorption lines are not black at the line center, 
  even within $40~\rm kpc$ from the central galaxy.
  The typical line width outside the central galaxy and its immediate neighborhood 
  is $\sim 200-300~\rm km/s$ FWHM reflecting the bulk inflow velocity in cold streams. 
  Our current simulations reproduce the \lya\ equivalent width distribution 
  observed in the circumgalactic medium of Lyman break galaxies. 
  Conversely, the predicted optical depth of metal lines is 
  suppressed by more than 5 orders of magnitude and the predicted 
  equivalent width of \SiIIl1260 is systematically lower than the observed values.
  An increase in the metallicity of the streams is not enough to match the observations. 
  Another mechanism is required, such as massive outflows beyond the
  existing winds in the current simulations.
\end{itemize}

From the systematic analysis of our simulations
and a careful comparison with observations, we conclude
that the theoretical predictions of cold gas accretion 
along filaments of the cosmic web are consistent with present day observations.
This conclusion is not driven by a null claim that the cold streams are 
undetectable in absorption, which would have been rather unsatisfactory. 
In fact, our current simulations, with all the caveats previously discussed, 
indicate that cold streams contribute significantly 
to the incidence and kinematics of absorbers observed at redshift $z\sim 2-3$. 

Progress can now be made in two directions. 
While this work offers a first quantitative analysis of detectability of 
cold streams in absorption, simulations representative of a larger spectrum of masses 
will facilitate robust statistical comparisons with observations. 
At the same time, improved modeling of outflows in simulations 
will allow a more detailed exploration of the interaction
between inflowing and outflowing gas. This is important to better assess the cross 
section distribution of neutral hydrogen, a key quantity for the visibility of cold streams in absorption.

Observationally, it should be possible to 
identify candidates of cold streams by searching the increasing spectroscopic data sets of 
quasars and galaxy pairs.  Despite the difficulties of measuring metallicity in LLSs,
one can investigate their kinematics in detail to disentangle between 
metal rich gas that is outflowing at velocities $> 600~\rm km/s$ from gas that 
is inflowing at velocities $\sim <200~\rm km/s$ with metallicity $\lesssim 0.01Z_\odot$.

Observed metal-poor LLSs are suggestive of the cold mode of inflow \citep[e.g.][]{proc10,rib11b}
and a confirmation of this phenomenon may arise if a nearby galaxy is observed to be fed by cold inflows. 
However, at low redshift the cold mode is expected to be 
limited to galaxies with stellar mass below $10^{11}$\msun~ \citep{dek06}, 
and to accrete from a wide angle at a lower column density.
This is significantly different from the cold, narrow, 
dense streams that are predicted at high redshift and
only direct evidence for cold inflow at high redshift would be a confirmation of these theoretical 
predictions. In future, provided reliable estimates from simulations
of observables such as the column density distribution, metallicities and kinematics, models of cold 
gas accretion can be also tested statistically against observations.

Following these paths, it may be possible to uncover 
the modes of gas accretion in high redshift galaxies, and their interplay with outflows.

\section*{Acknowledgments}

We thank the referee for comments and criticisms that help us to 
improve this paper. We are indebted to  J. Guedes for extensive help with the AMIGA halo finder and 
R. da Silva for helpful IDL tips. We acknowledge useful discussion with 
J. Hennawi, C.-A. Faucher-Giguere, S. Cantalupo, T. Goerdt, P. Madau, M. Rafelski, A. Sternberg, and  
A. Wolfe. MF acknowledges travel support from UC-HIPACC and 
thanks the CASS at UC San Diego for their hospitality. 
JXP is supported by NSF grant AST-0709235 and HST grant STScI HST-GO-11595.03-A.
AD and DC are supported by ISF grant 6/08, by GIF grant G-1052-104.7/2009, 
by a DIP grant. JRP and AD are supported by NSF grant AST-1010033.

\appendix

\section{A detailed comparison of the RT calculations}\label{rtcomp}

\begin{figure}
\includegraphics[angle=90,scale=0.32]{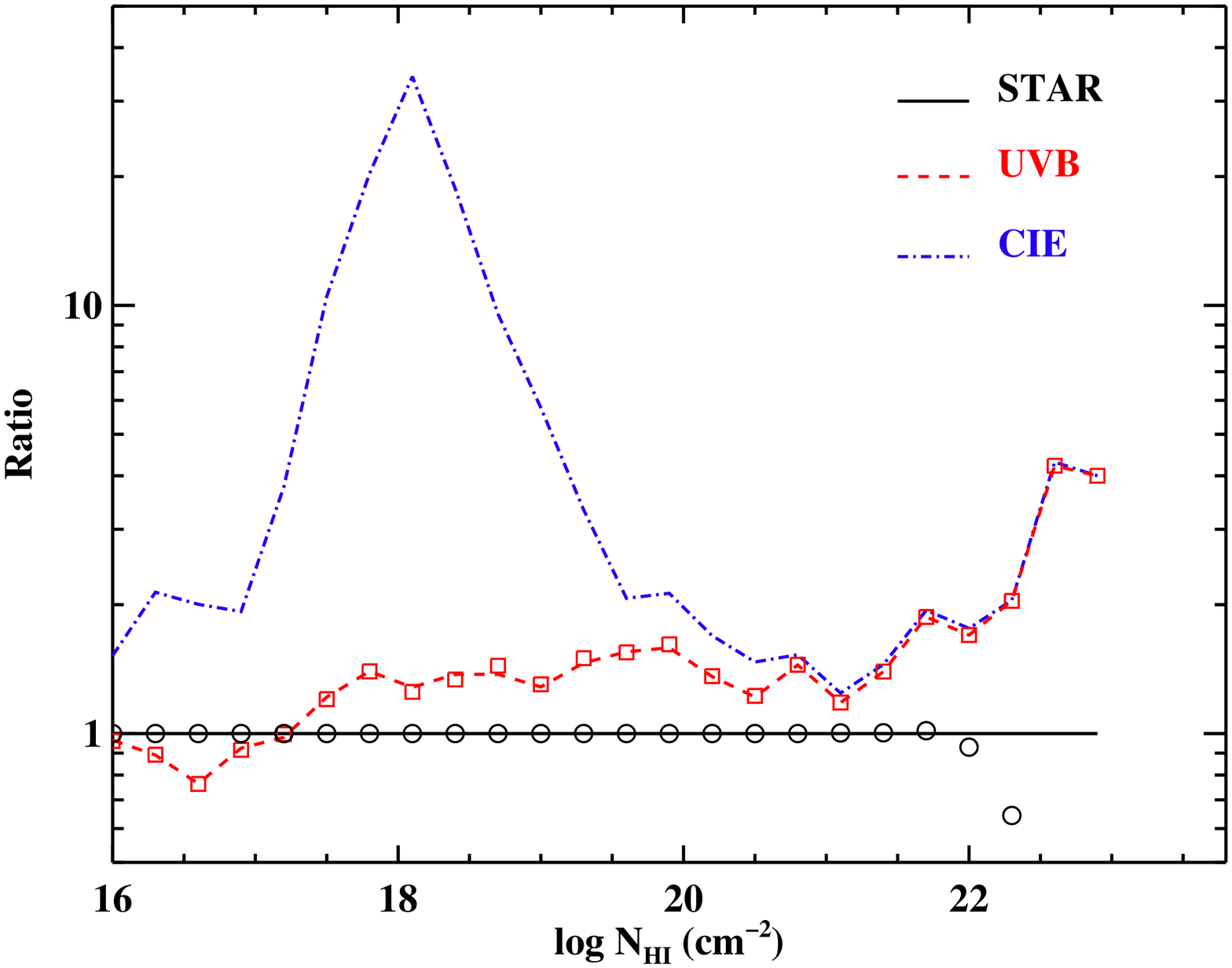}
\caption{Comparison of the neutral hydrogen column density in
the STAR model (black line), the CIE model (blue dash-dotted line), and the UVB model, 
with and without dust (red dashed line and open squares).
The STAR model corrected for molecular hydrogen is show with open circles. 
For each RT calculation, we plot the number of projected cells in a given interval of \NHI, 
in comparison to the STAR model. The CIE approximation holds only in highly optically thick 
regions. At  high column densities, neutral hydrogen is further ionized by local sources and 
depleted by molecules.}\label{xhi_all} 
\end{figure}

In this appendix, we provide a more detailed comparison of the results from the
different radiative transfer calculations. At first, we discuss the effects 
that the UVB and local sources have on the column density distribution. Then, we 
quantify the hydrogen density threshold above which the gas is self-shielded and 
we provide a numerical approximation useful to correct the CIE approximation 
in the absence of the UVB. Finally, we examine the impact that photoionization 
has on the neutral mass of galaxies and satellites.

\begin{figure*}
\begin{tabular}{c c}
\includegraphics[scale=0.32]{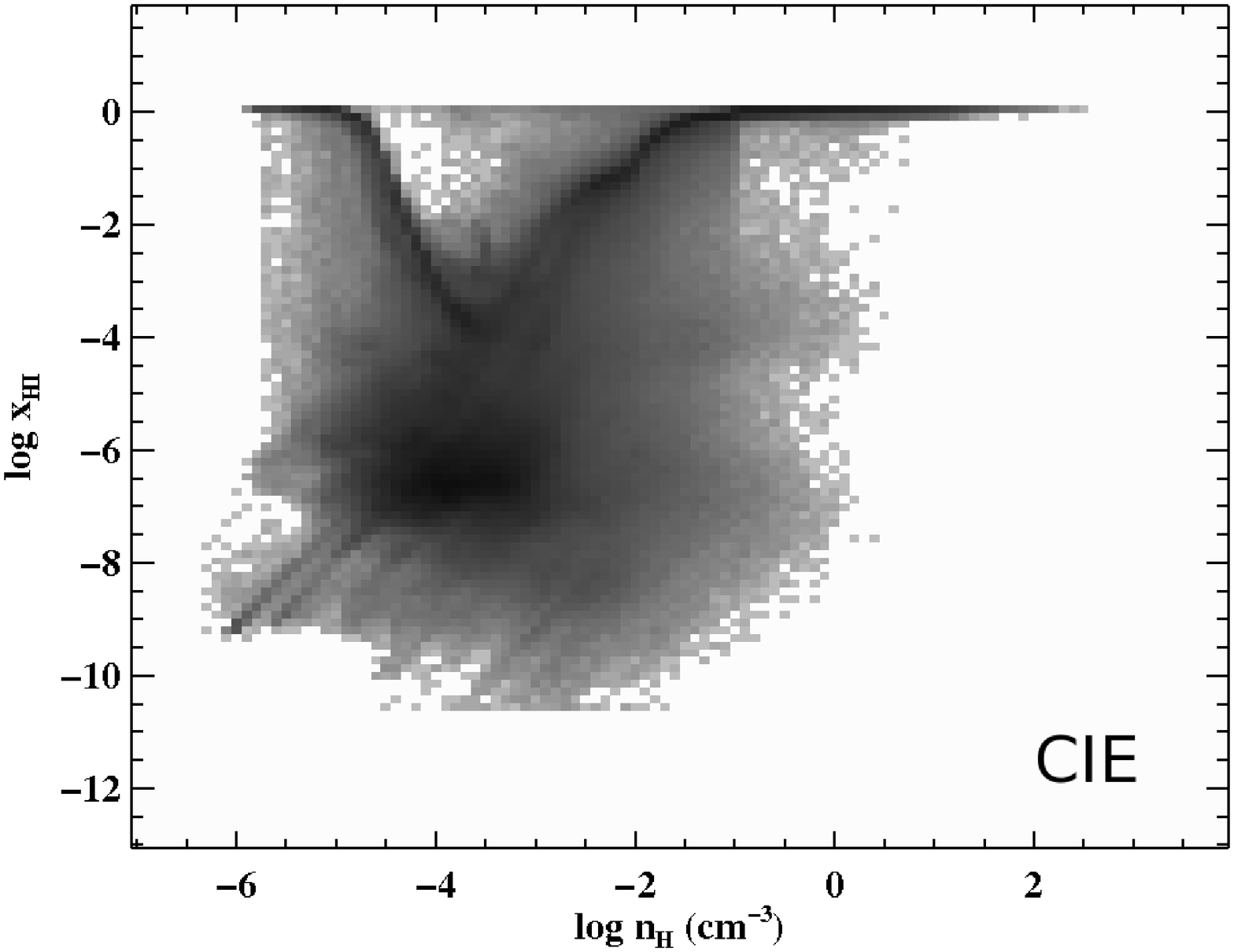}&
\includegraphics[scale=0.32]{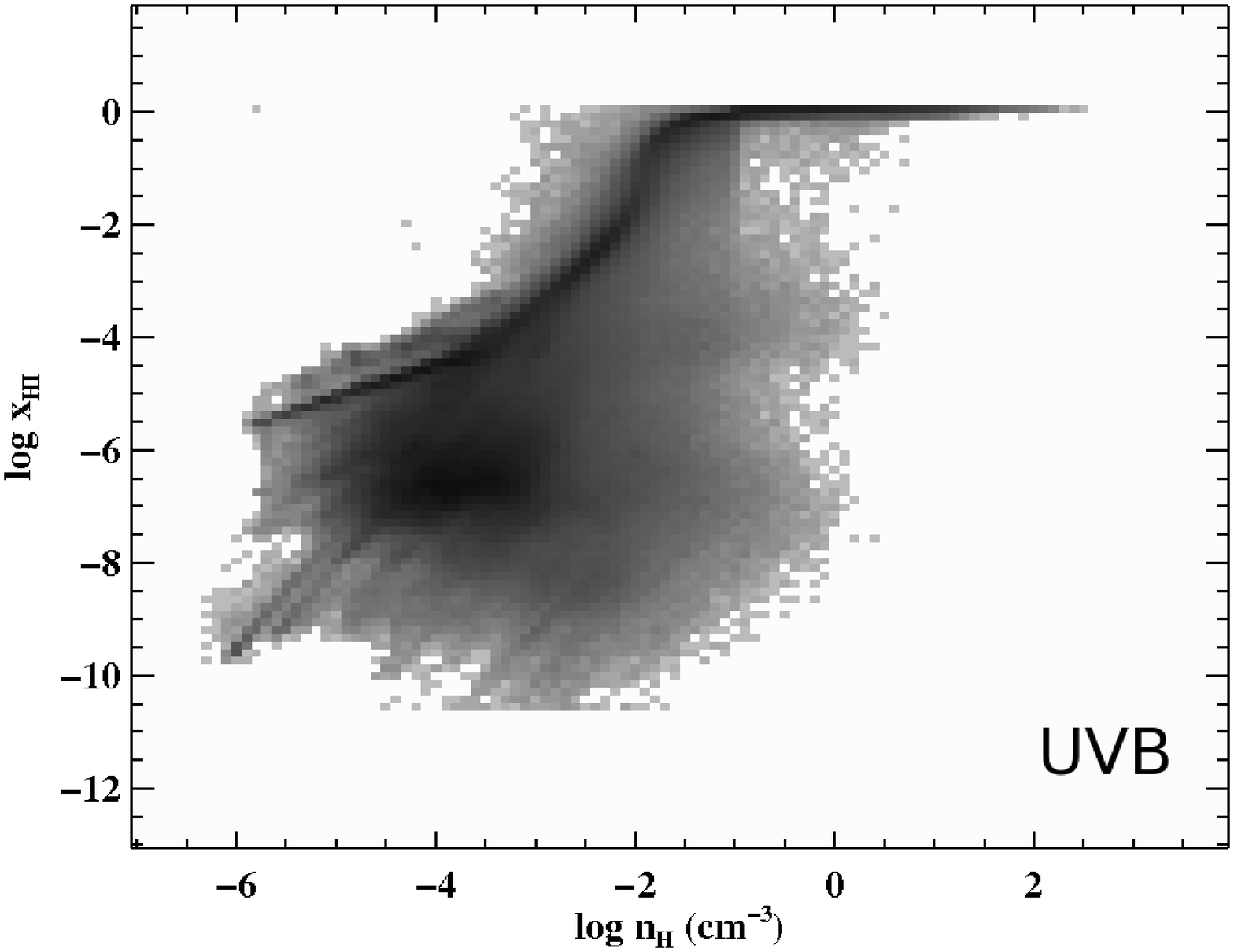}\\
\includegraphics[scale=0.32]{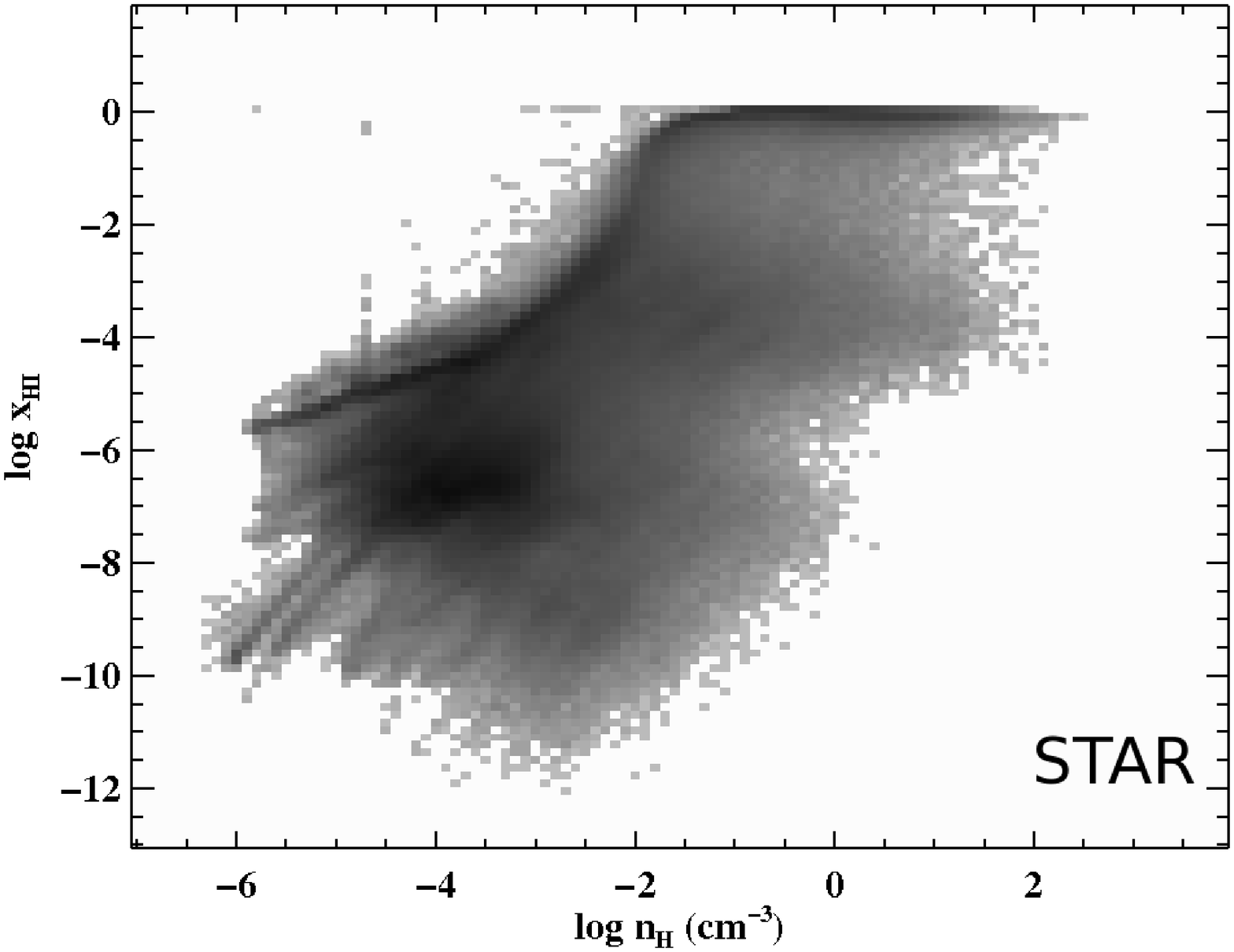}&
\includegraphics[angle=90,scale=0.32]{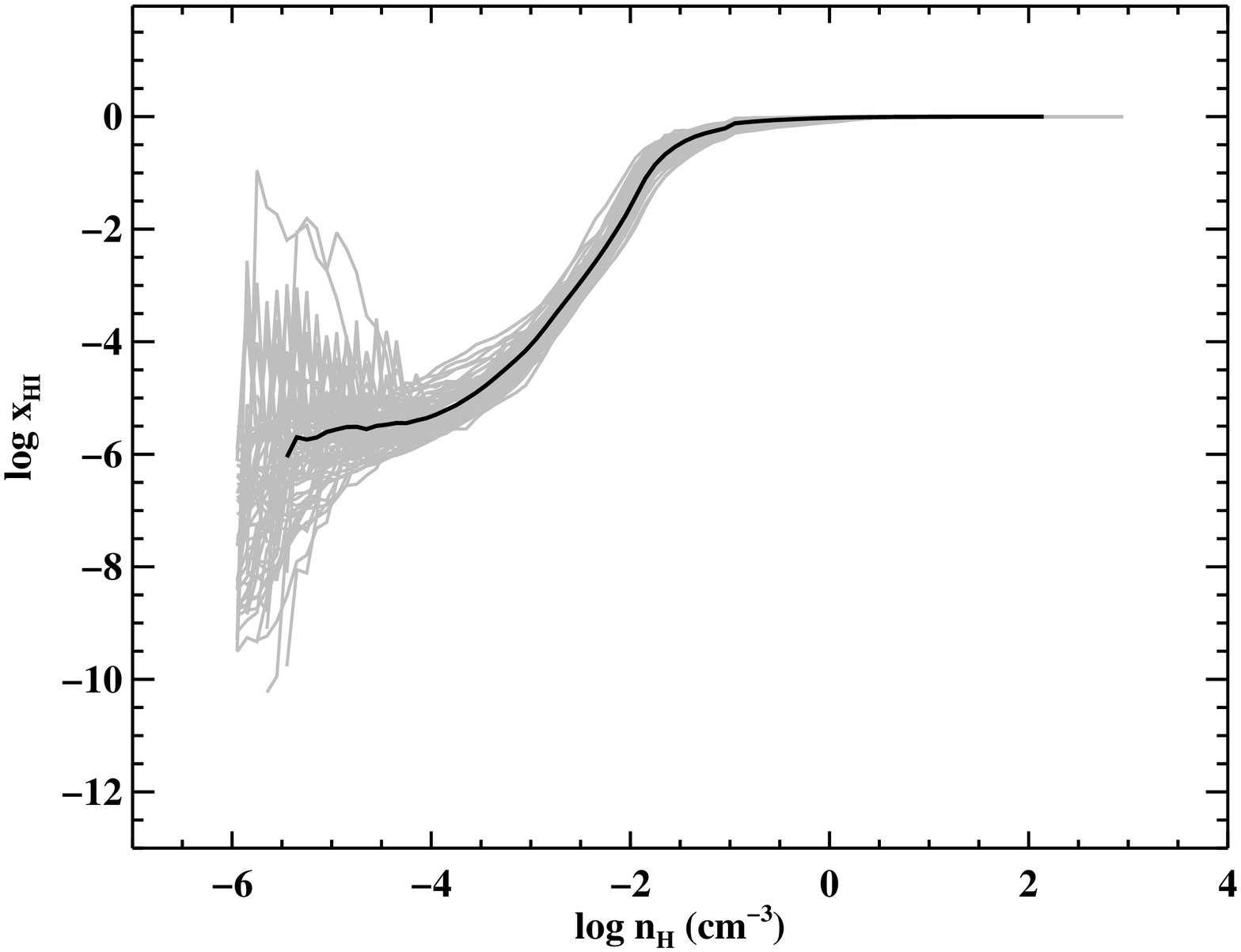}
\end{tabular}
\caption{Neutral fraction as a function of the hydrogen volume density for the 
CIE model (top left), the UVB model (top right) and the STAR model (bottom left) 
in MW3 at $z\sim 2.3$. The color scale is proportional to the number of AMR cells
in each bin, with the most populated region in dark gray. 
The arithmetic mean in bin of $\log n_{\rm H}$ for the UVB model 
in individual galaxies  is shown in the bottom right panel
(gray lines). The central values common to all the galaxies is also superimposed (black solid line).
Although with significant scatter, this distribution can be used as a crude approximation 
to the UVB model. RT with local sources is required to capture the intrinsic scatter in the 
neutral fraction above the $n_{\rm H}\sim 0.01-0.1~{\rm cm^{-3}}$, where gas becomes shielded from the UVB.}
\label{self_shi}
\end{figure*}

\subsection{Effects on the hydrogen column density distribution}

In Figure \ref{xhi_all}, we highlight the effect that different ionization sources have on the
\HI\ column density distribution by showing the number of AMR cells in bins of \NHI\ 
from an arbitrary projection of MW3 at $z=2.3$. The different RT models have been normalize 
with respect to the STAR model (black horizontal solid line).
The CIE model (blue dash-dotted line) offers an upper limit to the neutral fraction at 
all the column densities. The CIE approximation holds only in 
highly optically thick regions ($N_{\rm HI} \gtrsim 10^{20}$ \cmm),
while at lower column density overpredicts by a large amount the neutral fraction
in comparison to the UVB case (red dashed line). Photoionization cannot be neglected, 
particularly given the fact that streams are not entirely 
self-shielded, as evident from Figure \ref{mw3_ref_rt}. 

Comparing the UVB with the STAR model, it appears that radiation from local sources affects
primarily the high column densities ($N_{\rm HI} \gtrsim 10^{21}$ \cmm) which are completely 
self-shielded from the UVB. Since most of the ionizing photons are absorbed locally, 
the effect of local sources is modest below $N_{\rm HI} \sim 10^{20}$ \cmm\ and almost 
negligible below $N_{\rm HI} \sim 10^{18}$ \cmm. Note that quantities shown in Figure 
\ref{xhi_all} are normalized and not proportional to the area covered by gas at a given column density. 
In fact, radiation from local sources that leaks form the central disk can 
affect the \HI\ column density distribution function in the interval 
$10^{18}-10^{20}$ \cmm, when weighted by area. Dust produces little or no effect to the UVB model
(compare the red dashed line with open squares).

Although not included in the our RT calculation, in Figure \ref{xhi_all} we explore how molecules 
affect the shape of the hydrogen column density distribution. When atomic hydrogen is shielded from 
UV radiation, molecules can form, lowering the effective \HI\ column density.
To investigate the importance of this effect, we compute the molecular fraction following 
the \citet{mck10} formalism. This analytic model reproduces to first order the 
molecular gas fraction in nearby spirals \citep{kru09} and metal poor dwarfs \citep{fum10}
and approximates the results from numerical simulation \citep{krum10}.
In each AMR cell, we estimate the molecular gas fraction as a function of the total neutral 
gas column density $\Sigma_{\rm gas}=2 m_p N_{\rm H_2}+m_p N_{\rm HI}$ and metallicity. 
Since this model is designed to describe the molecular fraction in individual clouds, 
we correct for smearing due to resolution by assuming $\Sigma_{\rm cloud}=c\Sigma_{\rm gas}$ 
with $c=5$ for all the AMR cells with sizes above 200 pc and $c=1$ elsewhere. 
Although not a dominant correction, the inclusion of molecules lowers 
the \HI\ column densities above $N_{\rm HI}\sim 10^{22}$ \cmm\ (open circles). For this reason, 
the values of \NHI\ used in our analysis should be regarded as upper limits at the highest density. 
This column density threshold is in good agreement with the knee in the observed DLA column 
density distribution \citep{pro05,not09} that, indeed, is interpreted as the transition between atomic and 
molecular gas \citep[e.g.][]{zwa06}.

\begin{figure*}
\begin{tabular}{c c}
\includegraphics[angle=90,scale=0.32]{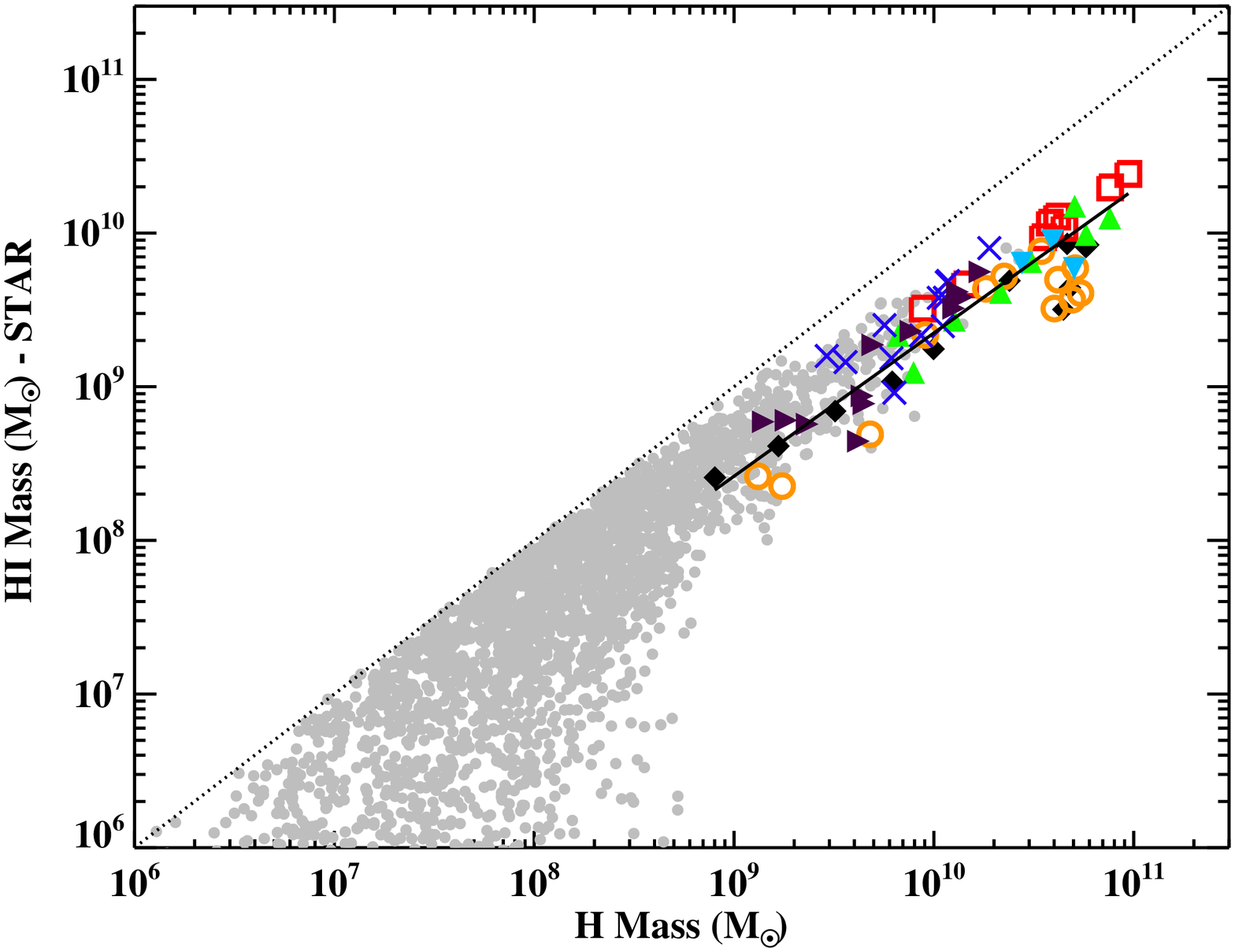}&
\includegraphics[angle=90,scale=0.32]{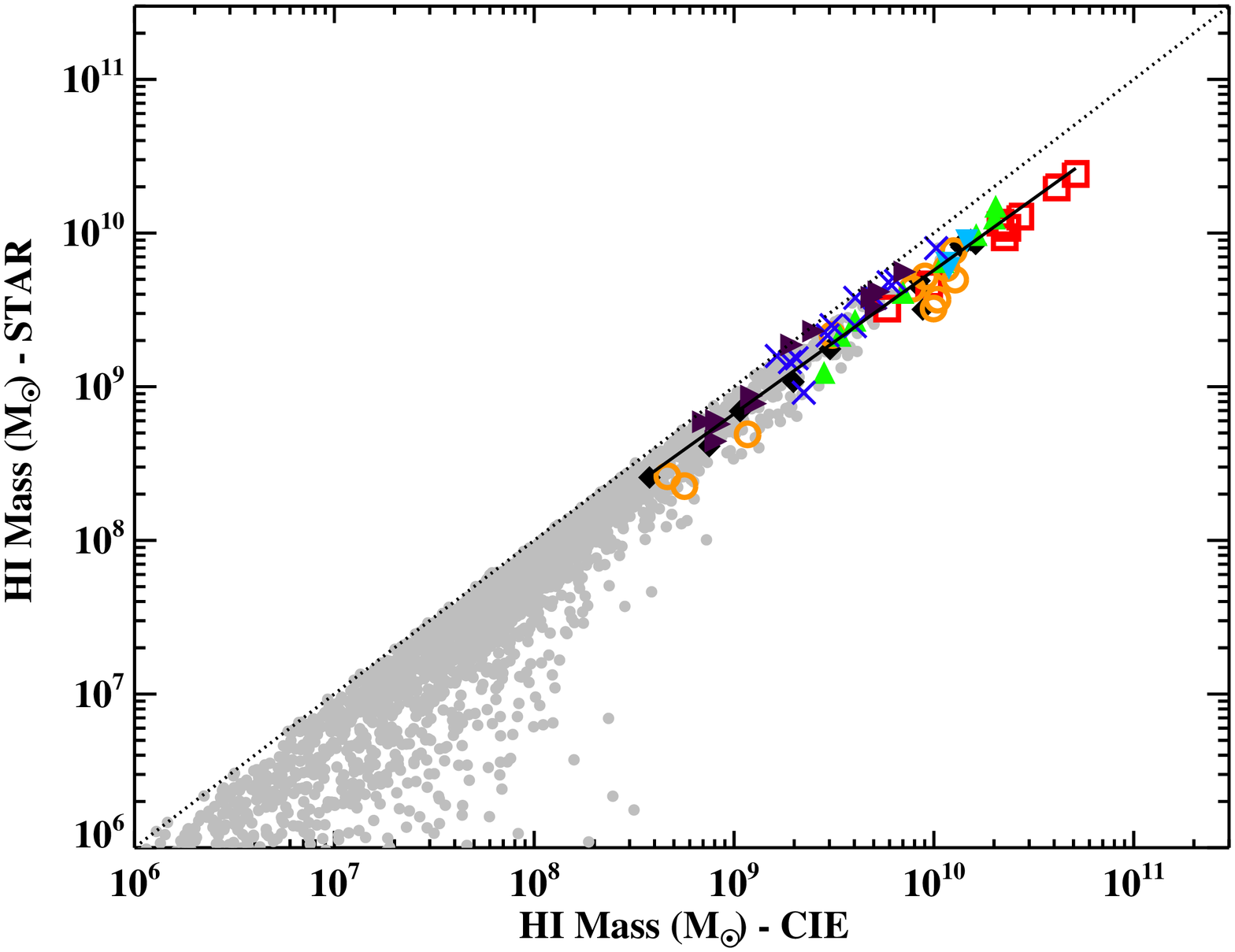}\\
\includegraphics[angle=90,scale=0.32]{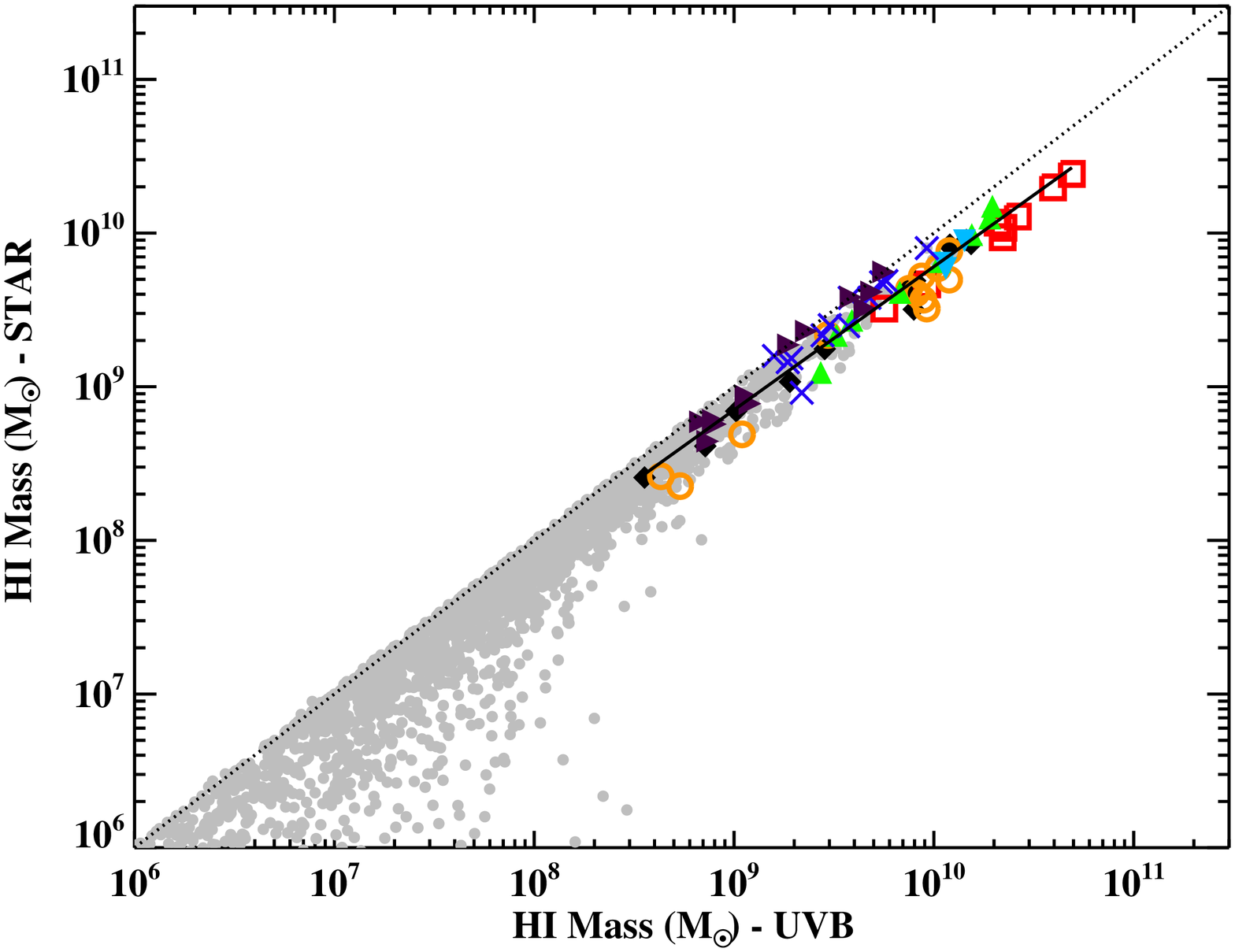}&
\includegraphics[angle=90,scale=0.32]{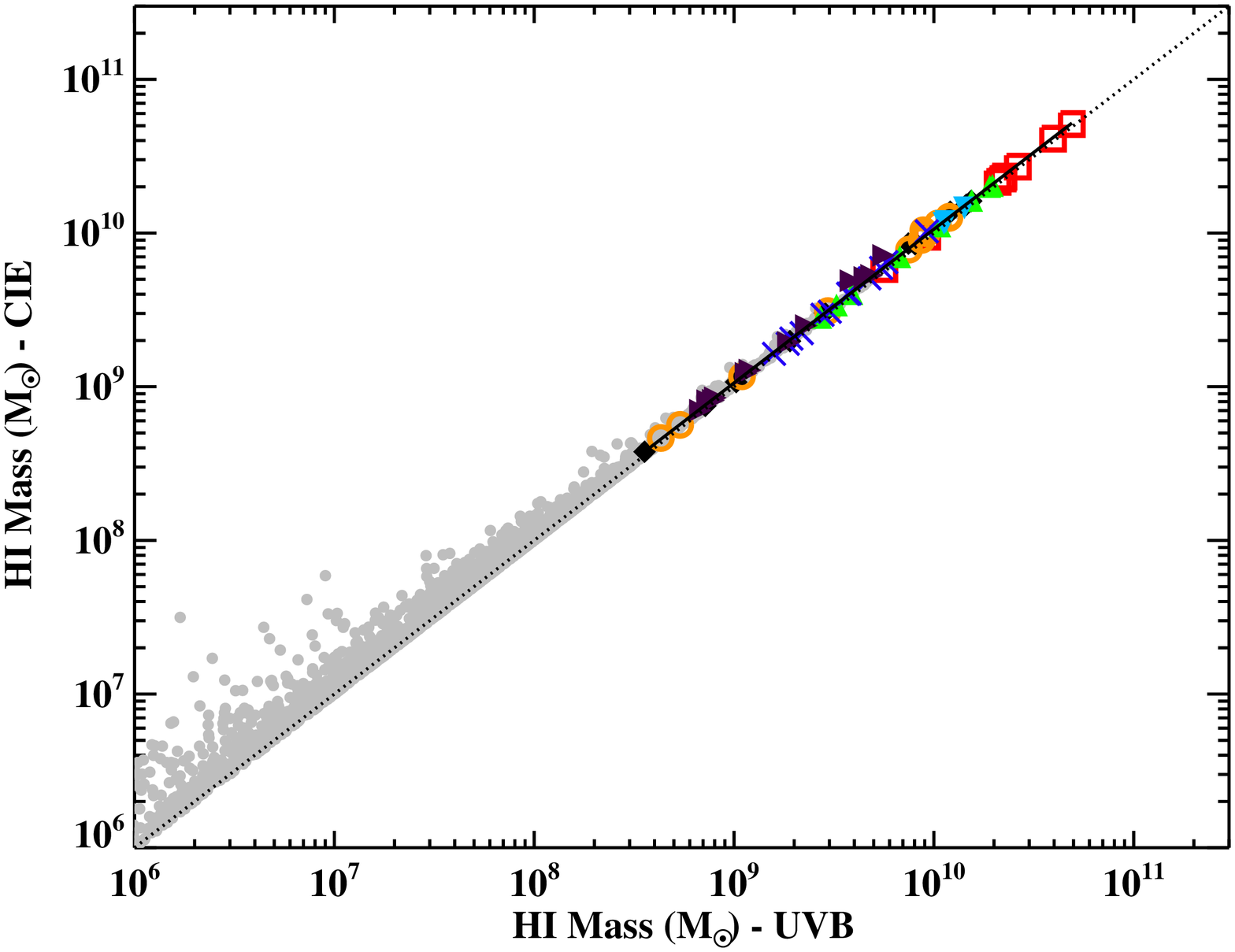}
\end{tabular}
\caption{Comparison of the neutral and total hydrogen mass enclosed within \rvir\ for 
the three RT models. In each panel, main halos are color coded as in Figure 
\ref{fig:galprop}. Satellites are displayed with gray circles. Dotted lines indicate
a 1:1 relation, while dashed lines show a robust linear regressions for the main halos only.
In massive galaxies, only 20-25\% of the total hydrogen mass is neutral, 
with local sources and electron collisions being responsible for most of the ionization. 
The UVB has no effects on the neutral mass due to nearly complete self-shielding at the highest 
hydrogen volume densities.}
\label{fig:himass}
\end{figure*}

\subsection{Self shielding and numerical approximations}

Figure \ref{self_shi} shows the neutral fraction $x_{\rm HI}$ as a function of total gas volume density $n_{\rm H}$,
for the CIE model (top left panel), the UVB model (top right panel), and the STAR model (bottom left panel),
again for MW3 at $z=2.3$.  In presence of the UVB, hydrogen is highly ionized 
below $n_{\rm H}=0.01$ cm$^{-3}$, whereas it is fully neutral above $n_{\rm H}=0.1$ cm$^{-3}$. 
A similar volume density for self-shielding has been consistently found with 
different numerical techniques \citep[e.g.][]{raz06,fau10}, justifying the use of a
fixed threshold above which the UVB can be suppressed while running the simulations.

Additional photo-heating, currently neglected in our RT calculations, 
will produce a smoother transition between ionized and shielded regions due the enhanced 
collisional ionization in partially shielded cells. 
Also, both the UVB and CIE models do not account for the presence of ionizing sources 
that increase the dispersion of $x_{\rm HI}$  (bottom left panel in Figure \ref{self_shi}).  
A significant fraction of the gas above $n_{\rm H}=0.01-0.1$ cm$^{-3}$ remains
fully neutral, but local fluctuations in the ionization state of up to 5 order of magnitude can be found.
Indeed, a locus of points with  $n_{\rm H}>0.01$ cm$^{-3}$ and  $x_{\rm HI} < 0.01$ is visible
in Figure \ref{self_shi}. Since this gas mostly resides in the central disk and in satellites 
where the majority of the star formation is present, the inclusion of ionizing photons 
seem to be an essential ingredient in models of galaxy formation.
Ionizing photons from stars are absorbed mostly in the inner part of the galaxy by 
dust and gas and there is only a minor contribution of the ionization 
below $n_{\rm H}=0.01$ cm$^{-3}$, in qualitative agreement with current 
constraints on the escape fraction of ionizing radiation. However, the degree of precision required in 
comparing  models and observations requires the inclusion of this RT effects 
on the neutral gas, commonly neglected or treated with some approximations.

Although below $n_{\rm H}=0.01$ cm$^{-3}$ the scatter in $x_{\rm HI}$  at a given 
volume density is large, most of the AMR cells accumulate in a well defined locus of points.
Due to a self-similarity in the probability distribution function of the hydrogen column density,
there is little scatter for different galaxies both with mass and redshift, as seen
from the bottom right panel of Figure \ref{self_shi}, where we plot the arithmetic mean 
of $x_{\rm HI}$ for the UVB model (gray lines for individual galaxies).
This invariance not only corroborates the assumption of a fixed threshold above which the UVB is 
suppressed, but also allows one to assume a lower limit on the hydrogen volume density in 
systems that host neutral gas, such as DLAs. 
Further, we can derive a crude approximation to the UVB model. 
In the CIE model, fully neutral or marginally ionized gas is found 
even below $n_{\rm H}=0.01$  cm$^{-3}$, with an error up to several order of magnitudes in 
comparison to the UVB model. Using the fact that the mean $x_{\rm HI}$ as a 
function of $n_{\rm H}$ is sensitive to the intensity of the UV background only, we derive
a relation  $x_{\rm HI}=x_{\rm HI}(n_{\rm H})$ that improves the CIE model in presence of a UVB. 
We obtain such relation by combining all the arithmetic means
computed in individual galaxies with a geometric mean (see Figure \ref{self_shi} and Table \ref{tab:gm}).

\subsection{Neutral hydrogen mass}\label{sec:himass}

For each RT model, we measure the total \HI\ mass enclosed in the virial radius 
of the main halos and satellites (see Table \ref{gasprop}). 
A comparison is provided in Figure \ref{fig:himass}. 
In the top panel, we show the relation between 
the total and the neutral hydrogen mass in the STAR model. 
In massive halos above $10^{9}$ \msun, less than $\sim 1/5$ 
of the hydrogen mass is neutral. A robust linear regression in logarithmic space 
for the main halos only gives $\log M_{\rm HI}=0.03+0.93\log M_{\rm H}$.
Less massive satellites cover a wider range in \HI\ mass, from almost neutral to highly ionized.
The CIE and UVB models predict nearly identical masses above $10^{9}$ \msun\ 
($\log M_{\rm HI,CIE}=0.02+1.00\log M_{\rm HI,UVB}$ for the main halos), while in the satellites
the CIE model predicts larger neutral hydrogen masses. \HI\ masses from the STAR model are systematically 
lower than those in the CIE or the UVB model above $10^{9}$ \msun, up to 
50\%-60\% above  $10^{10}$ \msun\ ($\log M_{\rm HI,STAR}=0.41+0.93\log M_{\rm HI,CIE}$ and 
$\log M_{\rm HI,STAR}=0.45+0.93\log M_{\rm HI,UVB}$ for the main halos). Satellites 
have a larger scatter. 

The explanation for this behavior naturally emerges from the previous considerations. 
In the main halos, most of the mass is in high density regions where the gas is self-shielded 
from the UVB. Photoionization from local sources and, to second order, electron collisions 
regulate the amount of mass in neutral hydrogen. For satellites, the observed scatter is due to the 
different SFRs in these systems and on the different distances from the central disks. 
Also, at lower masses, systems are more vulnerable to ionization 
from the UVB. This behavior is qualitatively in agreement with the simulations by \citet{gne10} who 
finds that the Lyman-Werner radiation from the UVB is negligible compared to the local radiation field
and that the ionizing radiation from the UVB plays a significant role only up to the edge of the \HI\ disks.
This result has implications for understanding the processes that shape the 
massive end of the \HI\ mass function and stresses further the necessity to account for local 
photoionization as an important feedback mechanism \citep[e.g.][]{can10}.

\begin{figure}
\includegraphics[angle=90,scale=0.32]{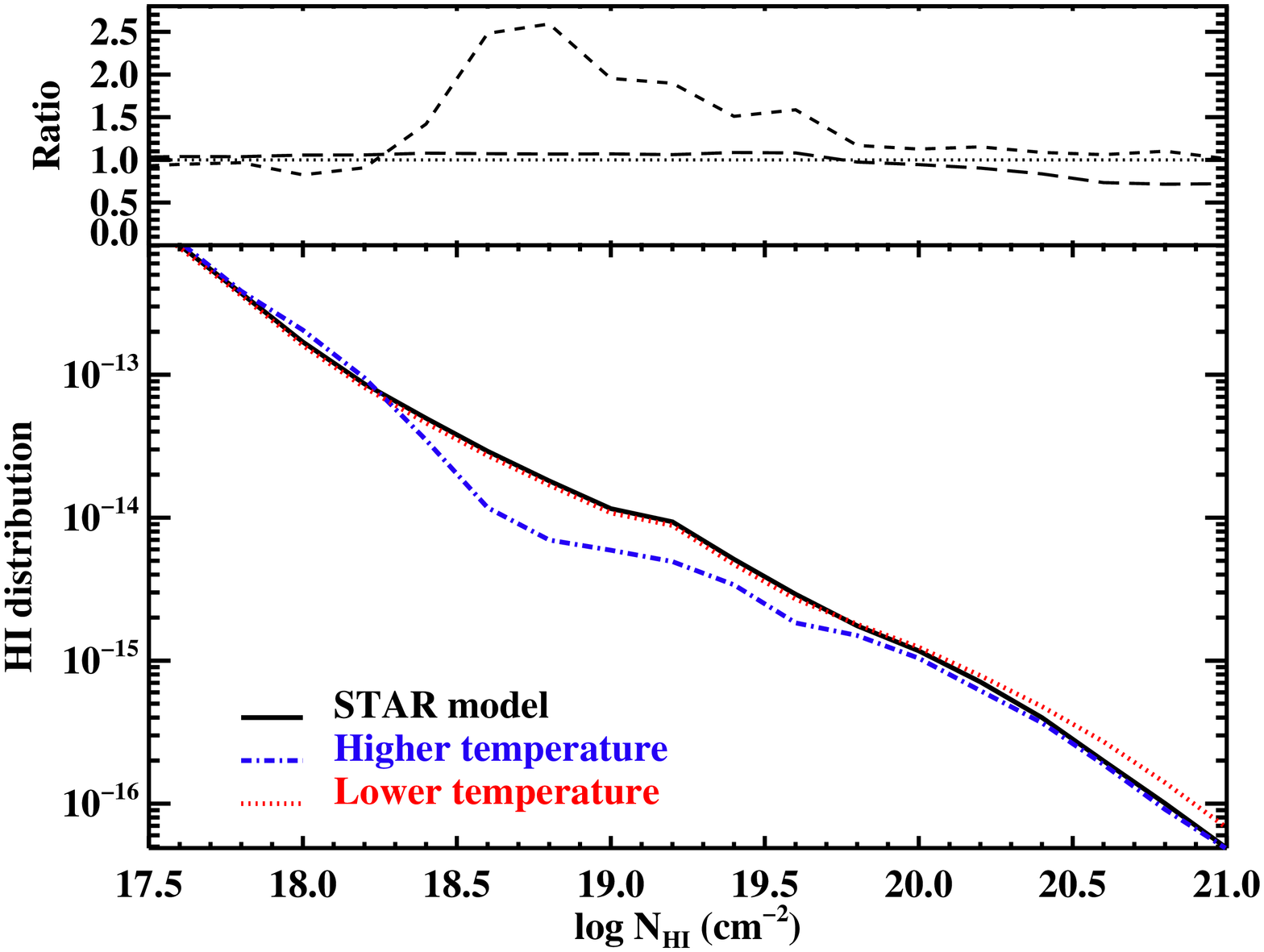}
\caption{Bottom panel: column density distribution function for the STAR model 
(black solid line), for a model with increased gas temperature (blue dash line)
and a model at lower temperature  (red dotted line).
Top panel: the ratio of the distribution for increased (decreased)
temperature and the STAR model is displayed with a dashed (long dashed) line. 
Identity between the models is marked with a dotted line.
Additional heating produces higher ionization in the interval $\log N_{\rm HI} = 18.3-19.8~\rm cm^{-2}$, but 
the covering factor of optically thick gas is nearly unperturbed, owing to a rising \fnx\ 
towards lower column densities. Lowering the temperature, little differences (less than a factor of two) 
are visible since the UV radiation is the dominant source of ionization in unshielded regions,
while shielded regions are already cold and neutral.}
\label{fig:photheat}
\end{figure}

\subsection{Uncertainties and future improvements}

Although our model significantly improves upon simpler RT calculations, 
few approximations contribute the final error budget. For example, these simulations do not include the 
presence of an active galactic nucleus and additional ionization from this harder spectrum
is neglected in this work.  Similarly, we include only photoionization at 912\AA\  due to stars and  
we do not considering harder SEDs that could enhance the ionization via heating. Similarly,
He recombination emission is neglected. Finally, we ignore the possibility that these galaxies could lie in
the proximity of other ionizing sources, such as QSOs  \citep[e.g.][]{can05}. We note the these effects would reduce 
further the neutral gas fraction and enhance the effects of ionizing radiation on the neutral 
volume and column density.

In the previous sections, we have clearly shown that the inclusion of both the UVB and local sources 
shapes the distribution of gas volume and column density at the low and high densities.
However, the exact amplitude of these effects is proportional to the intensity of the input 
radiation field which is subject to uncertainties, as discussed in Section \ref{sec:rtpost}. 
Heating from the UVB and  local sources is already 
included in these simulations, but during the post-processing we do not update the gas 
temperature \citep[see a discussion in][]{can10b}. And a sensible determination of 
the temperature has obvious implications for collision ionization and emission properties \citep{fau10}. 

We try to quantify the impact that an increase in temperature has on our results by
performing a simple test with MW3 at $z\sim2.3$, using the STAR model. 
After having increased by 50\% the temperature in photo-ionized cells with  $x_{\rm HI}<0.8$, 
we recompute the neutral fraction for CIE ($x_{\rm HI,cie}$). Then,  
we derive the projected \HI\ distribution function using an updated
neutral fraction $x_{\rm HI,new}=\min(\sqrt{x_{\rm HI,old}x_{\rm HI,cie}};x_{\rm HI,old};x_{\rm HI,cie})$,
defined in this way to favor high ionization.
Results are summarized in Figure \ref{fig:photheat}. In the bottom panel, we show the column density 
distribution function for the original STAR model (black solid line) and for the 
model at higher temperature (blue dash line). In the top panel we display the ratio of the two distributions.
Additional heating produces higher ionization in the interval $\log N_{\rm HI} = 18.3-19.8~\rm cm^{-2}$, 
yielding a flatter \fnx. The cross section of SLLSs is reduced by a factor of $\sim 2$,
increasing even further the discrepancy with observations.
The DLA cross section is decreased by no more than $20\%$ and the effect on the cumulative 
covering factor of LLSs is small, owing to a rising \fnx\ towards lower column densities.
This simple test suggests that a moderate increase is temperature does not have a major impact on the 
result presented in this paper. Conversely, if cooling is more effective than what assumed 
in these simulations, hydrogen is less ionized by electron collisions.
In Figure \ref{fig:photheat}, we shown the hydrogen distribution (red dotted line) for a model in which   
the temperature is reduced to 50\% of its value in all the cells with $n_{\rm H}>0.05$ \cmm, i.e. partially 
or totally shielded. 
Gas below this density is highly photoionized and the ionization state is 
insensitive to the gas temperature. In this case, the ionization fraction is assumed to be 
$x_{\rm HI,new}=x_{\rm HI,old}$ for  $n_{\rm H}<0.05$ \cmm\ and   $x_{\rm HI,new}=x_{\rm HI,cie}$
for $n_{\rm H}\ge 0.05$ \cmm. Since gas above $n_{\rm H} \sim 0.05$ \cmm\
is already at low temperature, little difference (less than a factor of two) is visible 
in the final \HI\ distribution.

\section{Tables}\label{app:table}
 
\begin{table*}
\caption{Summary of the baryonic and dark matter properties of the \ngal\ galaxies 
used in this study.}\label{galprop}
\centering
\begin{tabular}{c c c c c c c c c c c c c c c c c}
\hline
Redshift & &MW1 & MW2 & MW3 & MW4 & MW5 & MW8 & MW9 & & MW1 & MW2 & MW3 & MW4 & MW5 & MW8 & MW9 \\
\hline
& & \multicolumn{7}{c}{Virial Mass ($10^{11}$ \msun)} & & \multicolumn{7}{c}{Virial Radius (kpc)}\\
\hline
4.00 &\vline&	 0.11  &0.91 &  0.29 &  1.26 &  6.09  & 0.62  &  0.19  &\vline& 15.00  &   30.00  & 20.50  &  33.50  & 56.50  &   26.25  &   18.25  \\
3.55 &\vline&	 0.26  &1.51 &  0.37 &  1.73 &  9.58  & 0.73  &  0.33  &\vline& 21.75  &   39.00  & 24.50  &  41.00  & 72.25  &   30.50  &   23.50  \\
3.17 &\vline&	 0.53  &3.14 &  1.06 &  2.73 & 11.70  & 1.13  &  0.48  &\vline& 29.75  &   57.00  & 38.00  &  51.75  & 84.25  &   38.50  &   29.00  \\
2.85 &\vline&	 1.13  &4.19 &  1.62 &  4.23 &   -   &  1.52 &   1.22 &\vline& 41.75  &   64.50  & 47.25  &  64.50  &	-    &   46.00  &   42.75  \\
2.57 &\vline&	 1.68  &5.04 &  2.66 &  5.81 &   -   &  1.80 &   1.31  &\vline& 51.50  &   74.25  & 60.50  &  77.25  &   -    &  52.25  &   47.00  \\
2.33 &\vline&	 3.34  &5.92 &  3.55 &  9.21 &   -   &  2.31 &   1.29  &\vline& 68.75  &   83.25  & 70.50  &  96.50  &   -    &  60.75  &   50.00  \\
2.13 &\vline&	 6.49  &7.64 &  5.73 & 12.01 &   -   &  2.54 &   1.32  &\vline& 91.25  &   97.25  & 88.00  &  112.25 &   -    &  66.75  &   53.50  \\
1.94 &\vline&	 8.83  &9.12 &  8.06 & 15.17 &   -   &  2.79 &   1.75  &\vline& 106.75 &   108.75 & 104.00 &  129.00 &   -    &  73.00  &   62.50  \\
1.78 &\vline&	10.20  &-    &  9.27 & -     &   -   &  2.85 &   2.87  &\vline& 118.50 &     -   & 114.75 &     -    &	-    &   77.75  &   77.50  \\
1.63 &\vline&	11.47  &-    & 11.81 & -     &   -   &  2.85 &   3.00  &\vline& 129.50 &     -   & 131.25 &     -    &	-    &   81.75  &   82.75  \\
1.50 &\vline&	13.03  &-    & 12.59 & -     &   -   &  2.86 &   3.39 &\vline& 141.75 &     -	 & 140.25 &     -    &	-    &   85.50  &   90.50  \\
1.38 &\vline&	13.69  &-    & 13.52 & -     &   -   &  5.10 &   4.31 &\vline& 150.50 &     -	 & 150.00 &     -    &	-    &   108.00 &   102.25 \\
\hline
& & \multicolumn{7}{c}{Star Formation Rate (\sfr)} & & \multicolumn{7}{c}{Stellar Mass ($10^{10}$ \msun)}\\
\hline
4.00 &\vline&   0.94 &  37.79  &   3.74  & 21.19  &   49.49   &   4.23 &   2.02   &\vline&   0.03  & 	1.62 &    0.20  &   1.22  &   6.25  &  0.63  &   0.13 \\
3.55 &\vline&   4.61 &  58.25  &   2.70  & 28.66  &  140.61   &   5.70 &   2.88   &\vline&   0.17  & 	3.12 &    0.26  &   1.69  &  10.66  &  0.80  &   0.28 \\
3.17 &\vline&   4.42 & 148.23  &  14.44  & 15.77  &   76.14   &   4.48 &   3.12   &\vline&   0.31  & 	7.52 &    0.86  &   2.79  &  13.32  &  1.31  &   0.41 \\
2.85 &\vline&  14.76 & 157.48  &   5.54  & 38.11  & 	-     &  40.65 &   6.98   &\vline&   0.87  &   11.16 &    1.14  &   3.91  &    -    &  1.70  &   1.27 \\
2.57 &\vline&   9.65 & 104.36  &  25.95  & 48.22  & 	-     &   4.73 &   8.28   &\vline&   1.14  &   14.24 &    1.76  &   4.94  &    -    &  1.85  &   1.48 \\
2.33 &\vline&  35.71 &  99.77  &  30.49  & 44.21  & 	-     &   7.16 &   7.14   &\vline&   2.25  &   17.51 &    2.63  &   8.69  &    -    &  2.21  &   1.61 \\
2.13 &\vline&  52.47 & 143.23  &  46.77  & 73.75  & 	-     &   5.03 &   2.45   &\vline&   5.54  &   22.26 &    4.34  &  12.07  &    -    &  2.42  &   1.70 \\
1.94 &\vline&  62.03 & 187.86  &  71.46  & 77.22  & 	-     &  13.63 &   2.38   &\vline&   8.19  &   27.48 &    7.14  &  15.38  &    -    &  2.96  &   1.94 \\
1.78 &\vline& 108.52 &   -     &  84.00  & -	 &     -     &   6.26 &   5.44   &\vline&  10.57  &	-   &	 8.90  &  -	&    -    &  3.04  &   3.19 \\
1.63 &\vline&  20.71 &   -     &  50.19  & -	 &     -     &   3.42 &  17.68   &\vline&  11.43  &	-   &	12.10  &  -	&    -    &  3.09  &   3.46 \\
1.50 &\vline&  25.65 &   -     &  39.10  & -	 &     -     &   3.61 &   4.25   &\vline&  12.80  &	-   &	13.15  &  -	&    -    &  3.20  &   3.98 \\
1.38 &\vline&  23.19 &   -     &  45.64  & -	 &     -     &   5.31 &   6.77   &\vline&  14.19  &	-   &	14.38  &  -	&    -    &  6.42  &   5.49 \\
\hline								    
& & \multicolumn{7}{c}{Dark Matter Mass ($10^{11}$ \msun)} & & \multicolumn{7}{c}{Hydrogen Mass ($10^{10}$ \msun)}\\
\hline
4.00 &\vline&  0.10  & 0.66 &   0.25 &   1.07 &  5.19 &  0.53 &  0.16 &\vline&  0.080& 0.879& 0.132& 0.663& 2.761& 0.291& 0.140\\
3.55 &\vline&  0.23  & 1.06 &   0.33 &   1.49 &  8.13 &  0.61 &  0.29 &\vline&  0.167& 1.431& 0.174& 0.791& 3.921& 0.361& 0.182\\
3.17 &\vline&  0.46  & 2.04 &   0.93 &   2.32 &  9.86 &  0.95 &  0.41 &\vline&  0.320& 3.517& 0.480& 1.265& 5.041& 0.566& 0.232\\
2.85 &\vline&  0.98  & 2.66 &   1.42 &   3.62 &   -   &  1.29 &  1.05 &\vline&  0.619& 4.138& 0.906& 2.156&    - & 0.635& 0.494\\
2.57 &\vline&  1.46  & 3.24 &   2.30 &   5.01 &   -   &  1.55 &  1.11 &\vline&  0.996& 3.803& 1.823& 3.057&    - & 0.614& 0.447\\
2.33 &\vline&  2.88  & 3.72 &   3.07 &   7.84 &   -   &  2.01 &  1.09 &\vline&  2.390& 4.502& 2.249& 5.065&    - & 0.862& 0.419\\
2.13 &\vline&  5.54  & 4.65 &   4.95 &  10.23 &   -   &  2.20 &  1.11 &\vline&  3.990& 7.622& 3.453& 5.778&    - & 1.058& 0.437\\
1.94 &\vline&  7.54  & 5.43 &   6.92 &  12.88 &   -   &  2.39 &  1.48 &\vline&  4.648& 9.399& 4.181& 7.590&    - & 1.109& 0.767\\
1.78 &\vline&  8.70  &  -   &   7.98 &  -      &   -   &  2.43 &  2.42 &\vline&  4.449&    - & 4.009&	 - &	- & 1.136& 1.312\\
1.63 &\vline&  9.85  &	-   &  10.11 &  -      &   -   &  2.42 &  2.53 &\vline&  4.791&    - & 4.926&	 - &	- & 1.162& 1.255\\
1.50 &\vline& 11.17  &	-   &  10.77 &  -      &   -   &  2.42 &  2.86 &\vline&  5.763&    - & 5.097&	 - &	- & 1.177& 1.382\\
1.38 &\vline& 11.68  &	-   &  11.54 &  -      &   -   &  4.27 &  3.59 &\vline&  5.920&    - & 5.413&	 - &	- & 1.895& 1.701\\
\hline
\end{tabular}
\end{table*}

\begin{table*}
\caption{Summary of the neutral gas properties of the \ngal\ galaxies used in this 
study.}\label{gasprop}
\centering
\begin{tabular}{c c c c c c c c c c c c c c c c c}
\hline
Redshift & &MW1 & MW2 & MW3 & MW4 & MW5 & MW8 & MW9 & & MW1 & MW2 & MW3 & MW4 & MW5 & MW8 & MW9 \\
\hline
& & \multicolumn{7}{c}{\HI\ Mass ($10^{10}$ M$_\odot$) - UVB} & &\multicolumn{7}{c}{\HI\ Mass ($10^{10}$ M$_\odot$) - STAR}\\
\hline
4.00&\vline& 0.036& 0.565& 0.043& 0.326& 1.124& 0.158& 0.068&\vline& 0.026& 0.322& 0.026& 0.215& 0.637& 0.158& 0.059\\
3.55&\vline& 0.072& 0.926& 0.054& 0.271& 1.423& 0.186& 0.079&\vline& 0.041& 0.464& 0.023& 0.124& 0.891& 0.145& 0.060\\
3.17&\vline& 0.103& 2.210& 0.110& 0.389& 1.147& 0.302& 0.081&\vline& 0.069& 0.940& 0.049& 0.270& 0.589& 0.252& 0.057\\
2.85&\vline& 0.190& 2.631& 0.295& 0.676&   -  & 0.218& 0.187&\vline& 0.108& 1.281& 0.217& 0.410&   -  & 0.091& 0.187\\
2.57&\vline& 0.285& 2.085& 0.751& 1.071&   -  & 0.194& 0.120&\vline& 0.176& 1.169& 0.434& 0.651&   -  & 0.153& 0.078\\
2.33&\vline& 0.805& 2.230& 0.869& 1.963&   -  & 0.277& 0.074&\vline& 0.490& 1.081& 0.522& 1.494&   -  & 0.216& 0.044\\
2.12&\vline& 1.454& 3.950& 1.195& 1.548&   -  & 0.379& 0.115&\vline& 0.901& 1.983& 0.763& 0.977&   -  & 0.379& 0.087\\
1.94&\vline& 1.537& 4.902& 1.193& 1.902&   -  & 0.372& 0.230&\vline& 0.850& 2.427& 0.498& 1.252&   -  & 0.248& 0.230\\
1.78&\vline& 0.794&   -  & 0.922&   -  &   -  & 0.476& 0.488&\vline& 0.318&   -  & 0.323&   -  &   -  & 0.379& 0.415\\
1.63&\vline& 0.745&   -  & 0.885&   -  &   -  & 0.544& 0.452&\vline& 0.441&   -  & 0.369&   -  &   -  & 0.478& 0.324\\  			 
1.50&\vline& 1.207&   -  & 1.048&   -  &   -  & 0.579& 0.385&\vline& 0.810&   -  & 0.591&   -  &   -  & 0.487& 0.382\\  			 
1.38&\vline& 1.202&   -  & 0.818&   -  &   -  & 0.922& 0.560&\vline& 0.839&   -  & 0.407&   -  &   -  & 0.798& 0.559\\  			 
\hline								    
\end{tabular}					    
\end{table*}

\begin{table*}
\caption{Geometric mean of the $x_{\rm HI}$ vs $n_{\rm H}$ relation from the UVB model.}\label{tab:gm}
\centering
\begin{tabular}{c c c c c c c c c c c}
\hline
$\log n_{\rm H}$ (cm$^{-3}$)& $\log x_{\rm HI}$&\vline&$\log n_{\rm H}$ (cm$^{-3}$)&$\log x_{\rm HI}$&\vline&$\log n_{\rm H}$ (cm$^{-3}$)& $\log x_{\rm HI}$&\vline&$\log n_{\rm H}$ (cm$^{-3}$)& $\log x_{\rm HI}$\\
\hline
-4.750&-5.512E+00&\vline&-2.950&-3.959E+00&\vline&-1.150&-2.522E-01&\vline&0.650&-5.050E-03\\ 
-4.650&-5.554E+00&\vline&-2.850&-3.739E+00&\vline&-1.050&-2.097E-01&\vline&0.750&-4.217E-03\\ 
-4.550&-5.496E+00&\vline&-2.750&-3.506E+00&\vline&-0.950&-1.192E-01&\vline&0.850&-3.391E-03\\ 
-4.450&-5.475E+00&\vline&-2.650&-3.279E+00&\vline&-0.850&-1.012E-01&\vline&0.950&-2.770E-03\\ 
-4.350&-5.445E+00&\vline&-2.550&-3.054E+00&\vline&-0.750&-8.607E-02&\vline&1.050&-2.307E-03\\ 
-4.250&-5.448E+00&\vline&-2.450&-2.820E+00&\vline&-0.650&-7.321E-02&\vline&1.150&-1.850E-03\\ 
-4.150&-5.401E+00&\vline&-2.350&-2.579E+00&\vline&-0.550&-6.224E-02&\vline&1.250&-1.530E-03\\ 
-4.050&-5.360E+00&\vline&-2.250&-2.323E+00&\vline&-0.450&-5.388E-02&\vline&1.350&-1.256E-03\\ 
-3.950&-5.293E+00&\vline&-2.150&-2.050E+00&\vline&-0.350&-4.524E-02&\vline&1.450&-9.829E-04\\ 
-3.850&-5.212E+00&\vline&-2.050&-1.764E+00&\vline&-0.250&-3.801E-02&\vline&1.550&-6.561E-04\\ 
-3.750&-5.128E+00&\vline&-1.950&-1.432E+00&\vline&-0.150&-3.123E-02&\vline&1.650&-4.945E-04\\ 
-3.650&-5.020E+00&\vline&-1.850&-1.098E+00&\vline&-0.050&-2.520E-02&\vline&1.750&-3.221E-04\\ 
-3.550&-4.906E+00&\vline&-1.750&-8.472E-01&\vline& 0.050&-2.037E-02&\vline&1.850&-2.372E-04\\ 
-3.450&-4.775E+00&\vline&-1.650&-6.688E-01&\vline& 0.150&-1.579E-02&\vline&1.950&-1.611E-04\\ 
-3.350&-4.631E+00&\vline&-1.550&-5.363E-01&\vline& 0.250&-1.247E-02&\vline&2.050&-1.826E-04\\ 
-3.250&-4.479E+00&\vline&-1.450&-4.328E-01&\vline& 0.350&-1.015E-02&\vline&2.150&-8.635E-05\\ 
-3.150&-4.321E+00&\vline&-1.350&-3.546E-01&\vline& 0.450&-7.837E-03&\vline&	&	   \\	      
-3.050&-4.155E+00&\vline&-1.250&-2.978E-01&\vline& 0.550&-6.192E-03&\vline&	&	   \\
\hline									 	      			       
\end{tabular}					    
\end{table*}														 

\begin{table*}
\caption{DLAs and LLSs covering factor from the STAR model. The top table lists values for the streams and galaxies, while 
the bottom one lists values for the streams alone.}\label{tab:crsc}
\centering
\begin{tabular}{c c c c c c c c c c c c c c c c c}
\hline
Redshift & &MW1 & MW2 & MW3 & MW4 & MW5 & MW8 & MW9 & & MW1 & MW2 & MW3 & MW4 & MW5 & MW8 & MW9 \\
\hline
& & \multicolumn{7}{c}{DLA covering factor in \% at $2R_{\rm vir}$ (streams and galaxies)} & & \multicolumn{7}{c}{LLS covering factor in \% at $2R_{\rm vir}$ (streams and galaxies)}\\
\hline
4.00&\vline& 1.78  &  3.47&  1.09 &  2.68&  1.42  &  1.57   &  1.55 &\vline& 10.80 &12.54 &  6.70 &  12.19 &   12.43  & 11.20  &  11.72 \\
3.55&\vline& 1.03  &  3.65&  0.79 &  1.35&  1.22  &  0.85   &  1.02 &\vline&  7.92 & 9.02 &  6.24 &   8.56 &   11.11  &  9.35  &   8.27 \\
3.17&\vline& 0.94  &  1.95&  0.60 &  1.76&  1.08  &  0.88   &  0.76 &\vline&  8.03 & 5.08 &  2.35 &  13.12 &    8.68  &  4.71  &   6.41 \\
2.85&\vline& 0.79  &  2.65&  1.42 &  1.86&  -	  &  0.28   &  0.62 &\vline&  4.88 & 5.81 & 10.41 &  11.03 &	-     &  1.69  &   5.36 \\
2.57&\vline& 1.12  &  2.84&  0.88 &  1.60&  -	  &  0.31   &  0.29 &\vline& 10.25 & 7.13 &  6.52 &  10.30 &	-     &  4.08  &   2.98 \\
2.33&\vline& 1.37  &  2.22&  1.19 &  1.54&  -	  &  0.31   &  0.49 &\vline&  9.87 & 6.89 &  8.92 &   9.11 &	-     &  3.80  &   4.72 \\
2.12&\vline& 1.07  &  2.23&  0.75 &  1.00&  -	  &  0.59   &  0.82 &\vline&  6.64 & 6.79 &  5.74 &   7.03 &	-     &  4.03  &   5.06 \\
1.94&\vline& 0.62  &  2.07&  0.38 &  0.72&  -	  &  0.28   &  0.73 &\vline&  4.91 & 6.85 &  3.19 &   4.85 &	-     &  2.03  &   5.09 \\
1.78&\vline& 0.23  &	- &  0.17 & -	 &  -	  &  0.25   &  0.66 &\vline&  1.98 & -    &  1.83 &   -    &	-     &  1.63  &   4.20 \\
1.63&\vline& 0.36  &	- &  0.21 & -	 &  -	  &  0.51   &  0.30 &\vline&  4.25 & -    &  2.84 &   -    &	-     &  2.46  &   2.59 \\
1.50&\vline& 0.35  &	- &  0.27 & -	 &  -	  &  0.51   &  0.35 &\vline&  2.97 & -    &  2.82 &   -    &	-     &  2.87  &   5.04 \\
1.38&\vline& 0.26  &	- &  0.32 & -	 &  -	  &  0.48   &  0.44 &\vline&  1.84 & -    &  2.40 &   -    &	-     &  1.66  &   4.21 \\
\hline
& & \multicolumn{7}{c}{DLA covering factor  in \% at $2R_{\rm vir}$ (streams only)} & & \multicolumn{7}{c}{LLS covering factor in \%  at $2R_{\rm vir}$ (streams only)}\\
\hline
4.00&\vline& 1.47   & 2.94  & 0.82  &  2.41 &  0.77 &  0.95&  1.29 &\vline&  10.55  & 12.05 &  6.30 &  11.63&  11.82&	10.60&  11.42\\
3.55&\vline& 0.70   & 3.02  & 0.52  &  0.98 &  0.68 &  0.34&  0.74 &\vline&   7.46  &  8.28 &  5.78 &   7.95&  10.41&	 8.75&   7.83\\
3.17&\vline& 0.62   & 1.48  & 0.36  &  1.28 &  0.39 &  0.33&  0.57 &\vline&   7.51  &  4.41 &  1.87 &  12.45&   7.85&	 3.96&   6.09\\
2.85&\vline& 0.41   & 2.24  & 1.00  &  1.27 &  -    &  0.05&  0.37 &\vline&   4.17  &  5.23 &  9.67 &  10.24&	-   &	 1.13&   4.85\\
2.57&\vline& 0.69   & 2.40  & 0.32  &  0.91 &  -    &  0.03&  0.15 &\vline&   9.73  &  6.58 &  5.51 &   9.34&	-   &	 3.40&   2.49\\
2.33&\vline& 0.70   & 1.89  & 0.54  &  0.95 &  -    &  0.08&  0.15 &\vline&   8.99  &  6.42 &  7.88 &   8.29&	-   &	 3.13&   4.04\\
2.12&\vline& 0.50   & 1.83  & 0.32  &  0.50 &  -    &  0.15&  0.26 &\vline&   5.66  &  6.31 &  5.10 &   6.26&	-   &	 3.31&   4.21\\
1.94&\vline& 0.22   & 1.75  & 0.16  &  0.44 &  -    &  0.03&  0.37 &\vline&   4.23  &  6.41 &  2.60 &   4.38&	-   &	 1.56&   4.50\\
1.78&\vline& 0.07   & -     & 0.04  &  -    &  -    &  0.03&  0.34 &\vline&   1.62  &  -    &  1.44 &	-   &	-   &	 1.02&   3.42\\
1.63&\vline& 0.08   & -     & 0.07  &  -    &  -    &  0.24&  0.05 &\vline&   3.64  &  -    &  2.47 &	-   &	-   &	 1.98&   2.04\\
1.50&\vline& 0.07   & -     & 0.07  &  -    &  -    &  0.15&  0.12 &\vline&   2.30  &  -    &  2.31 &	-   &	-   &	 2.22&   4.30\\
1.38&\vline& 0.02   & -     & 0.14  &  -    &  -    &  0.28&  0.24 &\vline&   1.34  &  -    &  1.97 &	-   &	-   &	 1.19&   3.64\\
\hline
\end{tabular}					    
\end{table*}		

\begin{table*}
\caption{\fnx\ from the STAR model witin \rvir, 2\rvir\ and in the streams alone.}\label{tab:fnx}
\centering
\begin{tabular}{c c c c}
\hline
 log \NHI\ (\cmm) & \rvir & 2\rvir & Streams \\
\hline
   13.5 &$ 2.22\times 10^{-15}  $&$  8.18\times 10^{-15} $&$   8.30\times 10^{-15}$\\
   14.5 &$ 2.61\times 10^{-16}  $&$  1.73\times 10^{-15} $&$   1.74\times 10^{-15}$\\
   15.5 &$ 2.16\times 10^{-17}  $&$  1.35\times 10^{-16} $&$   1.36\times 10^{-16}$\\
   16.5 &$ 1.94\times 10^{-18}  $&$  5.98\times 10^{-18} $&$   6.06\times 10^{-18}$\\
   17.5 &$ 9.10\times 10^{-20}  $&$  3.03\times 10^{-19} $&$   3.00\times 10^{-19}$\\
   18.5 &$ 2.91\times 10^{-21}  $&$  9.72\times 10^{-21} $&$   9.52\times 10^{-21}$\\
   19.5 &$ 2.64\times 10^{-22}  $&$  7.70\times 10^{-22} $&$   7.58\times 10^{-22}$\\
   20.5 &$ 2.65\times 10^{-23}  $&$  6.43\times 10^{-23} $&$   5.46\times 10^{-23}$\\
   21.5 &$ 1.38\times 10^{-24}  $&$  2.55\times 10^{-24} $&$   1.17\times 10^{-24}$\\
   22.5 &$ 7.80\times 10^{-27}  $&$  1.26\times 10^{-26} $&$   1.78\times 10^{-28}$\\
   23.5 &$ 9.92\times 10^{-30}  $&$  1.24\times 10^{-29} $&$	   - $\\
   24.5 &$ 4.08\times 10^{-32}  $&$  4.08\times 10^{-32} $&$	   - $\\
\hline				  	        			  
\end{tabular}					    
\end{table*}														 

\begin{table*}
\caption{Average DLA cross sections within \rvir\ for all the halos at all redshifts.}\label{tab:dlacrss}
\centering
\begin{tabular}{c c c c c}
\hline
Halo Mass &\multicolumn{2}{c}{Total cross section} &\multicolumn{2}{c}{Central galaxy only} \\
 (\msun) &\multicolumn{2}{c}{(kpc$^2$)} &\multicolumn{2}{c}{(kpc$^2$)} \\
\hline
     &Mean &  Standard deviation &Mean&  Standard deviation \\
\hline
$3.6\times10^{10}$&46 & 24& 30& 21\\
$8.6\times10^{10}$&138& 93& 82& 58\\
$1.6\times10^{11}$&209&170&104& 86\\
$2.6\times10^{11}$&312&225&182&145\\
$3.9\times10^{11}$&723&396&251&184\\
$6.8\times10^{11}$&623&498&215&252\\
$1.1\times10^{12}$&522&252&216&163\\
\hline				  	        			  
\end{tabular}					    
\end{table*}

\begin{table*}
\caption{Equivalent widths for stacked Ly$\alpha$ and \SiIIl1260 
absorption profiles at different impact parameters.}\label{tab:ew}
\centering
\begin{tabular}{c c c c c c}
\hline
Radius & W(Ly$\alpha$) $z\sim 2.3$&W(\SiII) $z\sim 2.3$&W(Ly$\alpha$) $z\sim 3.2$&W(\SiII) $z\sim 3.2$&W(\SiII) enriched \\
(kpc) & (\AA) &  (\AA) &  (\AA) & (\AA) & (\AA) \\
\hline
0-10    &7.753 & 0.295 & 6.454&0.200&0.329\\
10-41   &2.053 & 0.050 & 1.300&0.023&0.082\\
41-82   &0.875 & 0.010 & 0.578&0.003&0.023\\
82-125  &0.679 & 0.006 & 0.390&0.001&0.014\\
125-200 &0.465 & 0.003 &    - &   - &0.006\\
\hline				  	        			  
\end{tabular}					    
\end{table*}

\input{journal.def}

\label{lastpage}

\end{document}